\documentclass[10pt,twocolumn,oneside,final]{IEEEtran}
\pdfoutput=1

\usepackage{cite}
\usepackage{graphicx}
\usepackage{amsmath}
\usepackage{times}
\usepackage{latexsym}
\usepackage{bm}
\usepackage{amssymb}
\usepackage[center,footnotesize]{caption2}
\usepackage{array}
\usepackage{fancyhdr}
\usepackage{cite,graphicx,amsmath,amssymb}
\usepackage{subfigure}
\usepackage{citesort}
\usepackage{psfrag}
\usepackage{multirow}

\usepackage[numbers,sort&compress]{natbib}

\usepackage{algorithm}               
\usepackage{algorithmic}             
\usepackage{multirow}                
\usepackage{amsmath}
\usepackage{xcolor}

\ifCLASSINFOpdf

\else

\fi

\makeatother

\begin{document}
\title{RFVTM: A Recovery and Filtering Vertex Trichotomy Matching for Remote Sensing Image Registration}

\author{\IEEEauthorblockN{Ming Zhao, Bowen An, Yongpeng Wu, \emph{Member IEEE}, Huynh Van Luong, Andr\'{e} Kaup, \emph{Fellow, IEEE}}

\thanks{This work was supported in part by National Natural Science Foundation of China under Grant 61302132, the Shanghai Educational Development Foundation under Grant 13CG51, and the Ministry of Transportation Applied Basic Research Projects under Grant 2014329810060. }

\thanks{M. Zhao is with the Department of Logistics Engineering, Shanghai Maritime University, Shanghai, 201306, China (e-mail: mingzhao@shmtu.edu.cn).}

\thanks{B. W. An is with the Department of Information Engineering, Shanghai Maritime University, Shanghai, 201306, China (e-mail: bwan@shmtu.edu.cn). }

\thanks{Y. P. Wu is with Institute for Digital Communications, Friedrich-Alexander University Erlangen-N$\ddot{u}$rnberg, Cauerstr. 7, 91058 Erlangen, Germany (email: yongpeng.wu@fau.de). }

\thanks{H. V. Luong is with the Chair of Multimedia Communications and Signal Processing, Friedrich-Alexander University Erlangen-N$\ddot{u}$rnberg, Cauerstr. 7, 91058 Erlangen, Germany (e-mail: huynh.luong@fau.de). }

\thanks{A. Kaup is with the Chair of Multimedia Communications and Signal Processing, Friedrich-Alexander University Erlangen-N$\ddot{u}$rnberg, Cauerstr. 7, 91058 Erlangen, Germany (e-mail: andre.kaup@fau.de). }

}
\maketitle

\begin{abstract}
Reliable feature point matching is a vital yet challenging process in feature-based image registration. In this paper, a robust feature point matching algorithm called Recovery and Filtering Vertex Trichotomy Matching (RFVTM) is proposed to remove outliers and retain sufficient inliers for remote sensing images.
A novel affine invariant descriptor called vertex trichotomy descriptor is proposed on the basis of that geometrical relations between any of vertices and lines are preserved after affine transformations, which is constructed by mapping each vertex into trichotomy sets.
The outlier removals in Vertex Trichotomy Matching (VTM) are implemented by iteratively comparing the disparity of corresponding vertex trichotomy descriptors.
Some inliers mistakenly validated by a large amount of outliers are removed in VTM iterations, and several residual outliers close to correct locations cannot be excluded with the same graph structures.
Therefore, a recovery and filtering strategy is designed to recover some inliers based on identical vertex trichotomy descriptors and restricted transformation errors. Assisted with the additional recovered inliers, residual outliers
can also be filtered out during the process of reaching identical graph for the expanded vertex sets.
Experimental results demonstrate the superior performance on precision and stability of this algorithm under various conditions, such as remote sensing images with large transformations, duplicated patterns, or inconsistent spectral content.
\end{abstract}

\providecommand{\keywords}[1]{\textbf{\textit{Index terms---}} #1}

\section{Introduction}
Image registration is a crucial preprocessing technology for image analysis, which has been widely applied in remote sensing, computer vision, medical imaging, and map search \cite{J_BZitova_2003_IVC}, \cite{J_GY_2014_ITCSVT}.
Regarding remote sensing applications in particular, image registration employed in geosciences is frequently used in a wide variety of disciplines, such as volcanology \cite{R1,R2,R3,R4}, structural geology and paleoseismology \cite{R5,R6,R7}, and soil microtopography \cite{R8}.
Image registration aims to align two or more images with overlapping scenes taken at different times, by different sensors, or from different viewpoints.
Feature point matching, which establishes reliable correspondences between feature points extracted from reference images and sensed images, is a representative and challenging step in feature-based registration techniques.

\begin{figure}[!tp]
\centering
 \setlength{\abovecaptionskip}{0pt}
 \setlength{\belowcaptionskip}{0pt}
 \setlength{\intextsep}{8pt plus 3pt minus 2pt}
  \subfigure[]{
    \label{fig:mini:subfig:a}
    \begin{minipage}[c]{0.4\textwidth}
      \centering
      \includegraphics[width=2.5in]{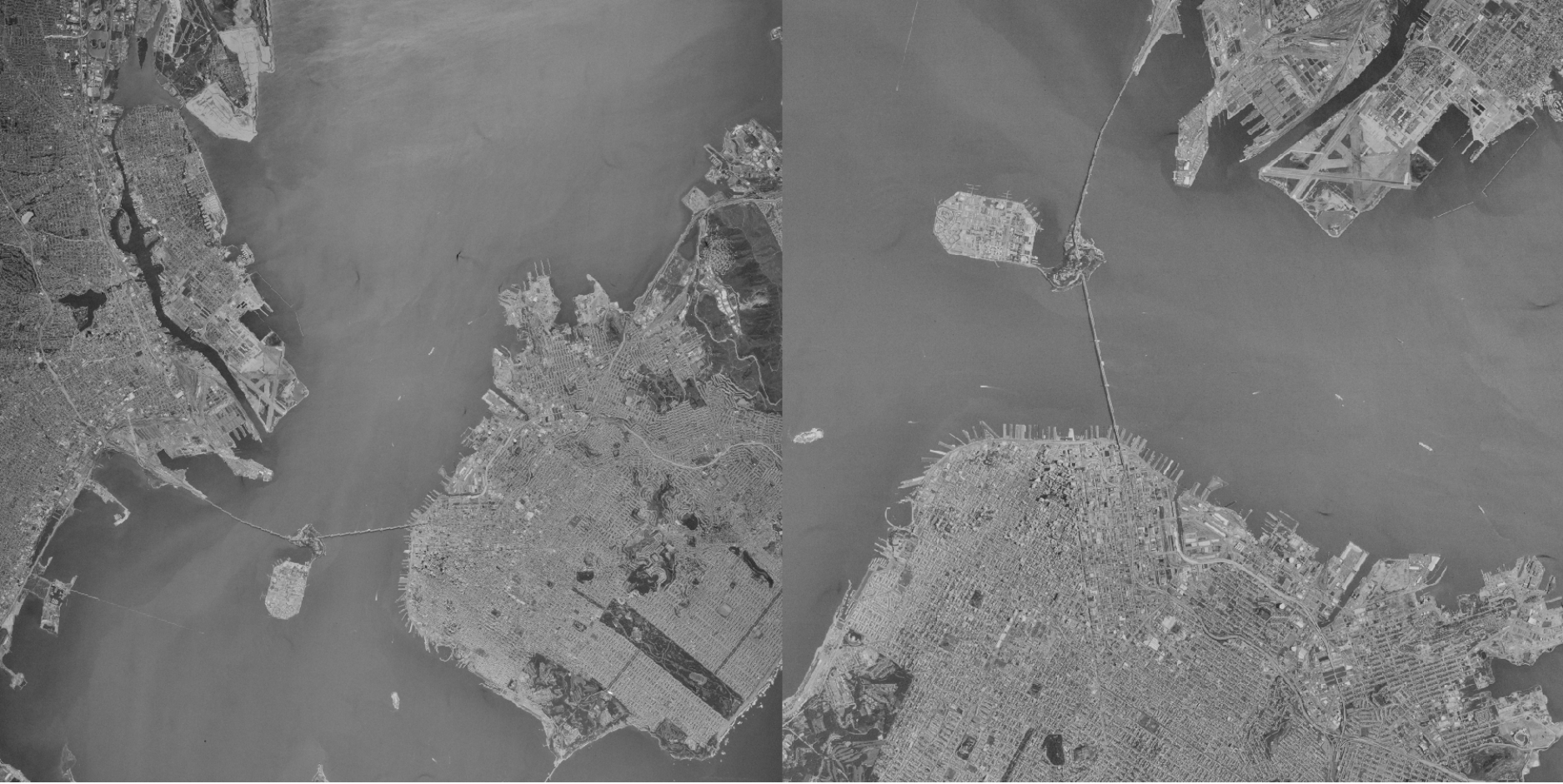}
    \end{minipage}}\\
  \subfigure[]{
    \label{fig:mini:subfig:b}
    \begin{minipage}[c]{0.4\textwidth}
      \centering
      \includegraphics[width=2.5in]{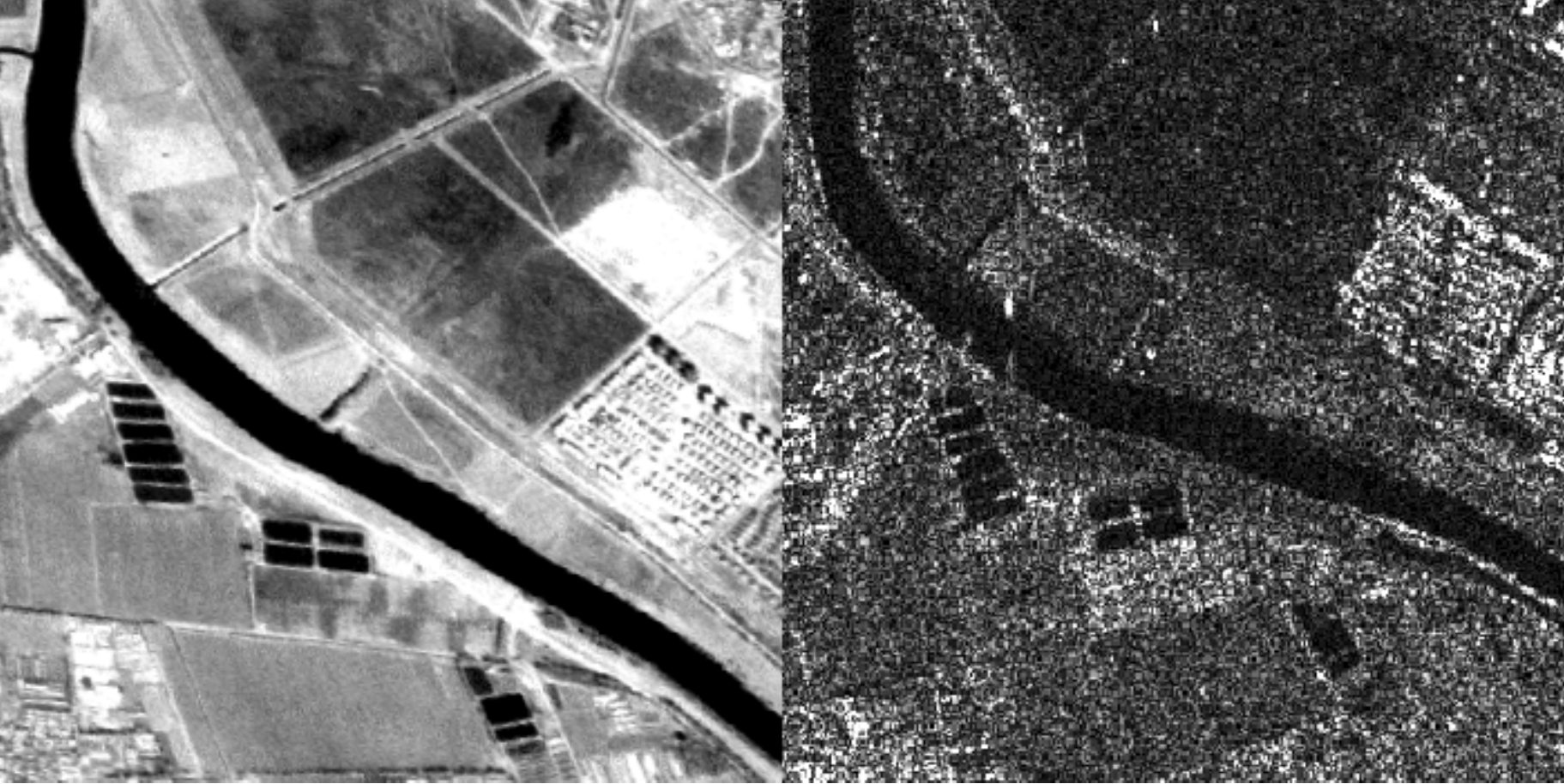}
    \end{minipage}}
  \captionstyle{normal}
  \caption{Examples of remote sensing images with large affine transformation. (a) Images with scale and rotations. (b) Images with shear deformation.}
  \label{fig-response}
\end{figure}

Several factors concerning the imaging conditions and sensor modalities may deteriorate the performance of feature point matching for remote sensing images.
First, large transformations between remote sensing images occur due to the great differences of remote sensors in resolutions, flight heights, or viewpoints.
Fig. \ref{fig-response} (a) and (b) illustrate examples of remote sensing images with rotation, scale deformation, and shear deformation.
Large transformations not only lead to the low overlapping area between images, but also cause great changes in geometrical features of corresponding areas, especially in the case of shear deformations.
Therefore, outliers are inevitably established from the non-overlapped areas.
Secondly, outliers occur very easily in feature point matching which only relies on local descriptors, especially for the images with duplicate or similar patterns.
Last but not least, inconsistent spectral content often exists in multispectral and multimodal images, where the intensities of pixels in the same region may be quite different. It is difficult to obtain reliable and sufficient correspondences according to the inconsistent image intensities. Therefore, it is necessary to explore proper approaches to overcome the problems caused by large affine transformations, duplicate patterns and inconsistent spectral content in feature point matching.

In this paper, a novel feature point matching approach named Recovery and Filtering Vertex Trichotomy Matching (RFVTM) is proposed for remote sensing images. Based on the global spatial relations of feature points, an affine invariant descriptor called vertex trichotomy is proposed, due to the fact that the geometrical relations between any vertices and lines are preserved after affine transformations. Therefore, vertex trichotomy descriptor is invariant under affine transformations, including rigid deformations (i.e., rotation, scaling, and translation) and non-rigid deformations (i.e., shear). This makes it superior to most existing graph descriptors which are only invariant to rigid transformation \cite{J_ZXLiu_2012_ITGRS,J_ZLS_2014_ITGRS,J_MZ_2015_IGRSL,J_MI_2012_ITGRS,J_WAr_2009_IVC}. To estimate the similarity of corresponding points, vertex trichotomy descriptor is constructed by mapping each vertex into trichotomy sets according to the spatial relations between vertices and trichotomy vectors.
Candidate outliers in Vertex Trichotomy Matching (VTM) are determined by iteratively comparing the disparity of corresponding vertex trichotomy descriptors.
It is noted that the similarity of corresponding graph structures is validated by all of vertices, which enables the outlier removal more reliable.
However, some inliers mistakenly validated by a large amount of outliers are easily to be removed for most of graph-based feature point matching methods.
Besides, several residual outliers cannot be excluded with pseudo isomorphic graphs, i.e., the corresponding graphs have the same graph descriptor but still include outliers that close to correct locations.
To deal with this, we design a recovery and filtering strategy to recover inliers from candidate outliers based on identical vertex trichotomy descriptors and restricted transformation errors. Assisted with the recovered inliers, both of the outliers among the recovered candidates and the outliers in the residual sets can be filtered out during the process of reaching identical graph for the expanded vertex sets.

The remainder of this paper is organized as follows: Section II gives a brief literature review for feature point matching. Section III presents problem formulations for graph-based feature point matching, where an affine invariant descriptor called vertex trichotomy is defined. Section IV introduces Vertex Trichotomy Matching to determine candidate outliers and a robust strategy to achieve inlier recovery and outlier filtering. Section V illustrates experimental results with representative remote sensing image pairs, and performance evaluation of the proposed algorithms is presented. Finally, Section VI presents the concluding remarks.
The notations used in this paper are summarized in Table \ref{table-summery}.

\begin{table}[!h]
\centering
  \captionstyle{normal}
  \setlength{\abovecaptionskip}{0pt}
  \setlength{\belowcaptionskip}{10pt}
\caption{SUMMERY OF NOTATIONS USED IN THIS PAPER}
\newsavebox{\tablebox}
\begin{lrbox}{\tablebox}
\begin{tabular}{l l}
\hline
${P_R} = \{ {p_i}\}$, ${P_S} = \{ {p_j'}\}$ & Extracted feature points from reference and sensed images \\
${\bf{C}}(i,j)$ &  Correspondence matrix of feature points \\
$T(\cdot )$ & Affine transformation \\
${\theta ^ * }$ &  Affine transformation coefficients \\
$\overline E\left( {{P_R},{P_S},{\bf{C}}}\right)$ & Global transformation error \\
${G_R}( P_R,{\bf{C}})$, ${G_S}( P_S,{\bf{C}})$ & Graph structures of reference and sensed images \\
$V = \{v_i\}$, $V' = \{v_i', \}$ & Initial corresponding vertices \\
$\phi(\cdot)$ & Graph descriptor\\

$\vec E\left( {i,j} \right)$ & Trichotomy vector starts from $v_i$  to $v_j$ \\
$\phi(G|\vec E)$, $\phi(G'|\vec E')$ & Corresponding  vertex trichotomy descriptor \\
$T_ +  \left( {i,j} \right)$, $T_ o  \left( {i,j} \right)$, $T_ -\left( {i,j} \right)$ & Three trichotomy sets for $\vec E\left( {i,j} \right)$ \\
${{\bf{M}}_{i \mapsto j}}[k]$, ${{\bf{M}}'_{i \mapsto j}}[k]$ & Reference and sensed trichotomy matrices\\
$\Delta {{\bf{M}}_{i \mapsto j}}$ & Disparity matrix \\
$(v_{d^{out}},v_{d^{out}}')$ & Candidate outliers \\
$\{V_{res},V_{res}'\}$ & Residual vertex sets \\
$\{V_{del},V_{del}'\}$ & Deleted vertex sets \\
$\{V_{rec},V_{rec}'\}$ & Recovered vertex sets \\

$O(\cdot)$ & Time complexity\\

\hline

\end{tabular}
\end{lrbox}
\scalebox{0.85}{\usebox{\tablebox}}
 \label{table-summery}
\end{table}

\section{LITERATURE REVIEW}\label{sec:model}
Numerous previous approaches of image registration have been developed.
One representative class of image registrations focusses on finding the matching information by comparing global similarity metrics for all pixels in the images, such as mutual information \cite{R10} or correlations \cite{R12}.
Another class of algorithms transforms images to a new domain, and then applies these global similarity metrics to the transformed images, where global features are more prominent \cite{R9}, \cite{R11}.
Yet another class of algorithms includes feature-based methods, which seeks the correspondence between the features of images. The adopted features must be salient, distinct, and stable, which can be regions, curves, or points.
Regarding feature point based registration in particular, feature point matching is a crucial step with the objective of establishing corresponding points and removing outliers, which can be broadly grouped into two main categories: feature point matching based on local feature similarity and feature point matching based on relations between feature points.
This section presents a brief review for these representative work of feature point matching.


\subsection{Feature Point Matching Based on Local Feature Similarity}
Feature point matching strategies based on local feature similarity are commonly adopted to find reliable corresponding points. The local features around points are compared to measure the similarity between the correspondences.
A basic idea for local feature similarity is to calculate the intensities within a close neighborhood, which is commonly known as an intensity patch.
However, besides high time complexity, this idea is sensitive to illumination changes and inconsistent spectral content in multispectral/multimodal images. Therefore, it may fail in registering remote sensing images acquired with different modalities or illumination conditions. Scale-invariant feature transform (SIFT) \cite{J_DGL_2004_IJCV} is a classical feature descriptor to extract distinctive invariant features from images, which is capable of performing reliable matches across a substantial range of affine distortion, addition of noise, and changes in illumination or 3-D viewpoints.
Various adapted versions of SIFT have been proposed to improve the performance of SIFT.
Principal component analysis (PCA)-SIFT is obtained by applying PCA on normalized gradient neighborhood.
Gradient location and orientation histogram \cite{J_KM_2005_ITPAMI} is computed on a log-polar grid and upon 17 sectors.
The compact feature vector of PCA-SIFT is derived from projecting the gradient histogram into the eigenspace, and the size of the resulting vector is significantly reduced with PCA \cite{J_FD_2015_ITGRS}.
Bilateral filter SIFT (BF-SIFT) \cite{J_SHW_2012_IGRSL} establishes feature matches for SAR images by replacing the Gaussian scale space with an anisotropic one. Despite of reducing false matches, BF-SIFT decreases image resolution and loses details of images. Adaptive binning (AB-SIFT) \cite{J_AS_2015_ITGRS} considers both locations and gradient orientations with an adaptive histogram quantization, which is able to increase the descriptor performance against local distortions.
Besides, harmonic analysis transforms have begun to be applied to the filed of SIFT-based feature point matching methods \cite{R11}.
For example, kernel affine invariant SIFT (KA-SIFT) \cite{KASIFT} matches the feature points detected from the different sub-images in the corresponding layer, which are obtained by the shearlet decomposition and affine-SIFT (ASIFT) algorithm \cite{ASIFT}.
Instead of building gradient histograms as SIFT-like descriptors, speeded-up robust feature (SURF) \cite{C_HB_PECCV} utilities the sums of Haar wavelet responses, and estimates the principal orientations within a sliding orientation window.
In contrast to gradient-based matching, ARRSI \cite{J_AW_2007_ITGRS} utilizes the phase-congruency moment-based patches as the local feature descriptor. The principle moments of phase congruency provide the representation of corner feature points. The local similarity between point candidates is determined by finding the correlation of maximum moments of phase congruency within a neighborhood centered on the candidates.
Generally, these local feature descriptors are available to obtain feature points in uncomplicated applications, such as homologous images with consistent and obvious features.
The main issue with them is in their high sensitivity to the local characteristics of features \cite{J_MI_2012_ITGRS}.
They are not well suited for the registration of remote sensing images with monotonous backgrounds, similar patterns, or inconsistent intensities of the same scene in multispectral and multimodal images \cite{J_ZXLiu_2012_ITGRS}.

\subsection{ Feature Point Matching Based on Relations Between Feature Points}
Numerous approaches for feature point matching have been motivated by the relations between extracted feature points to reject outliers.
Spatial-based models are commonly utilized as the relation constraints to match feature points.
The classic approach motivated by spatial transformation models is Random Sample Consensus (RANSAC) \cite{J_MAFichler_1081_CACM}, which estimates parameters of a transform model from two sets of feature points, and simultaneously distinguishes inliers from the outliers. However, it is a nondeterministic algorithm which does not work well when outliers account for a high proportion.
Another simple and fast method called Iterative Closest Point (ICP) \cite{J_PB_1992_ITPAMI} utilizes the closest form solution to assign a binary correspondence in each iteration. The transformation is then refined by the estimate of this correspondence. Under the assumption of initializing an adequate set of poses, it can converge to a global minimum for rigid transformations. Unfortunately, such an assumption is no longer valid in the case of non-rigid transformations, especially for large deformations.
To overcome the limitation of ICP, thin plate spline robust point matching (TPS-RPM) \cite{J_HC_2003_CVIU} solves both the corresponding matches and projection transformation parameters through deterministic annealing and soft-assignment for spatial mapping and outlier rejection.
In contrast with these above-mentioned methods, many probabilistic algorithms are developed by adopting the relations of density-based models. These methods attempt to represent feature point sets with density-based models, and match two density-based models instead of feature points.
The representative Gaussian mixture model (GMM) is employed in \cite{J_AM_2010_ITPAMI}, \cite{J_BJ_2011_ITPAMI} to represent the input point sets. A probabilistic modeling framework for rigid and non-rigid point set registration is proposed in \cite{J_BJ_2011_ITPAMI}. Point set registration is reformulated as the problem of aligning two Gaussian mixtures, where a statistical discrepancy measure between the two corresponding mixtures is minimized. Coherent Point Drift (CPD) algorithm \cite{J_AM_2010_ITPAMI} can be viewed as a special case in the generic framework presented in \cite{J_BJ_2011_ITPAMI}.
However, these probabilistic methods cannot perform well on the two point set with high percentages of outliers \cite{J_ZXLiu_2012_ITGRS}.

Recently, the concept of graph structures has been adopted to represent a higher level spatial relation between feature points. The similarity of the graph structures has been exploited to perform feature point matching.
Representative work in \cite{J_SB_2002_ITPAMI},  \cite{J_YZ_2006_ITPAMI},  \cite{J_MZ_2013_IGRSL}, \cite{J_MZ_2014_JIMW}, and \cite{J_WAr_2009_IVC}  explore the graphs of neighbor structures for feature point matching.
Shape context descriptor \cite{J_SB_2002_ITPAMI} captures the distribution of feature points by normalizing the histogram of vectors originating from one point to all other sample points.  The solution that minimizes the overall shape context distances is the optimal match between feature point sets under scaling and translation transformations.
Another simple graph matching interpretation proposed in \cite{J_YZ_2006_ITPAMI} preserves local neighborhood, under the assumptions that the neighborhood structures of points are generally well preserved under non-rigid transformation.
The edges of neighborhood structures for each point are constructed with their nearest neighbors. The optimal solution of this method is implemented by maximizing the number of matched edges of two neighborhood structures. However, this method may not be adequate to deal with significant changes between local neighborhood structures caused by large rotations and noise.
Graph Transformation Matching (GTM) \cite{J_WAr_2009_IVC} relies on finding consensus K Nearest Neighbor (KNN) graphs constructed with the restriction of average structures.
This method iteratively eliminates dubious matches by selecting the maximal disparities of edges connecting KNN points.
However, GTM has difficulties in obtaining reliable matches when outliers have the same local neighbor structures, or inliers have different neighbor structures interrupted by existing outliers.
Restricted Spatial Orders Constraints (RSOC) \cite{J_ZXLiu_2012_ITGRS} integrates the two-way spatial order constraints and the transformation error restrictions into KNN point matching. However, the convergence rates and accuracy depend on transformation models and the initial parameter settings. Also, the cyclic string matching for spatial orders is time consuming.
Weighted Graph Transformation Matching (WGTM) algorithm \cite{J_MI_2012_ITGRS} utilizes the angular distances between edges that connect a feature point to its KNN as the weight. The angular distance in WGTM is only invariant with respect to scales and rotations, but will be variant to shear deformations.
Delaunay Triangulation Matching \cite{J_MZ_2015_IGRSL} establishes Delaunay Triangulations according to the random incremental method \cite{C_YM_PWCSE}. Then Delaunay edge matrices are utilized to measure the connection relations between the vertices of triangles. The vertices with the maximum distinction of Delaunay edge connections are selected as candidate outliers.
Histogram of TAR Sample Consensus (HTSC) \cite{J_ZLS_2014_ITGRS} adopts the histogram of triangle area representation to find out correct corresponding triangles with three matched vertices from the feature point candidates.

Although extensive work as mentioned above has been done, feature point matching is still a challenging task, particularly for remote sensing images with large transformations, duplicated patterns, and inconsistent spectral content.
It is noted that most of graph structures cannot keep invariant with respect to shear deformations, which often occur in remote sensing images \cite{J_MI_2012_ITGRS}.
Besides that, each feature point in most existing graph descriptors is only validated by local graph structures with a small number of points, such as KNN adopted in GTM \cite{J_WAr_2009_IVC}, ROSC \cite{J_ZXLiu_2012_ITGRS}, and WGTM \cite{J_MI_2012_ITGRS}, triangles adopted in DTM \cite{J_MZ_2015_IGRSL} and HTAR \cite{J_ZLS_2014_ITGRS}. Therefore, inliers with a large amount of outlier in their local graphs are easily to be mistakenly removed in feature point matching.
In this paper, we explore an affine invariant graph descriptor and a recovery and filtering strategy for robust and reliable feature point matching.

\section{PROBLEM FORMULATION AND VERTEX TRICHOTOMY DESCRIPTOR}\label{sec:precoder}
In this section, we approach the problem of feature point matching for image registration as a graph matching problem solved by finding matched graphs that minimize global transformation errors.
Compared to most of local feature descriptors, the proposed vertex trichotomy descriptor is constructed by mapping each vertex into trichotomy sets according to the spatial relations between vertices and trichotomy vectors.
The similarity of corresponding graph structures is validated by all of vertices, which enables the outlier removal more reliable.

\subsection{ Problem Formulation for Graph based Point Matching }
Assuming that we have two images in which feature points have been extracted, denoted as ${P_R} = \{ {p_i}|i = 1,2,3,....{N_R}\}$ and ${P_S} = \{ {p_j'}|j = 1,2,3,....{N_S}\}$ respectively, and any feature point from one set corresponds to at most one feature point from the other set.
Define a correspondence matrix by ${\bf{C}}(i,j)\in \{0,1\}$  such that ${\bf{C}}(i,j)=1$  when $p_i$ corresponds to $p_j'$. Otherwise, ${\bf{C}}(i,j)=0$.
Feature point matching can be treated as finding an optimal correspondence matrix ${\bf{C}}(i,j)$ to minimize the global  transformation error $\overline E \left( {{P_R},{P_S},{\bf{C}}} \right)$. Here, $\overline E \left( {{P_R},{P_S},{\bf{C}}} \right)$  is defined by

\begin{equation} \label{eqn:achievable_2}
\overline E \left( {{P_R},{P_S},{\bf{C}}} \right) = \sqrt {\frac{{\sum\limits_{i = 1}^{{N_R}} {\sum\limits_{j = 1}^{{N_S}} {{\bf{C}}\left( {i,j} \right){{\left\| {T\left( {{p_i},\theta } \right) - p_j'} \right\|}^2}} } }}{{\sum\limits_{i = 1}^{{N_R}} {\sum\limits_{j = 1}^{{N_S}} {{\bf{C}}(i,j)} } }}}
\end{equation}%
where transformation parameters $\theta $  are estimated by at least three corresponding pairs of feature points under affine transformation $T(\cdot )$. Therefore, the optimal match $\tilde{\bf{C}}$  is the solution to the following optimization problem:

\begin{equation} \label{eqn:achievable_3}
{\tilde{\bf{C}}} = \mathop {\arg \min }\limits_{\bf{C}} \overline E \left( {{P_R},{P_S},{\bf{C}}} \right).
\end{equation}%

Graph structures are established by a higher level geometrical relations between corresponding feature points. Any feature points $(p_i,p_j')$ exist as vertices of graphs when $p_i$ corresponds to $p_j'$ , i.e., $\{(p_i,p_j')|{\bf{C}}(i,j)= 1,1 \le i \le {N_R},1 \le j \le {N_S}\} $. Graph matching is trying to find two matched graphs ${G_R}( P_R,{\bf{C}})$  and ${G_S}( P_S,{\bf{C}})$ from the reference image and the sensed image respectively, such that corresponding vertices of graphs are matched in pairs.
Graph descriptor $\phi(\cdot)$ is explored in graph based methods to describe the features of graph structures, such as K Nearest Neighbours (KNN) \cite{J_WAr_2009_IVC}, Histogram of Triangle Area Representation (HTAR) \cite{J_ZLS_2014_ITGRS}.
If all of the corresponding vertices of graphs are real matches, the descriptors of ${G_R}( P_R,{\bf{C}})$ and ${G_S}( P_S,{\bf{C}})$ should be identical.
Therefore, the optimal problem of feature point matching can be simplified as finding two matched graphs that minimize global transformation error as follows:

\begin{multline} \label{eqn:achievable_4}
{\tilde{\bf{C}} = \mathop {\arg \min }\limits_{\bf{C}} \overline E \left( {{G_R},{G_S},{\bf{C}}} \right), \ s.t. \ {\rm{ }}\phi \left( {{G_R}\left( {{P_R},{\bf{C}}} \right)} \right)} =\\
 { \phi \left( {{G_S}\left( {{P_S},{\bf{C}}} \right)} \right).}
\end{multline}%

\subsection{ Affine Invariant Graph Descriptor Based on Vertex Trichotomy }
An affine transformation is composed of linear transformations including rotation, scaling, translation, shear deformations, and finite combinations thereof. Based on the consensus that affine transformation preserves collinearity and parallelity \cite{J_ZLS_2014_ITGRS}, the followings can be derived:

\begin{itemize}
\item[1)] All points lying on a line initially still lie on a line after affine transformation.
\item[2)] Any points lying on the same side of a line remain the same side after affine transformation.
\end{itemize}

We assume there are two initial sets of match correspondences between the two images: $V = \{v_1,v_2, \ldots, v_i, \ldots ,v_N \}$  and $V' = \{v_1',v_2', \ldots, v_i', \ldots ,v_N' \}$  of size $N \le \min \left\{ {{N_R},{N_S}} \right\}$, where $v_i$ matches $v_i'$.
The proposed vertex trichotomy descriptor $\phi(G|\vec E) = \left( {T_ +  ,T_o ,T_ -  } \right)$ is created according to the following definitions:

\begin{itemize}
\item[1)] Define a trichotomy vector starts from $v_i$  to $v_j$  as $\vec E\left( {i,j} \right)$, $\forall 1\le i,j\le N$.
\item[2)] Define $T_ +  \left( {i,j} \right)$, $T_ o  \left( {i,j} \right)$, and $T_ -\left( {i,j} \right)$ as three trichotomy sets for $\vec E\left( {i,j} \right)$. Any of the vertices $v_k$ belongs to one of these trichotomy sets relying on the spatial relations between $v_k$ and $\vec E\left( {i,j} \right)$, which can be obtained by computing the determinant of the following matrix

\begin{equation} \label{eqn:achievable_4}
\det \left( {v_i ,v_j ,v_k } \right) = \left| {\begin{array}{*{20}c}
   {x_{v_i } } & {x_{v_j } } & {x_{v_k } }  \\
   {y_{v_i } } & {y_{v_j } } & {y_{v_k } }  \\
   1 & 1 & 1  \\
\end{array}} \right|
\end{equation}%
where $\left( {x_{v_i } ,y_{v_i } } \right)$, $\left( {x_{v_j } ,y_{v_j } } \right)$, and $\left( {x_{v_k } ,y_{v_k } } \right)$ are the coordinates of $v_i$, $v_j$, and $v_k$ respectively. $v_k  \in T_ +  \left( {i,j} \right)$, when ${\det \left( {v_i ,v_j ,v_k } \right) > 0}$;  $v_k  \in T_o \left( {i,j} \right)$, when ${\det \left( {v_i ,v_j ,v_k } \right) = 0}$;  Otherwise,  $v_k  \in T_ -  \left( {i,j} \right)$, when ${\det \left( {v_i ,v_j ,v_k } \right) < 0}$.

\end{itemize}

We establish the corresponding descriptors $\phi(G|\vec E) = \left( {T_ +  ,T_o ,T_ -  } \right)$ and $\phi(G'|\vec E') = \left( {T_ + ' ,T_o' ,T_ - '} \right)$ according to the above definitions.
$\phi (\left. G \right|\vec E) = \phi (\left. {G' } \right|\vec E' )$ can be obtained when all of the vertices are correctly matched in pairs.
If this is not the case (in general), the disparity of corresponding graph descriptors exist between the two graph structures.

\begin{figure}[!h]
\centering
 \setlength{\abovecaptionskip}{0pt}
 \setlength{\belowcaptionskip}{0pt}
 \setlength{\intextsep}{8pt plus 3pt minus 2pt}
  \subfigure[]{
    \label{fig:mini:subfig:a}
    \begin{minipage}[c]{0.45\textwidth}
      \centering
      \includegraphics[width=2.5in]{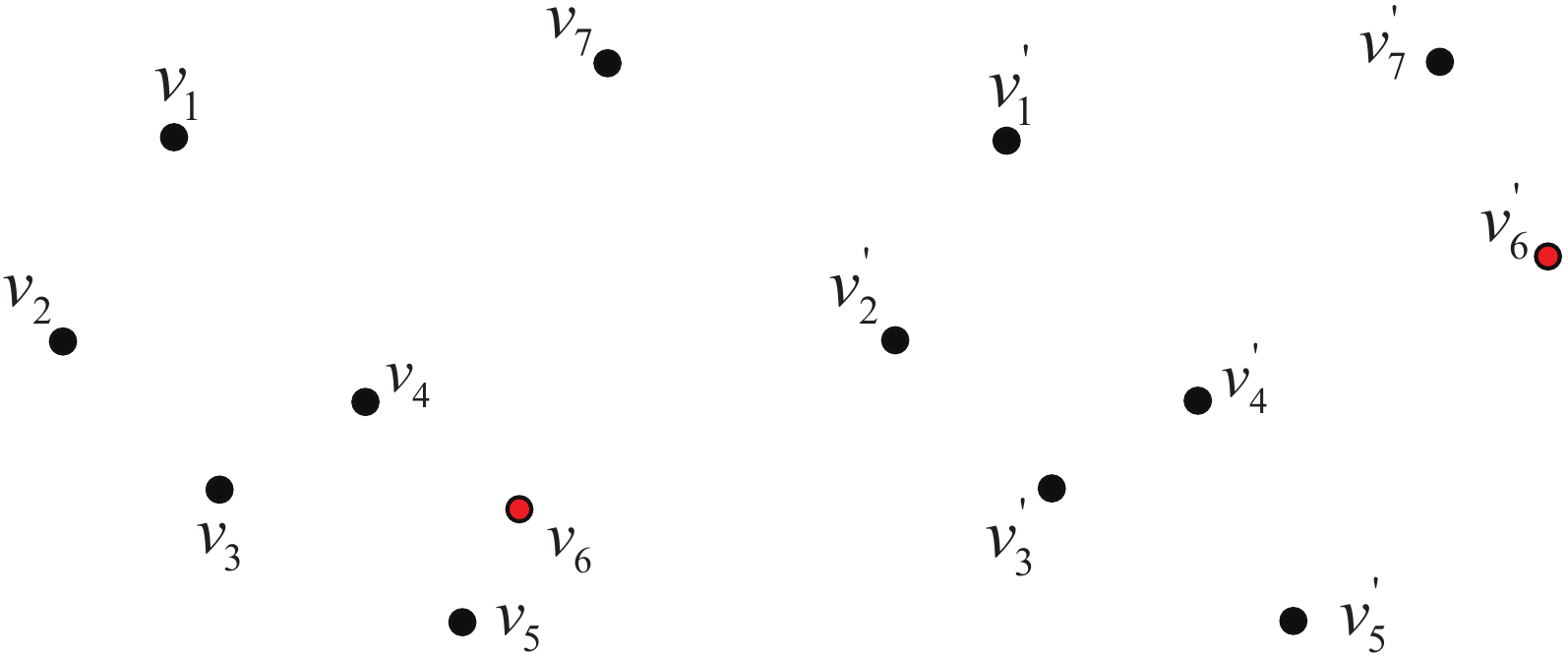}
    \end{minipage}}
  \subfigure[]{
    \label{fig:mini:subfig:b}
    \begin{minipage}[c]{0.45\textwidth}
      \centering
      \includegraphics[width=2.5in]{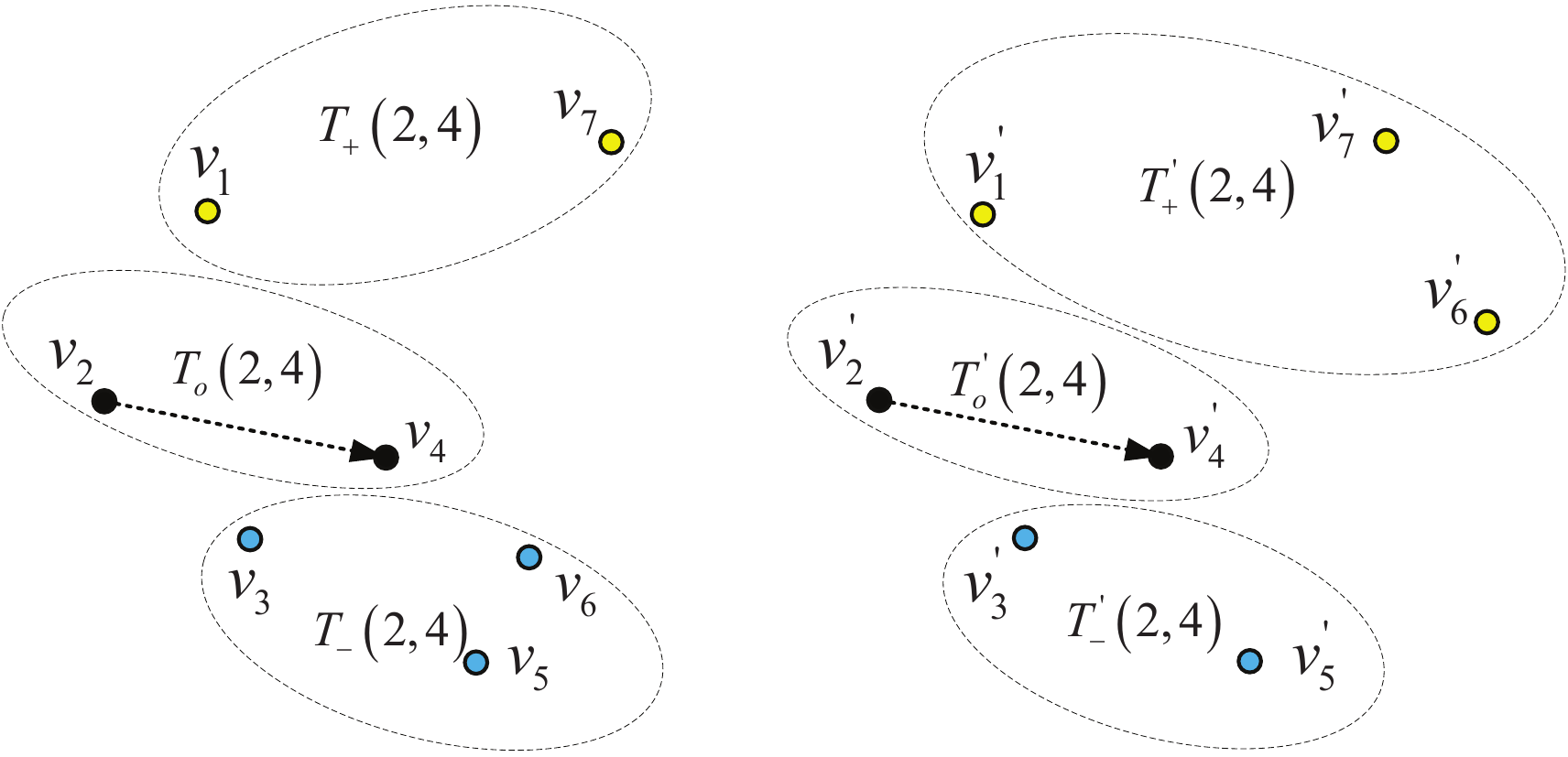}
    \end{minipage}}\\
  \captionstyle{normal}
  \caption{A demonstration of vertex trichotomy descriptor for $V=\{v_1,v_2,v_3,v_4,v_5,v_6,v_7\}$ and $V'=\{{v_1}',{v_2}',{v_3}',{v_4}',{v_5}',{v_6}',{v_7}'\}$. (a) Initial vertex sets with one outlier $(v_6,{v_6}')$. (b) $\phi (\left. G \right|\vec E(2,4))$ and $\phi (\left. G' \right|\vec E'(2,4))$.}
  \label{fig-visio}
\end{figure}

Fig. \ref{fig-visio} shows an example of vertex trichotomy descriptor for two corresponding vertex sets  without any transformation.
The initial corresponding vertex sets  $V=\{v_1,v_2,v_3,v_4,v_5,v_6,v_7\}$ and $V'=\{{v_1}',{v_2}',{v_3}',{v_4}',{v_5}',{v_6}',{v_7}'\}$ with one outlier $(v_6,v_6')$ are demonstrated in Fig. \ref{fig-visio} (a).
The vertex trichotomy descriptor of graphs is constructed as $\phi (\left. G \right|\vec E(i,j)) = \left( {T_ +  (i,j),T_o (i,j),T_ -  (i,j)} \right)$ and $\phi (\left. G' \right|\vec E'(i,j)) = \left( {T_ +'  (i,j),T_o' (i,j),T_ -'  (i,j)} \right)$, $\forall 1\le i, j\le 7$.
The vertex trichotomy descriptor with $i=2,j=4$ is shown in Fig. \ref{fig-visio} (b), which are depicted as $\phi (\left. G \right|\vec E(2,4))=(\{v_1,v_7\},\{v_2,v_4\},\{v_3,v_5,v_6\}$ and $\phi (\left. G' \right|\vec E'(2,4))=(\{v_1',v_6',v_7'\},\{v_2',v_4'\},\{v_3',v_5'\})$.
$\phi (\left. G \right|\vec E(2,4))$ does not exactly match to $\phi (\left. G' \right|\vec E'(2,4))$ due to the existence of outlier $(v_6,v_6')$. These disparities of corresponding graph descriptors are utilized by vertex trichotomy matching in the following section.

\section{SEARCHING FOR AN OPTIMAL SOLUTION}\label{sec:precoder}
\begin{figure*}[htb]
\centering
 \setlength{\abovecaptionskip}{0pt}
 \setlength{\belowcaptionskip}{0pt}
 \setlength{\intextsep}{8pt plus 3pt minus 2pt}
  \subfigure[]{
    \begin{minipage}[c]{0.4\textwidth}
      \centering
      \includegraphics[width=2.9in]{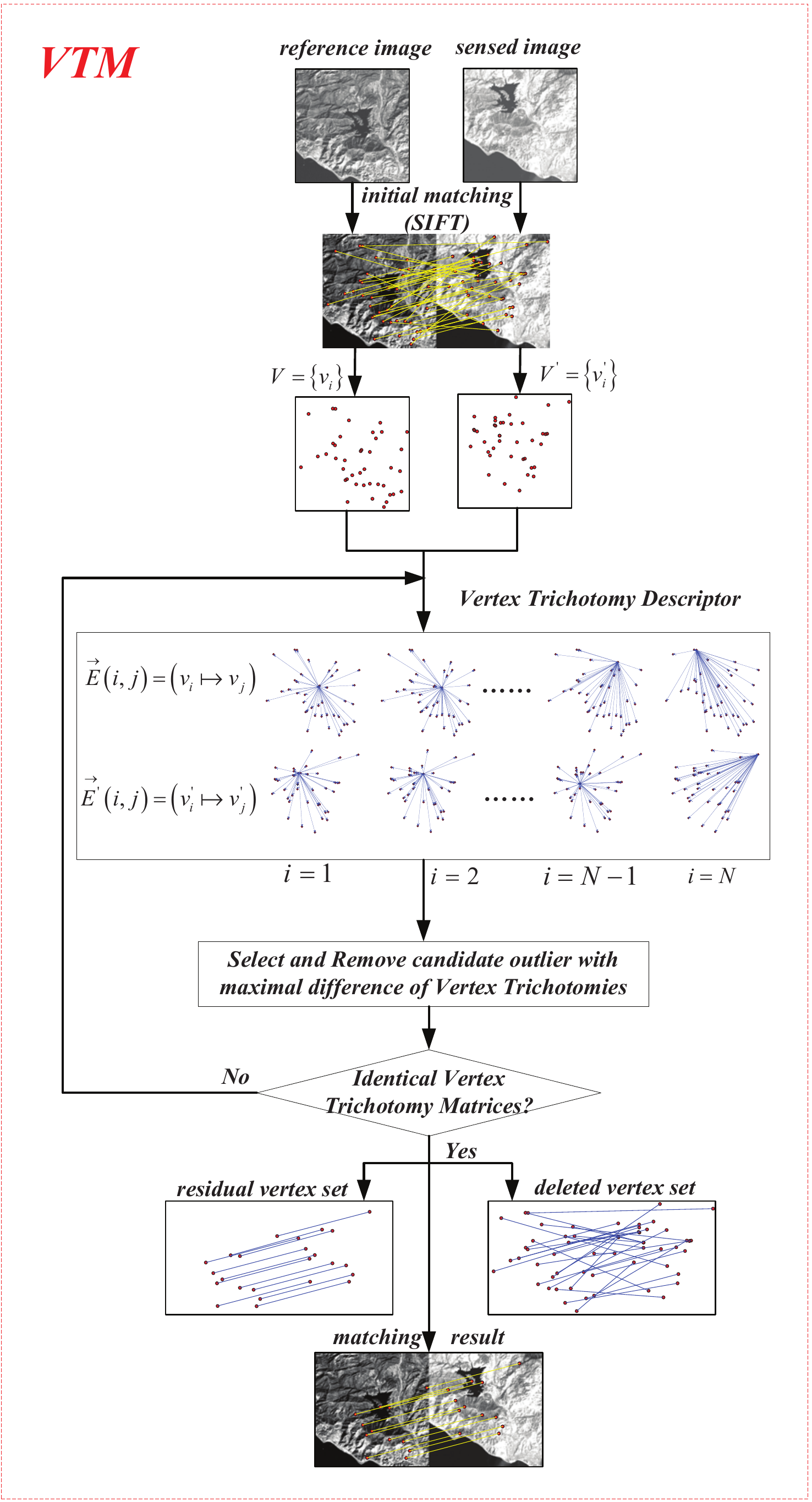}
    \end{minipage}}
  \subfigure[]{
    \begin{minipage}[c]{0.4\textwidth}
      \centering
      \includegraphics[width=2.9in]{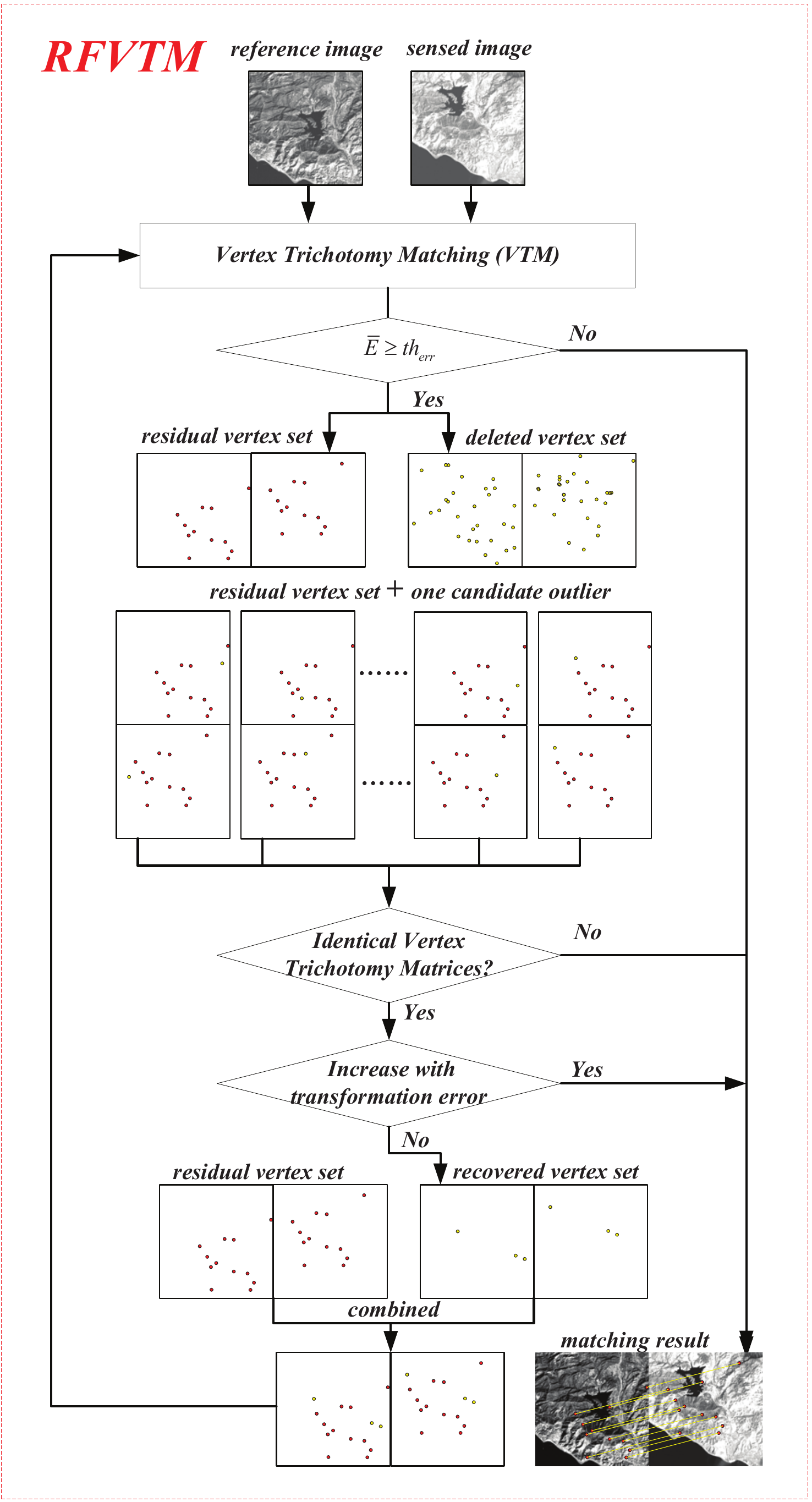}
    \end{minipage}}\\
  \captionstyle{normal}
  \caption{The frameworks of VTM and RFVTM.}
  \label{fig-framework}
\end{figure*}

In this section, the searching process is introduced to obtain an optimal solution of finding two matched graphes with the identical vertex trichotomy descriptors that minimize global transformation error. The initial one-to-one correspondences between two images as the input of the searching process are established by the classical local feature similarity method SIFT \cite{J_DGL_2004_IJCV}. First, Vertex Trichotomy Matching (VTM) is proposed, which utilizes the disparity of vertex trichotomy descriptors iteratively to determine candidate outliers. Then, inlier recovery and outlier filtering strategy for VTM (RFVTM) is proposed with two criteria restrictions of identical vertex trichotomy descriptors and reductive transformation errors. The end of this section discusses approaches to reduce the time complexity in the searching process. The frameworks of VTM and RFVTM are shown in Fig. \ref{fig-framework}.

\subsection{Vertex Trichotomy Matching (VTM)}
The Vertex Trichotomy Matching method is proposed to find candidate outliers based on the disparity of vertex trichotomy descriptors, which are validated by all of the corresponding vertices between two graphs.
The corresponding vertices with the maximum accumulated disparities between corresponding descriptors are selected as candidate outliers.
The determination of candidate outliers iterates through the following steps:
\begin{itemize}
\item[1)]

The trichotomy matrix denoted by ${{\bf{M}}_{i \mapsto j}}[k]$  represents the attribution of $v_k$  according to $\phi(G|\vec E(i,j)) = \left( {T_ +(i,j)  ,T_o(i,j) ,T_ -(i,j) } \right)$ and $\phi(G'|\vec E') = \left( {T_ + ' ,T_o' ,T_ - '} \right)$.
${{\bf{M}}_{i \mapsto j}}[k]=0$  when $v_k$ belongs to $T_ o(i,j)$; Otherwise, ${{\bf{M}}_{i \mapsto j}}[k]=-1$  when $v_k \in {T_ -(i,j)}$ ; ${{\bf{M}}_{i \mapsto j}}[k]=1$  when $v_k \in {T_ +(i,j)}$.  Accordingly, the corresponding trichotomy matrix ${{\bf{M}}'_{i \mapsto j}}[k]$  is derived from $\phi(G'|\vec E'(i,j)) = \left( {T_ + '(i,j) ,T_o'(i,j) ,T_ -'(i,j)} \right)$.

\item[2)]
Select the candidate outlier $\left( {{v_{{d^{out}}}},v_{{d^{out}}}'} \right)$ from the matching set. The difference between corresponding graphs is approximated by computing the accumulated disparity matrix $\Delta {{\bf{M}}_{i \mapsto j}}$:

\begin{equation} \label{eqn:achievable_6}
\Delta {{\bf{M}}_{i \mapsto j}} = \left\{ {\begin{array}{*{20}{c}}
{\sum\limits_{k = 1}^N {\left( {{{\bf{M}}_{i \mapsto j}}\left[ k \right] \oplus {{\bf{M}}'}_{i \mapsto j}\left[ k \right]} \right),i \ne j} }\\
{{\rm{\qquad\qquad\quad 0\qquad\qquad}\quad\quad}, i = j}
\end{array}} \right.
\end{equation}%
where $(\cdot)\oplus (\cdot)$  denotes the logical operation exclusive or (i.e., XOR) that outputs true only when the two inputs (in this case ${{\bf{M}}_{i \mapsto j}}\left[ k \right]$ and ${{\bf{M}}'}_{i \mapsto j}\left[ k \right]$) differ from each other.
The outlier $\left( {{v_{{d^{out}}}},v_{{d^{out}}}'} \right)$ is selected yields the maximal difference of vertex trichotomy descriptors as follows:

\begin{equation} \label{eqn:achievable_7}
{d^{out}} = \mathop {\arg \max }\limits_{j = 1,2,...,N} \sum\limits_{i = 1}^N {\Delta {{\bf{M}}_{i \mapsto j}}}.
\end{equation}%

\item[3)]
Once the candidate outliers are identified, all references to ${d^{out}}$ in the corresponding trichotomy matrices ${\bf{M}}_{i\mapsto j}[ k]$ and ${\bf{M}}'_{i\mapsto j}[ k]$, $\forall i,j,k = {d^{out}}$ should be removed.
A new iteration begins with the decrement of residual vertices until $\Delta {\bf{M}}_{i \mapsto j}=0 ,\forall i,j$, which indicates that there is no difference between the corresponding vertex trichotomy descriptors, and no candidate outlier needs to be removed.
The details of VTM to determine candidate outliers are outlined in Algorithm 1.
\end{itemize}

\begin{algorithm}[htb]         
\caption{VTM}             
\label{alg:Framwork}
\begin{algorithmic}[1]
\REQUIRE ~~\\
    The corresponding vertex sets $V=\{v_k\}$  and  $V'=\{v_k'\}$ of size $N$ initialized by SIFT, where $v_k$ matches $v'_k$;
\ENSURE ~~\\
    The residual vertex sets $\{V_{res},V'_{res}\}$;
\STATE Initialization: $\left\{ {{V_{res}},V_{res}'} \right\} = \left\{ {V,{V'}} \right\}$;
\FOR{each $(v_i,v_i'),(v_j,v_j'),(v_k,v_k') \in \{V_{res},V'_{res}\}$}
    \STATE ${{\bf{M}}_{i \mapsto j}}[k]$ $\leftarrow$ CreateTrichotomyMatrix$(v_i,v_j,v_k)$;
    \STATE ${{\bf{M}}'_{i \mapsto j}}[k]$ $\leftarrow$ CreateTrichotomyMatrix$(v_i',v_j',v_k')$;
\ENDFOR
\FOR{each $k$ from 1 to {\bf sizeof} $(V_{res})$ }
    \STATE $\Delta{\bf{M}}$ $\leftarrow$  SumDisparityMatrix$({{\bf{M}}_{i\mapsto j}}[k],{{\bf{M}}'}_{i\mapsto j}[k])$;
\ENDFOR
\WHILE {$\Delta {\bf{M}} \ne {\bf{0}}$}
    \STATE $d^{out}$ $\leftarrow$  FindMaxRowSumDiff$(\Delta {{\bf{M}}_{i \mapsto j}})$;
    \IF{{\bf sizeof$(d^{out})>0$}}
        \STATE $\{V_{res},V'_{res}\}$ $\leftarrow$  DeleteOutliers$(\{V_{res},V'_{res}\},d^{out})$;
        \STATE ${{\bf{M}}_{{d^{out}} \mapsto j}}[k] = 0$; ${{\bf{M}}_{i \mapsto j}}[{{d^{out}}}]--$;
        \STATE ${{\bf{M}}'_{{d^{out}} \mapsto j}}[k] = 0$;
        ${{\bf{M}}'_{i \mapsto j}}[{{d^{out}}}]--$;
        \STATE $\Delta {\bf{M}}$ $\leftarrow$  SumDisparityMatrix$({{\bf{M}}_{i\mapsto j}}[k],{{\bf{M}}'}_{i\mapsto j}[k])$;
    \ELSE
        \RETURN $\{V_{res},V'_{res}\}$;
    \ENDIF

\ENDWHILE
\end{algorithmic}
\end{algorithm}

\subsection{Inlier Recovery and Outlier Filtering Strategy for VTM (RFVTM)}
Despite being invariant with respect to affine transformations, vertex trichotomy matching could still fail to remove few stubborn outliers, namely false matches closed to correct locations but mistakenly supported by the identical trichotomy sets during VTM iterations.
These stubborn outliers cannot be picked out from the residual vertices without any addition of fresh matches.
One the other hand, a certain amount of inliers are inevitably removed as candidate outliers when mistakenly supported by a large number of outliers in the previous iterations.
Therefore, a restricted inlier recovery and outlier filtering strategy (RFVTM) are proposed to recover inliers and filter outliers.
First, we pick up the candidate outliers into the residual set one by one. The following two coarse constraints are imposed for these candidate outliers, which can be recovered, i.e.,
1) The residual vertices should have identical vertex trichotomy descriptors with one addition of the candidate outlier being checked;
2) The matching accuracy on the basis of current transformation parameters will not decrease with one addition of candidate outlier being checked.
Then, VTM is re-implemented to the expanded vertex sets containing residual vertices and all of recovered candidate outliers, so that both of the outliers among the recovered candidates and the outliers in the residual sets are capable of being removed with the updated vertex trichotomy descriptors.
The proposed strategy is described as follows:

\begin{itemize}
\item[1)] The residual vertices with exactly the same vertex trichotomy descriptors are left in the residual vertex sets $\{V_{res},V_{res}'\}$, while the candidate outliers $(v_{d^{out}},v_{d^{out}}')$ resulting from VTM are assembled into the deleted vertex sets $\{V_{del},V_{del}'\}$.
    The transformation errors $E(k)$ and $\bar E$ respectively provide individual and average measures of matching accuracy on the basis of current affine transformation coefficients ${\theta ^ * }$,
    \begin{equation} \label{eqn:achievable_8}
    E\left( k, {\theta ^ * }\right) = {\left\| {T\left( {{v_k},{\theta ^ * }} \right) - v_k'}
    \right\|^2}
    \end{equation}%

    \begin{equation} \label{eqn:achievable_8}
    \bar E = \sqrt {\frac{1}{{N_s }}\sum\limits_{k = 1}^{N_s } {\left\| {T\left( {v_k ,\theta ^ *  } \right) - v_k' } \right\|^2 } }
    \end{equation}%

    where ${\theta ^ * }$ are estimated by the matched vertices in  $\{V_{res},V_{res}'\}$ through the common model parameter estimation approach Least Square Method (LSM) \cite{J_SU_1991_ITPAMI}. Here, $N_s$ represents the size of residual vertex sets.

\item[2)] Check candidate outliers individually whether they are likely to be arbitrarily deleted in the previous iterations.
    Each time there is only one candidate outlier from $\{V_{del},V_{del}'\}$  to be checked with the residual vertices. These candidate outliers with a high probability of being inliers are restored into the recovered vertex sets $\{V_{rec},V_{rec}'\}$ when both of the following criterions are satisfied.

    \begin{itemize}
    \item[a)] The corresponding vertex trichotomy descriptors should remain identical with one additional candidate $\left( {{v_{{d^{out}}}},v_{{d^{out}}}'} \right)$.
        The corresponding trichotomy matrices of residual vertices have already become the same after VTM iterations. Therefore, we only need to consider if the corresponding trichotomy matrices related to the additional candidate outlier are identical, i.e.,

\begin{multline} \label{eqn:achievable_9}
{\left\{{\begin{array}{*{20}c}
   {\!\!\!\!\!{\bf{M}}_{i \mapsto j} \left[ {d^{out} } \right] = {\bf{M}}_{i \mapsto j}' \left[ {d^{out} } \right]}  \\
   \begin{array}{l}
 {\!\!\!\!\!\bf{M}}_{d^{out}  \mapsto j} \left[ i \right] = {\bf{M}}_{d^{out}  \mapsto j}' \left[ i \right] \\
 {\!\!\!\!\!\bf{M}}_{i \mapsto d^{out} } \left[ j \right] = {\bf{M}}_{i \mapsto d^{out} }' \left[ j \right] \\
 \end{array} \\
\end{array}}\right.\!\!\!\!\!\!\!, \!\forall (v_i ,v_i' ),(v_j ,v_j' )} \\
{\in \left\{ {V_{res}, V_{res}' } \right\}}.
\end{multline}%

    \item[b)] The individual transformation error should not increase with one additional candidate $\left( {{v_{{d^{out}}}},v_{{d^{out}}}'} \right)$ in terms of the current transformation coefficients ${\theta ^ * }$  obtained in Step 1) , i.e.,
            \begin{equation} \label{eqn:achievable_10}
            E\left( {{d^{out},{\theta ^ * }}} \right) \le {E_{\max }}
            \end{equation}%
            where ${E_{\max }} = \max ( {{{\left\| {T\left( {{v_k},{\theta ^ * }} \right) - v_k'} \right\|}^2}} ),\forall \left( {{v_k},v_k'} \right) \in \left\{ {{V_{res}},V_{res}'} \right\}$.

    \end{itemize}

\item[3)]
All of vertices in $\left\{ {{V_{rec}},V_{rec}'} \right\}$  are recovered into $\left\{ {{V_{res}},V_{res}'} \right\}$.
In order to remove the outliers existing in the recovered sets and the residual sets, VTM is re-implemented with the expanded vertex sets to achieve identical vertex trichotomy descriptors. Accordingly, both of vertices in $\{V_{res},V_{res}'\}$ and $\{V_{del},V_{del}'\}$ are updated.
A new iteration of RFVTM begins with the updated $\{V_{res},V_{res}'\}$ and $\{V_{del},V_{del}'\}$ until $\bar E$  reaches to a preset matching accuracy $th_{err}$ (set to 0.5 in this paper), or no candidate outlier needs to be recovered.
The searching process of RFVTM is summarized in Algorithm 2.

\end{itemize}

\begin{algorithm}[htb]         
\caption{RFVTM}             \label{alg:Framwork}
\begin{algorithmic}[1]
\REQUIRE ~~\\
    The corresponding vertex sets $V=\{v_k\}$  and  $V'=\{v_k'\}$ of size $N$ initialized by SIFT, where $v_k$ matches $v'_k$;
\ENSURE ~~\\
    The residual vertex sets $\{V_{res},V'_{res}\}$;
\STATE Initialization: $\left\{ {{V_{res}},V_{res}'} \right\} = \left\{ {V,{V'}} \right\}$, ${th_{err}}$, $\bar E = \inf $;
\WHILE {$\bar E \ge {th_{err}}$}
    \STATE $\{{{V_{rec}},V_{rec}'}\}=\emptyset$; $\{{{V_{del}},V_{del}'}\}=\emptyset$; \STATE ${\bf{M}},{\bf{M}}'$ $\leftarrow$ CreateTrichotomyMatrix$(\{V_{res},V'_{res}\})$;
    \STATE $\Delta {\bf{M}}$ $\leftarrow$ SumDisparityMatrix$({{{\bf{M}}_{i\mapsto j}}[k],{{\bf{M}}'}_{i\mapsto j}[k]})$;
    \WHILE {$\Delta {\bf{M}} \ne 0$}
        \STATE ${d^{out}}$ $\leftarrow$ FindMaxRowSumDiff$(\Delta {{\bf{M}}_{i \mapsto j}})$;
        \IF{{\bf sizeof$(d^{out})>0$}}
            \STATE $\{{{V_{del}},V_{del}'}\}$ $\leftarrow$ AddVertices$(\{{{V_{del}},V_{del}'}\},d^{out})$;
            \STATE $\{{{V_{res}},V_{res}'}\}$ $\leftarrow$ DeleteOutliers$(\{{{V_{res}},V_{res}'}\},d^{out})$;
            \STATE Update ${\bf{M}}$, ${\bf{M}'}$ and $\Delta {\bf{M}}$;
        \ELSE
            \STATE \bf{break};
        \ENDIF
    \ENDWHILE
    \STATE $( {{\theta ^*},{E_{\max }},\bar E})$ $\leftarrow$ EstimateLSM $(\{V_{res},V_{res}'\})$;
    \FOR{each $i$ from 1 to {\bf{sizeof} $(V_{del})$}}
        \STATE $d^{out}$ $\leftarrow$ $(V_{del}[i],V_{del}'[i])$;
        \STATE ${\bf{M}},{\bf{M}}'$$\leftarrow$CreateTrichotomyMatrix$(\{V_{res},V_{res}'\},d^{out} )$;
        \IF{$( {\bf{M}} =  = {{\bf{M}}}' )$ and $\left( {E\left( {d^{out},{\theta ^ * }} \right) \le {E_{\max }}} \right)$}
            \STATE $\{V_{rec},V_{rec}'\}$ $\leftarrow$ AddVertices$(\{V_{rec},V_{rec}'\},d^{out})$;
        \ENDIF
    \ENDFOR
        \STATE $\left\{ {{V_{res}},V_{res}'} \right\} = \left\{ {{V_{res}},V_{res}'} \right\} \cup \left\{ {{V_{rec}},V_{rec}'} \right\}$;
\ENDWHILE
\end{algorithmic}
\end{algorithm}

\subsection{Complexity Analysis and Optimization}
The time complexities breakdown for the first iteration of VTM and RFVTM with initial sets of $n$ correspondences are as follows:
\begin{itemize}
\item[1)] Creating vertex trichotomy descriptor: $O\left( {C_n^2\left( {n - 2} \right)} \right) \to O\left( {{n^3}} \right)$
\item[2)] Removing  vertices with maximum disparity of vertex trichotomy matrix: $O\left( n \right)$
\item[3)] Estimating transform errors with respect to the residual vertices: $O\left( n \right)$
\item[4)] Checking candidate outliers: $O\left( {{n^2}} \right) + O\left( n \right)$
\end{itemize}
where $C_n^k$ denotes the number of $k$-combinations from a given set of $n $ elements, i.e., $C_n^k  = \frac{{n(n - 1)(n - 2) \cdots (n - k + 1)}}{{k(k - 1)(k - 2) \cdots 1}}$.

The iterative time depends on the number of initial points and proportions of outliers.
Candidate outliers are removed and residual correspondences are decreased during the iterations of VTM.
Besides removing candidate outliers, RFVTM provides inlier recovery with the recovery criterion mentioned above.
Therefore, vertex trichotomy descriptor requires frequently updating along with outlier removals and inlier recoveries.
These can be simplified by decreasing or increasing the values of corresponding trichotomy matrices related to the vertex which need to be removed or recovered, rather than reconstructing the vertex trichotomy descriptors in each iteration.

\begin{table*}[htb]
\centering
  \captionstyle{normal}
  \setlength{\abovecaptionskip}{0pt}
  \setlength{\belowcaptionskip}{10pt}
\caption{SPECIFICATIONS OF TYPICAL IMAGE PAIRS FROM IMAGE DATASET}
\begin{lrbox}{\tablebox}
\begin{tabular}{|c|c|c|c|c|c|c|}
\hline
Pairs & Spectrum & Sensor & Segmented size & Resolution & Date & Descriptions\\
\hline
ImgSp1-1 &	VI & High altitude aerial image from USC & 512$\times$512 & 20m & 1977 & Simulated: rotate $120^ \circ$  and scale 2.2, seacoast of San Diego\\
\hline
ImgSp1-2 &	VI & High altitude aerial image from USC & 1024$\times$1024 &20m & 1979 & Simulated: sheared with h=0.1 and v=0.1, runway of an airport\\
\hline
ImgSp2-1 &	VI & Landsat TM Band 2 & 1890$\times$1890 & 30m &	1988 &	duplicated pattern\\
\cline{2-6}
& VI & Landsat TM Band 2 & 1890$\times$1890 & 30m & 1988 &\\
\hline
ImgSp2-2 & VI & Experiment Satellite Band 1 & 512$\times$512 & 20m & 2008 & duplicated pattern\\
\cline{2-6}
& SWIR & Experiment Satellite Band 2 & 400$\times$400 & 50m & 2008 & Shanghai Oriental Pearl TV Tower, China\\
\hline
ImgSp3-1 & VI & SPOT HRV band XS1 (0.50-0.59m) & 278$\times$278 & 20m & 1990 & different spectral images from multispectral imagery\\
\cline{2-6}
& SWIR & SPOT HRV band XS3 (0.79-0.89m) & 278$\times$278 & 20m & 1990 &\\
\hline
ImgSp3-2 & VNIR & ASTER L1B Band 1  & 512$\times$512 & 15m & 1999 & different spectral images from different sensors \\
\cline{2-6}
& SAR & PALSAR fine mode & 512$\times$512 & 18m & 2006 & with speckle noise, Tokyo bay, Japan\\
\hline
\end{tabular}
\end{lrbox}
\scalebox{0.85}{\usebox{\tablebox}}
\label{table-dataset}
\end{table*}

\begin{figure*}[htp]
\centering
 \setlength{\abovecaptionskip}{0pt}
 \setlength{\belowcaptionskip}{0pt}
 \setlength{\intextsep}{8pt plus 3pt minus 2pt}
  \subfigure[]{
    \label{fig:mini:subfig:a}
    \begin{minipage}[c]{0.3\textwidth}
      \centering
      \includegraphics[width=2.2in]{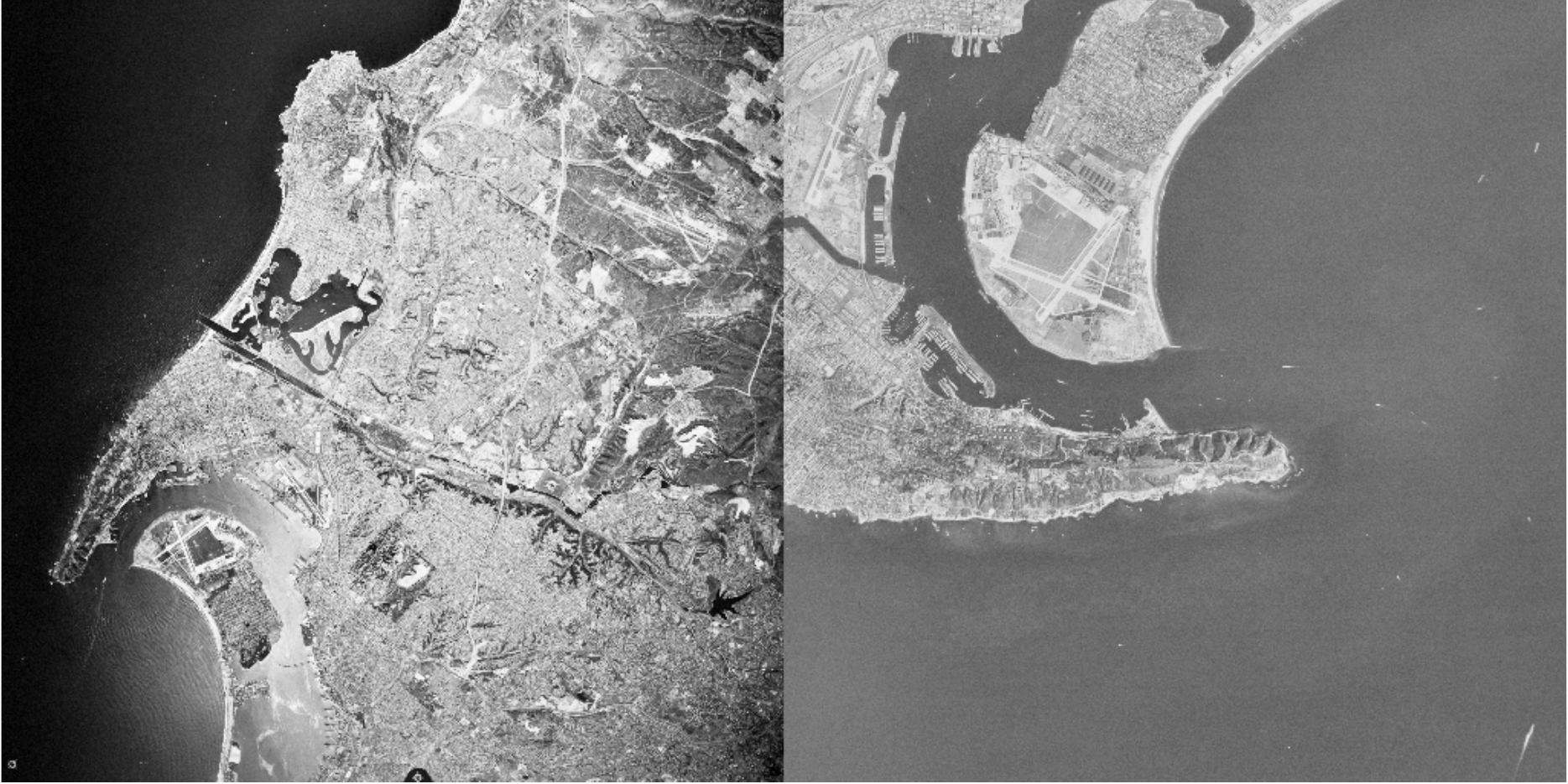}
    \end{minipage}}
  \subfigure[]{
    \label{fig:mini:subfig:b}
    \begin{minipage}[c]{0.3\textwidth}
      \centering
      \includegraphics[width=2.2in]{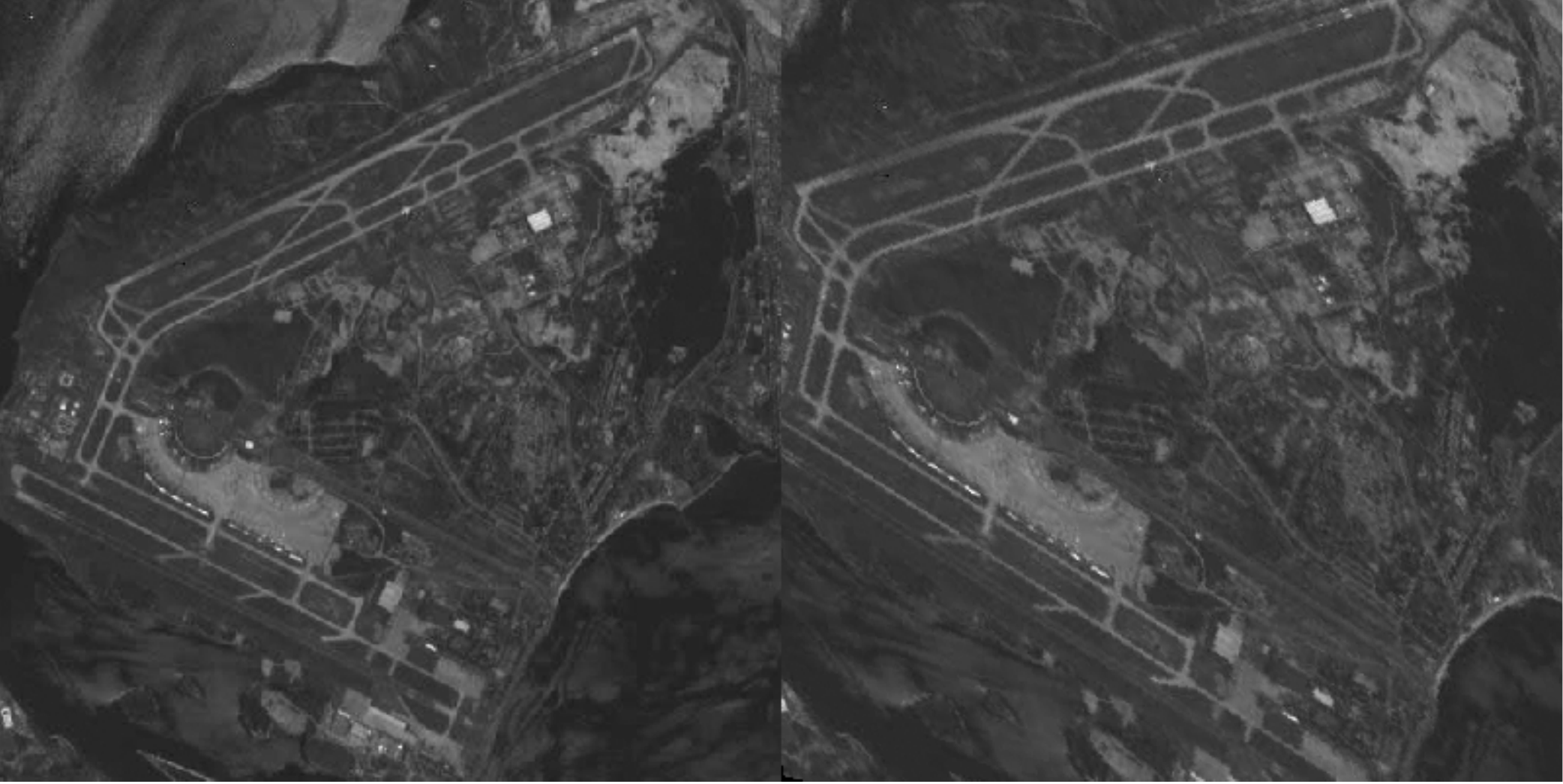}
    \end{minipage}}
  \subfigure[]{
    \label{fig:mini:subfig:c}
    \begin{minipage}[c]{0.3\textwidth}
      \centering
      \includegraphics[width=2.2in]{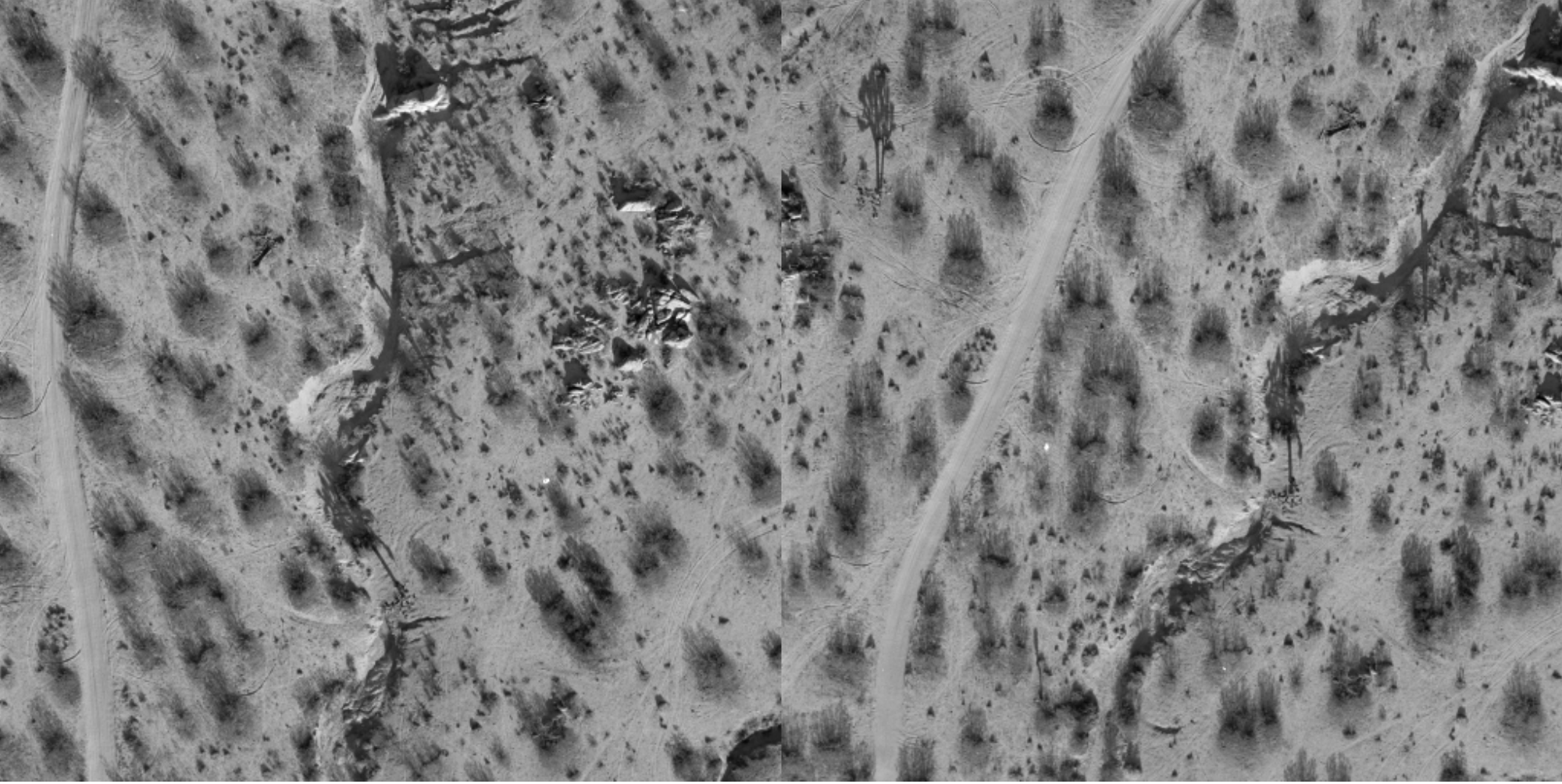}
    \end{minipage}}\\
  \subfigure[]{
    \label{fig:mini:subfig:a}
    \begin{minipage}[c]{0.3\textwidth}
      \centering
      \includegraphics[width=2.2in]{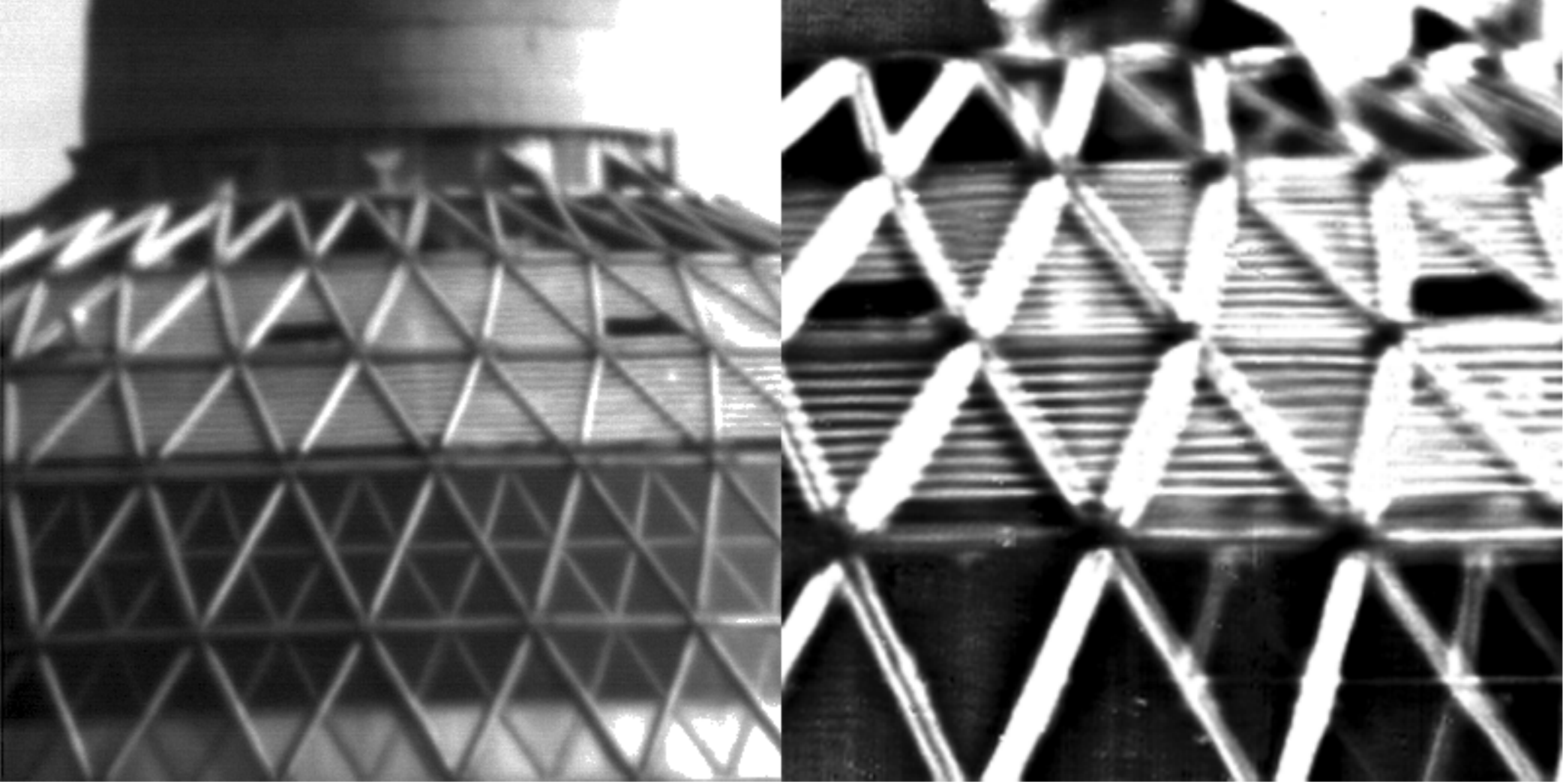}
    \end{minipage}}
  \subfigure[]{
    \label{fig:mini:subfig:b}
    \begin{minipage}[c]{0.3\textwidth}
      \centering
      \includegraphics[width=2.2in]{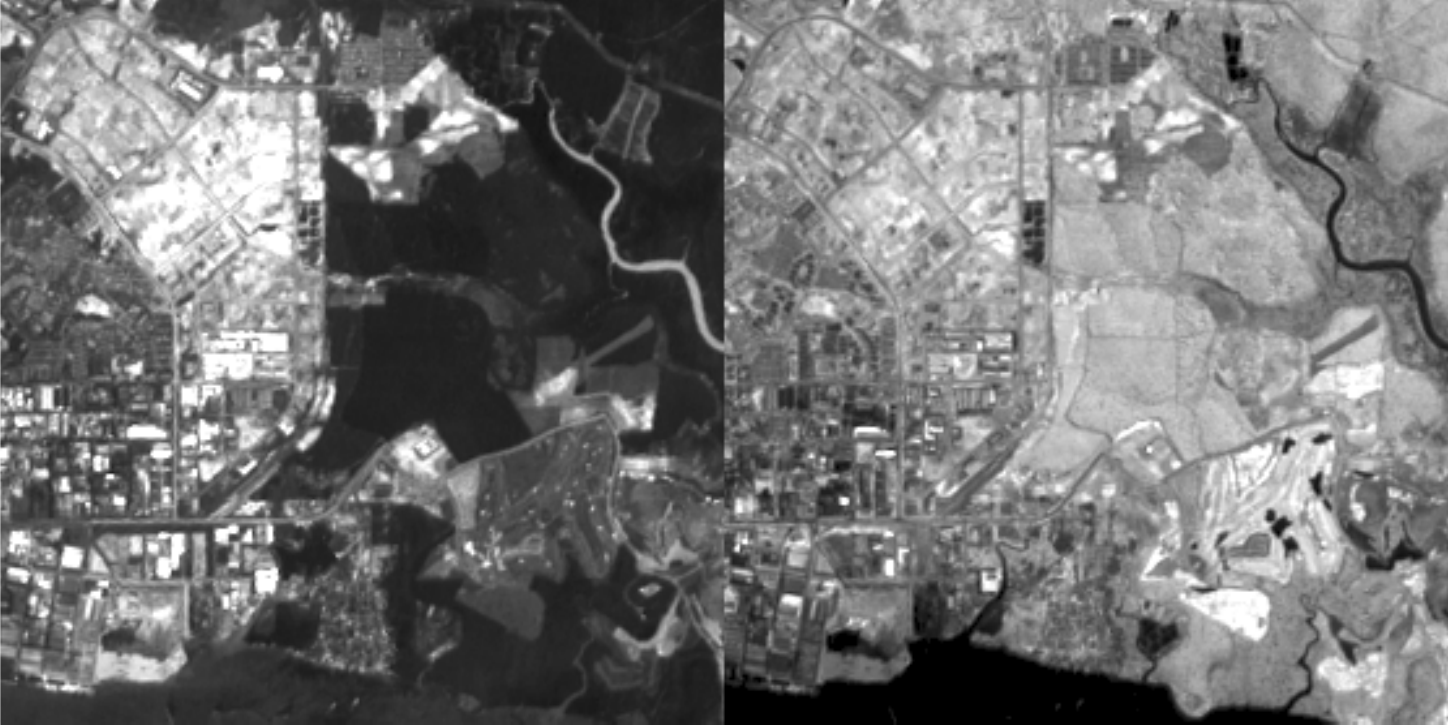}
    \end{minipage}}
  \subfigure[]{
    \label{fig:mini:subfig:c}
    \begin{minipage}[c]{0.3\textwidth}
      \centering
      \includegraphics[width=2.2in]{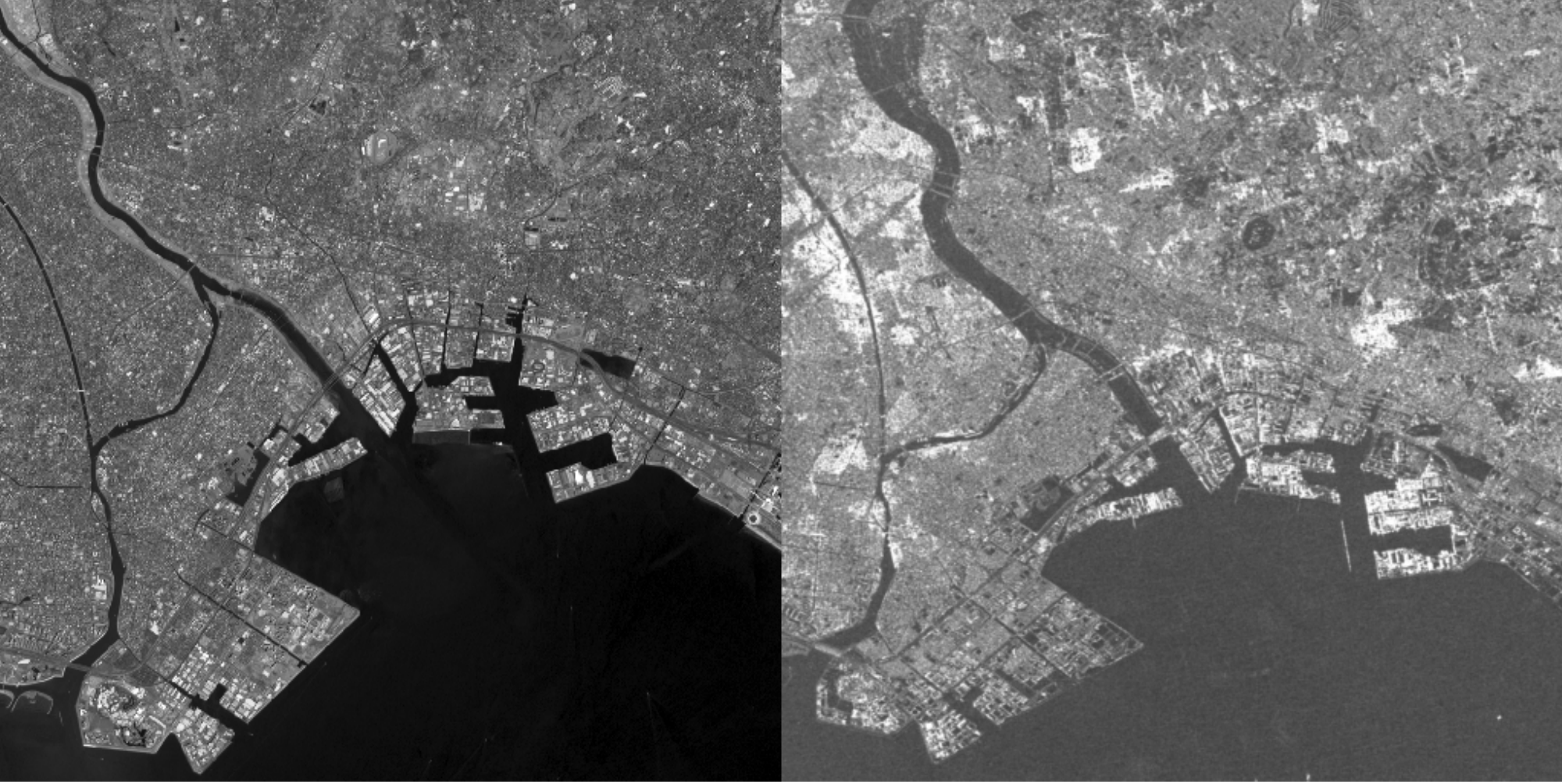}
    \end{minipage}}\\
  \captionstyle{normal}
  \caption{Typical image samples from dataset. (a) ImgSp1-1. (b) ImgSp1-2. (c) ImgSp2-1. (d) ImgSp2-2. (e) ImgSp3-1. (f) ImgSp3-2.}
  \label{fig-dataset}
\end{figure*}

Similar to most graph based matching methods, the most time consuming step in VTM and RFVTM is creating vertex trichotomy descriptor.
The time complexity of this step grows in terms of $O\left( {C_n^2\left( {n - 2} \right)} \right) = O\left( {\frac{{n\left( {n - 1} \right)\left( {n - 2} \right)}}{2}} \right)$.
In order to reduce the time cost in this step, a subdivision for initial correspondences is explored, which is implemented by randomly splitting the initial vertices of size $n$  into $m$  groups distributed over the image.
Then, matching iterations for  $n$ vertices are converted into matching $\frac{n}{m}$ correspondences for  $m$ groups respectively, and the time complexity is reduced to $O\left( {C_{\frac{n}{m}}^2\left( {\frac{n}{m} - 2} \right)m} \right) = O\left( {\frac{{n\left( {n - m} \right)\left( {n - 2m} \right)}}{{2{m^2}}}} \right)$.
The runtime analysis is provided in detail in Section V-D.

\section{ EXPERIMENT AND ANALYSIS}\label{sec:precoder}
Experiments in this section are conducted to validate the effectiveness of the proposed algorithms\footnote{The proposed algorithms of VTM and RFVTM are implemented in MATLAB Release 2013b and tested using images with bmp file format (.bmp).} in a laptop with 2-GHz CPU and 8-GB RAM (Intel Core i5).
First, image datasets and evaluation criterion are presented respectively. Secondly, experimental results of VTM and RFVTM are compared with RANSAC and GTM in terms of their accuracy, specificity, precision and recall values. Finally, we discuss the sensitivity of VTM and RFVTM with respect to different numbers of inliers and subdivisions.

\subsection{Data Set}
Fifty remote sensing image pairs are selected as the testing dataset on the basis of representative problems in remote sensing image registration. These image pairs are distributed into three image sets: 1) ImgSet1: 20 images with simulated large affine transformations; 2) ImgSet2: 20 images with duplicate patterns; 3) ImgSet3: 10 multispectral/multimodal images with inconsistent spectral content.
Table \ref{table-dataset} depicts the specifications of typical image pairs in each three image sets, which are also demonstrated in Fig. \ref{fig-dataset}.

\subsection{Evaluation Criterion}
The accuracy, specificity, precision, and recall are explored as evaluation criterions for the matching results \cite{J_MI_2012_ITGRS}, \cite{J_KM_2005_ITPAMI}, \cite{J_GJB_2009_CVIU}. The total matches in the initial sets are composed of residual correct matches (RC), residual false matches (RF), deleted false matches (DF), and deleted correct matches (DC).

\begin{itemize}
\item[1)] $r_{acc}$ is the ratio of correctly identifying matches: ${r_{acc}} = \frac{{RC + DF}}{{RC + DF + DC + RF}}$.
\item[2)] $r_{spe}$  is the proportion of those false matches correctly identified: ${r_{spe}} = \frac{{DF}}{{DF + RF}}$.
\item[3)] $r_{pre}$ is the ratio between the residual correct matches and the residual matches: ${r_{pre}} = \frac{{RC}}{{RC + RF}}$.
\item[4)] $r_{rec}$ is the proportion of residual correct matches in the initial correct matches: ${r_{rec}} = \frac{{RC}}{{RC + DC}}$.
\end{itemize}

Twenty control point pairs are manually selected to estimate affine transformation models for each image pair beforehand. The corresponding point belongs to correct matches if the point in one image transformed by the affine transformation model lies within 2 pixels from its matching point in the other image. Otherwise, it belongs to false matches.

\begin{figure*}[!tp]
\centering
 \setlength{\abovecaptionskip}{0pt}
 \setlength{\belowcaptionskip}{0pt}
 \setlength{\intextsep}{8pt plus 3pt minus 2pt}
  \subfigure[]{
    \label{fig:mini:subfig:a}
    \begin{minipage}[c]{0.3\textwidth}
      \centering
      \includegraphics[width=2.2in]{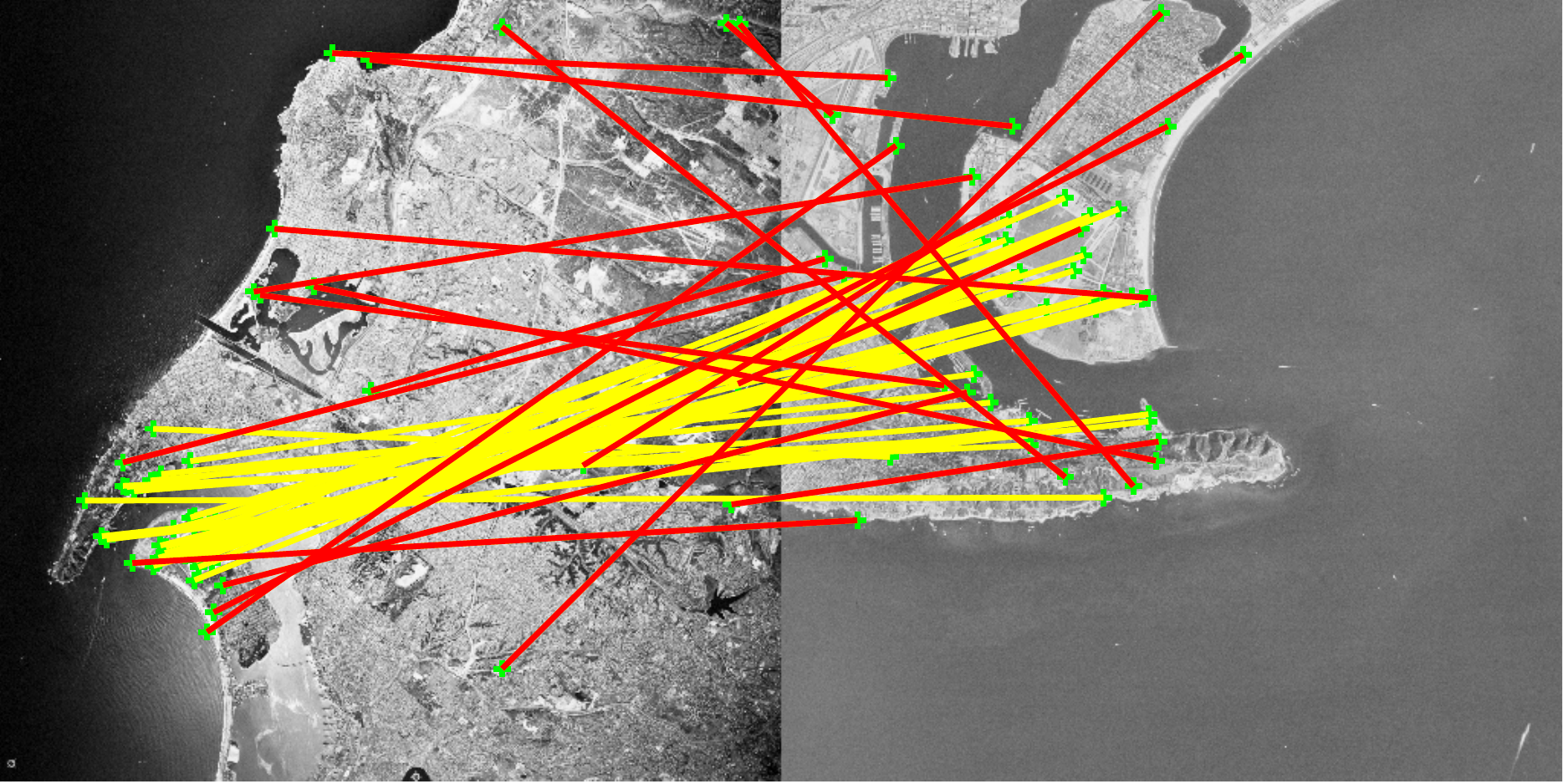}
    \end{minipage}}
  \subfigure[]{
    \label{fig:mini:subfig:b}
    \begin{minipage}[c]{0.3\textwidth}
      \centering
      \includegraphics[width=2.2in]{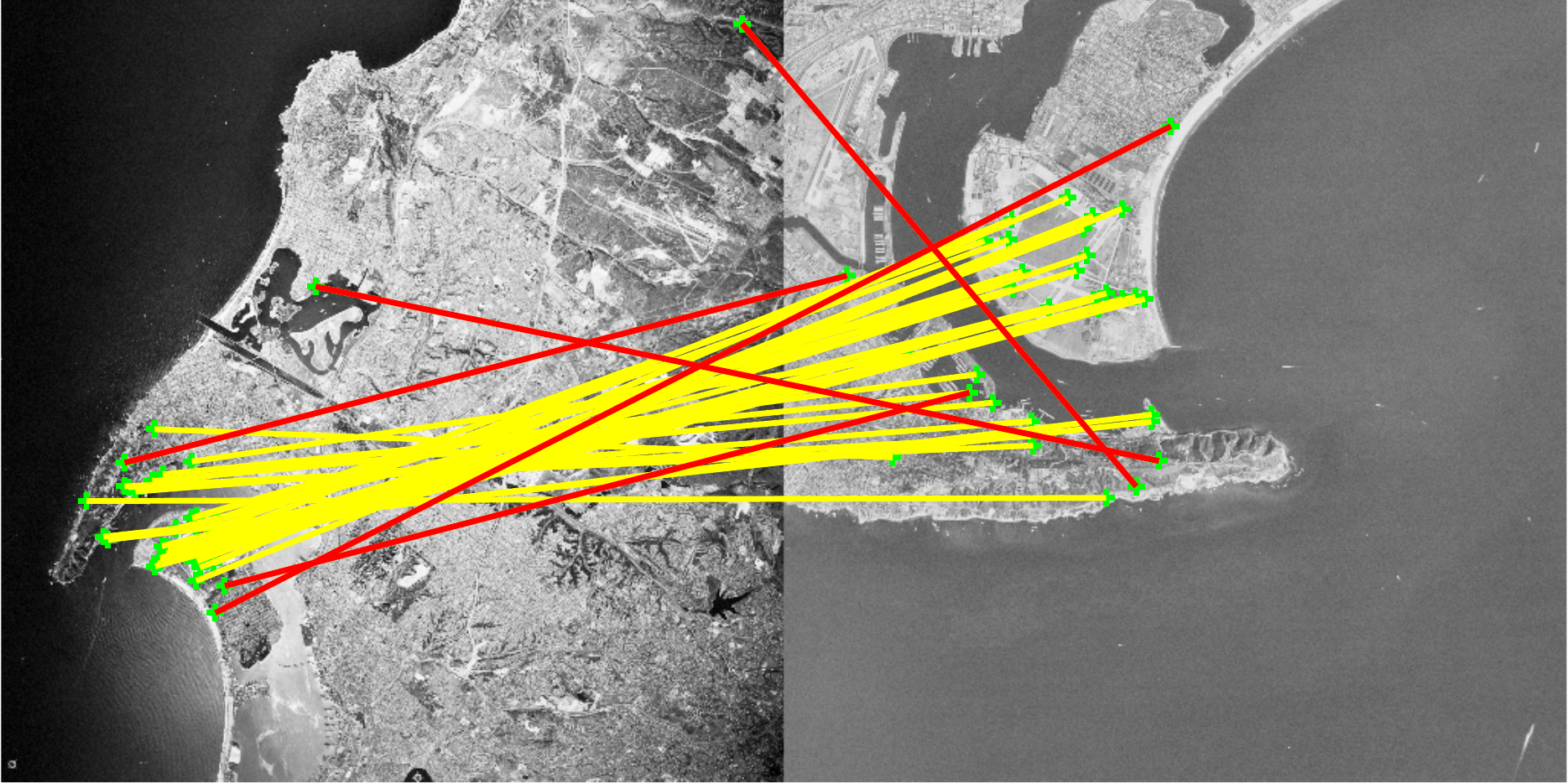}
    \end{minipage}}
  \subfigure[]{
    \label{fig:mini:subfig:c}
    \begin{minipage}[c]{0.3\textwidth}
      \centering
      \includegraphics[width=2.2in]{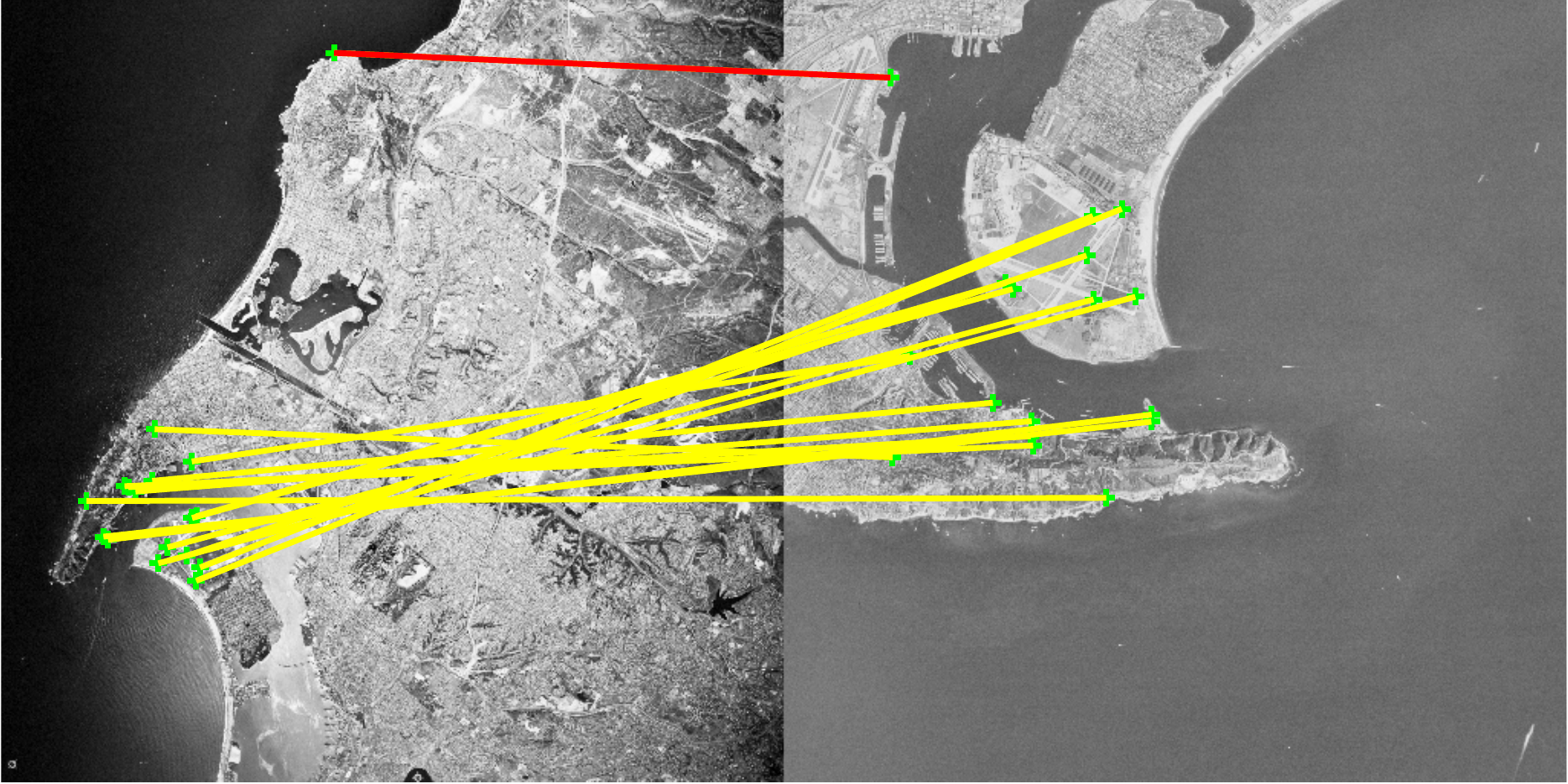}
    \end{minipage}}\\
  \subfigure[]{
    \label{fig:mini:subfig:a}
    \begin{minipage}[c]{0.3\textwidth}
      \centering
      \includegraphics[width=2.2in]{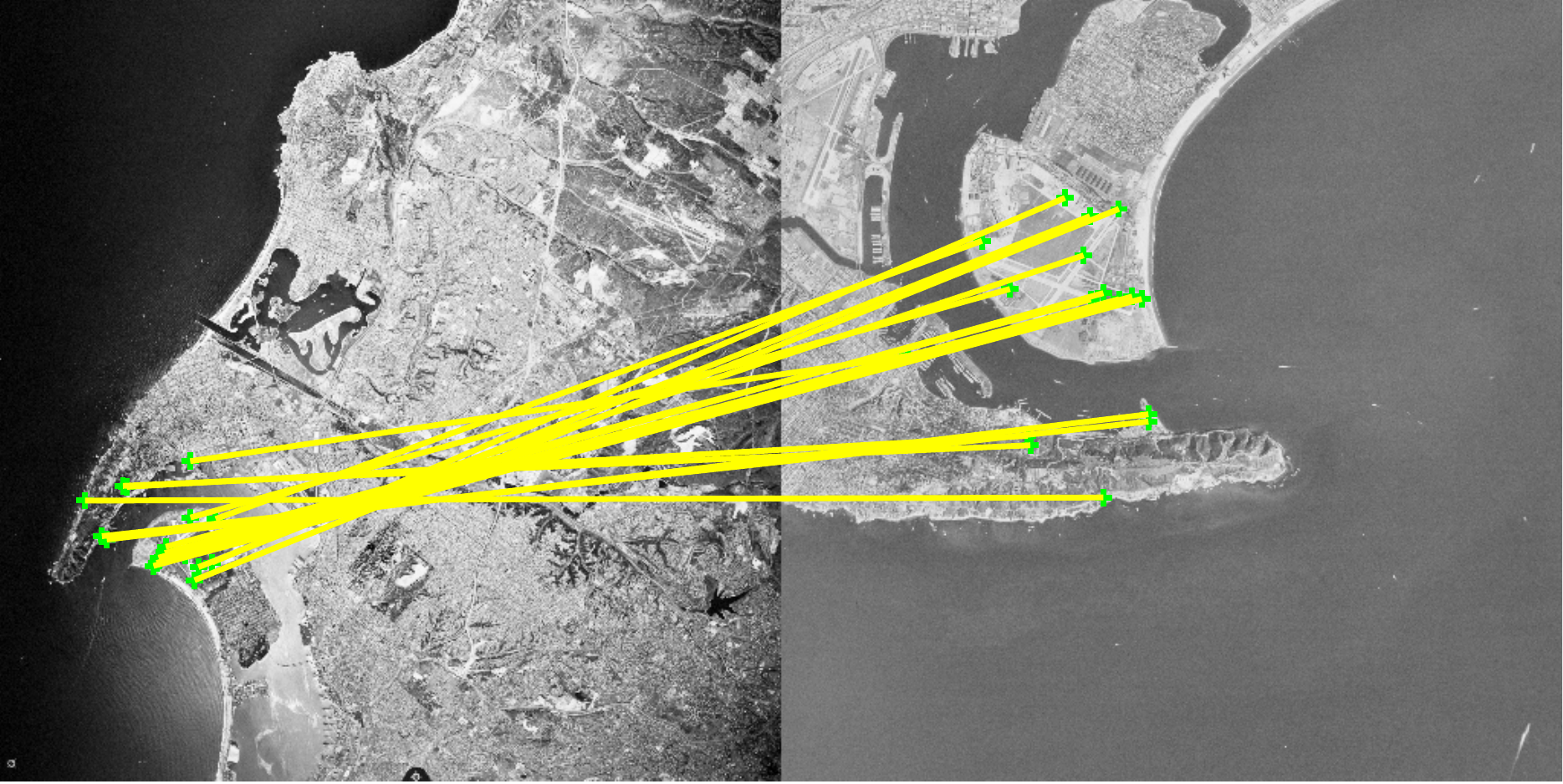}
    \end{minipage}}
  \subfigure[]{
    \label{fig:mini:subfig:a}
    \begin{minipage}[c]{0.3\textwidth}
      \centering
      \includegraphics[width=2.2in]{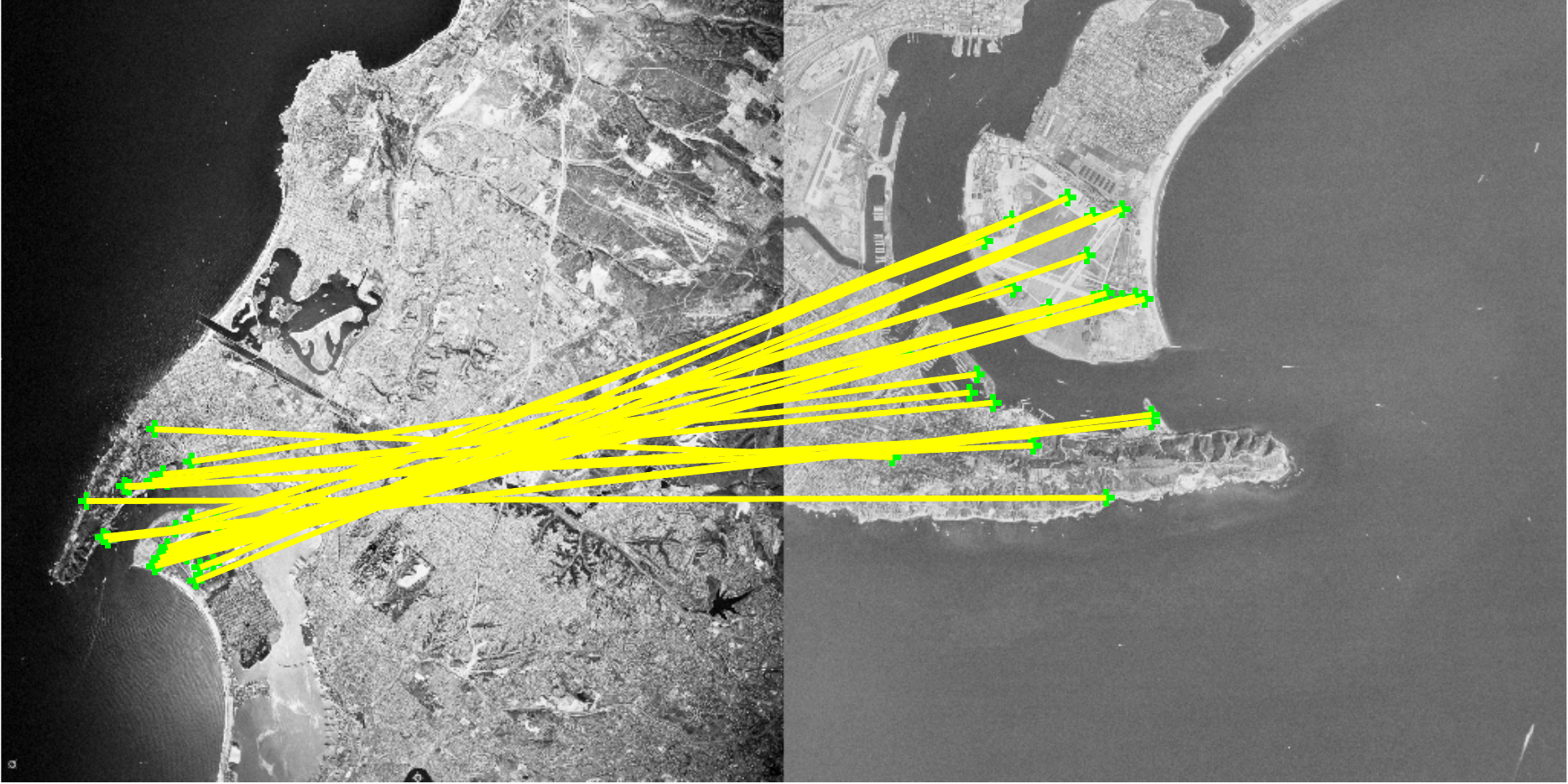}
    \end{minipage}}\\
  \captionstyle{normal}
  \caption{Examples of matching results for ImgSp1-1 with rotation of $120^ \circ$  anti-clockwise and scale factor of 2. (a) SIFT. (b) RANSAC. (c) GTM. (d) VTM. (e) RFVTM.}
  \label{fig-im1}
\end{figure*}

\begin{figure*}[!htb]
\centering
 \setlength{\abovecaptionskip}{0pt}
 \setlength{\belowcaptionskip}{0pt}
 \setlength{\intextsep}{8pt plus 3pt minus 2pt}
  \subfigure[]{
    \label{fig:mini:subfig:a}
    \begin{minipage}[c]{0.3\textwidth}
      \centering
      \includegraphics[width=2.2in]{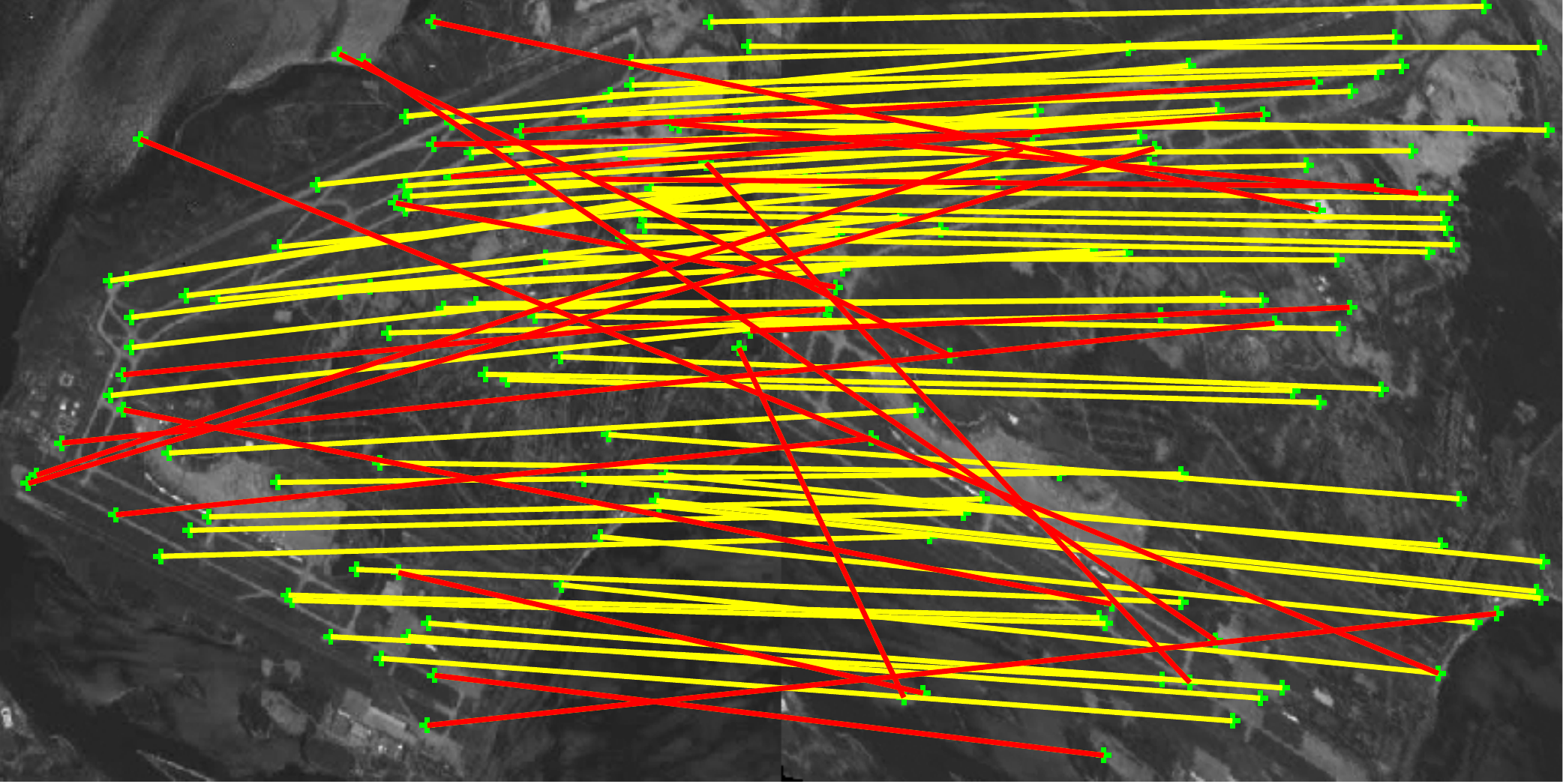}
    \end{minipage}}
  \subfigure[]{
    \label{fig:mini:subfig:b}
    \begin{minipage}[c]{0.3\textwidth}
      \centering
      \includegraphics[width=2.2in]{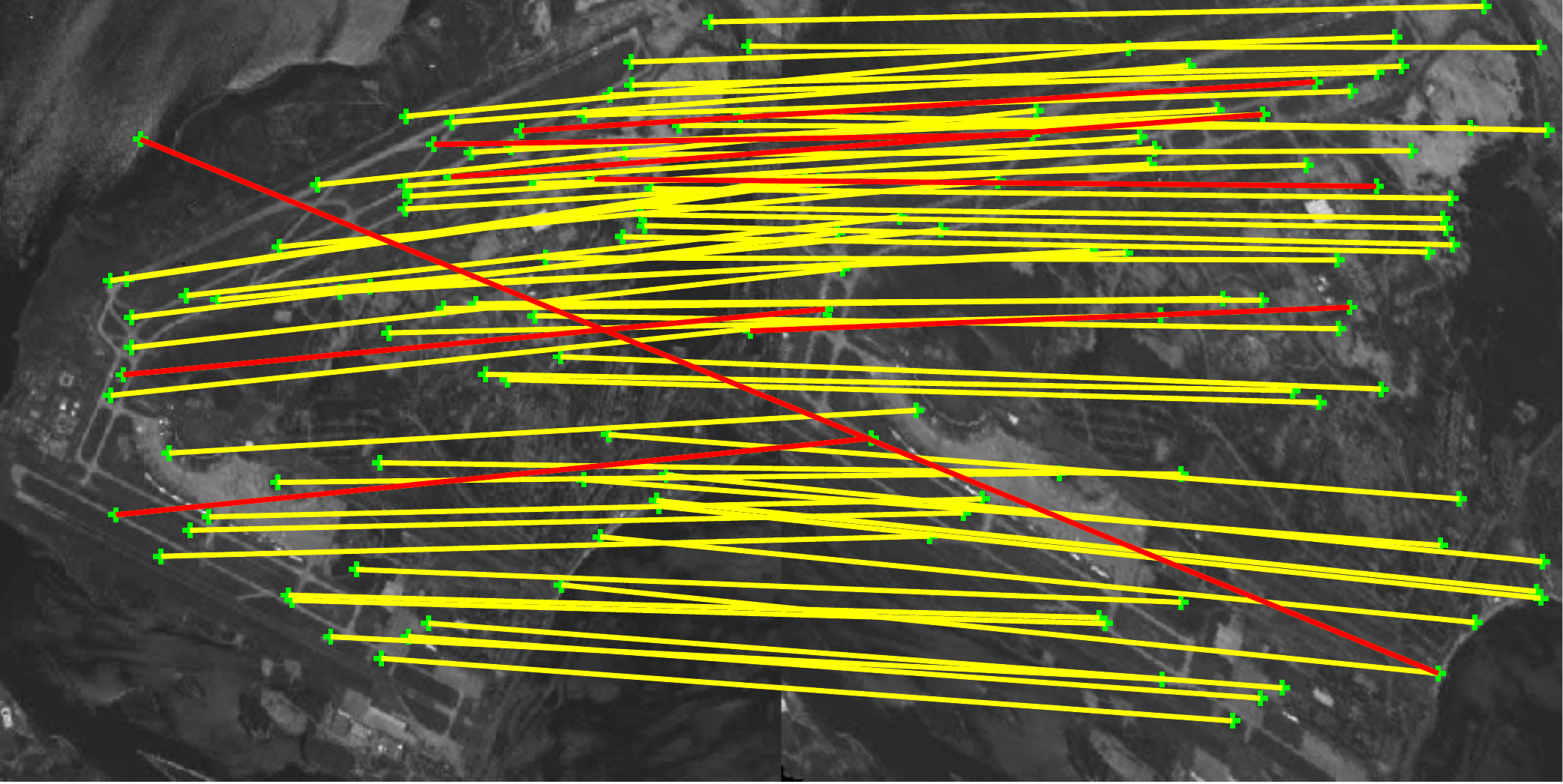}
    \end{minipage}}
  \subfigure[]{
    \label{fig:mini:subfig:c}
    \begin{minipage}[c]{0.3\textwidth}
      \centering
      \includegraphics[width=2.2in]{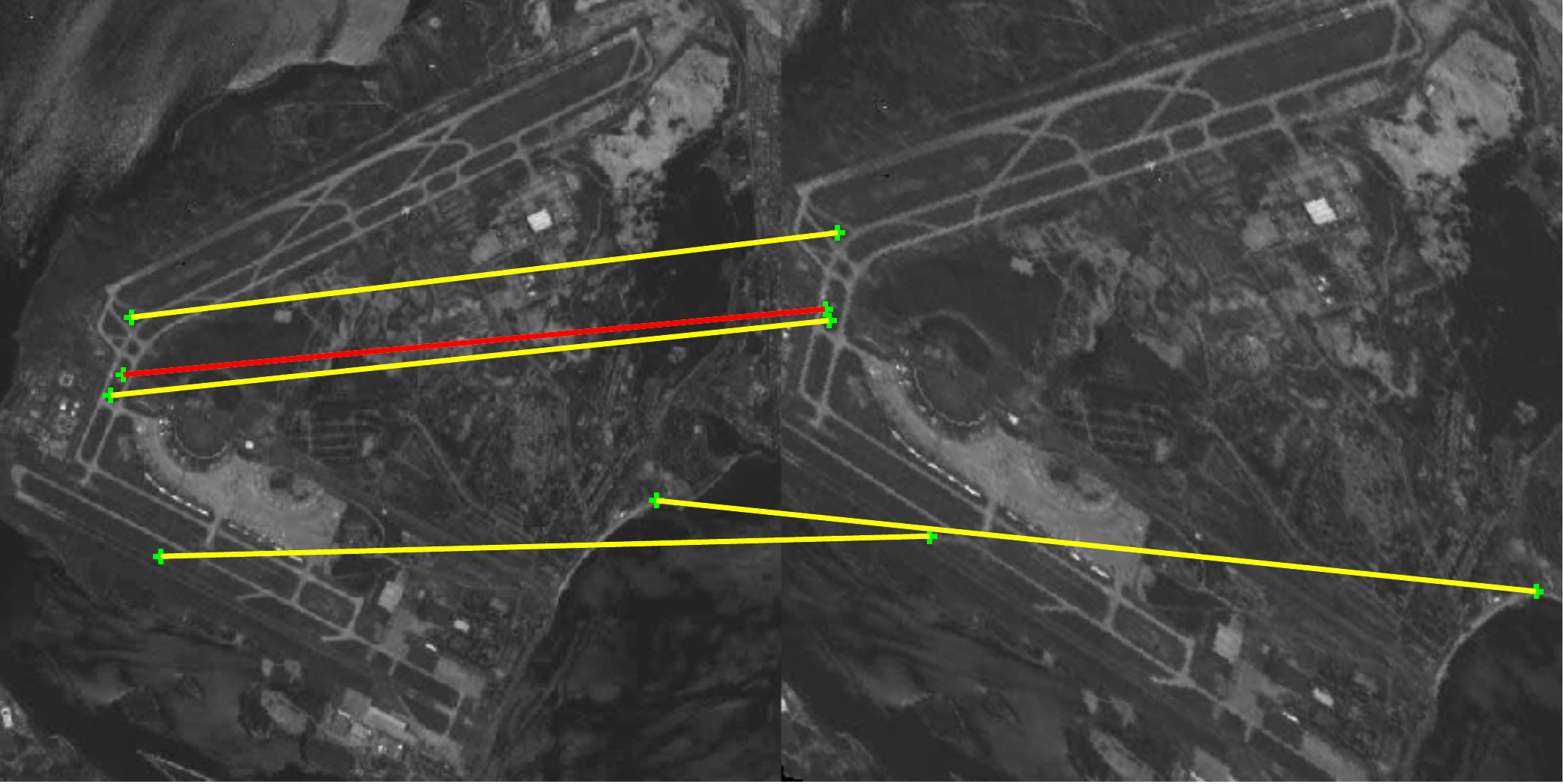}
    \end{minipage}}\\
  \subfigure[]{
    \label{fig:mini:subfig:a}
    \begin{minipage}[c]{0.3\textwidth}
      \centering
      \includegraphics[width=2.2in]{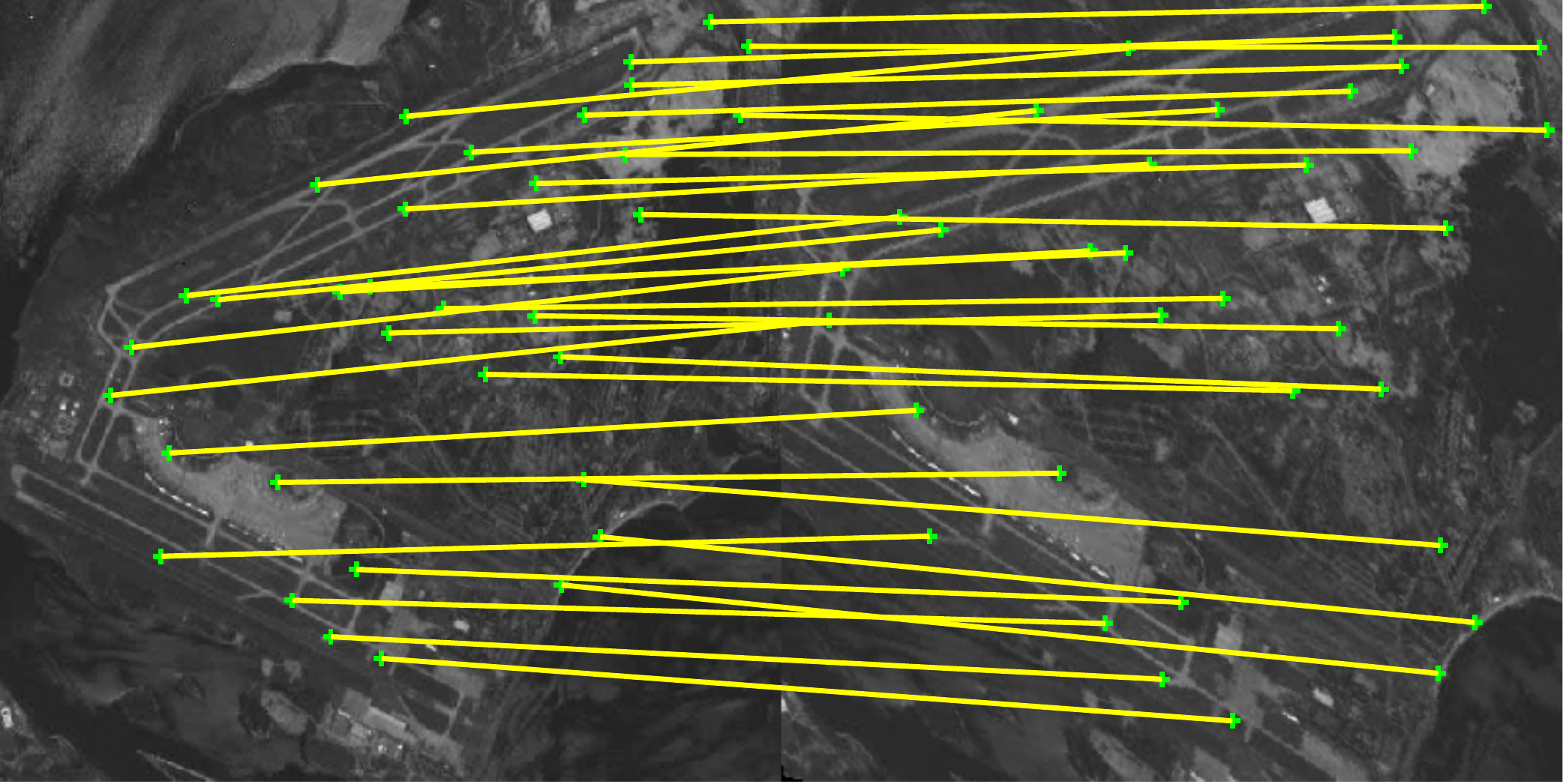}
    \end{minipage}}
  \subfigure[]{
    \label{fig:mini:subfig:a}
    \begin{minipage}[c]{0.3\textwidth}
      \centering
      \includegraphics[width=2.2in]{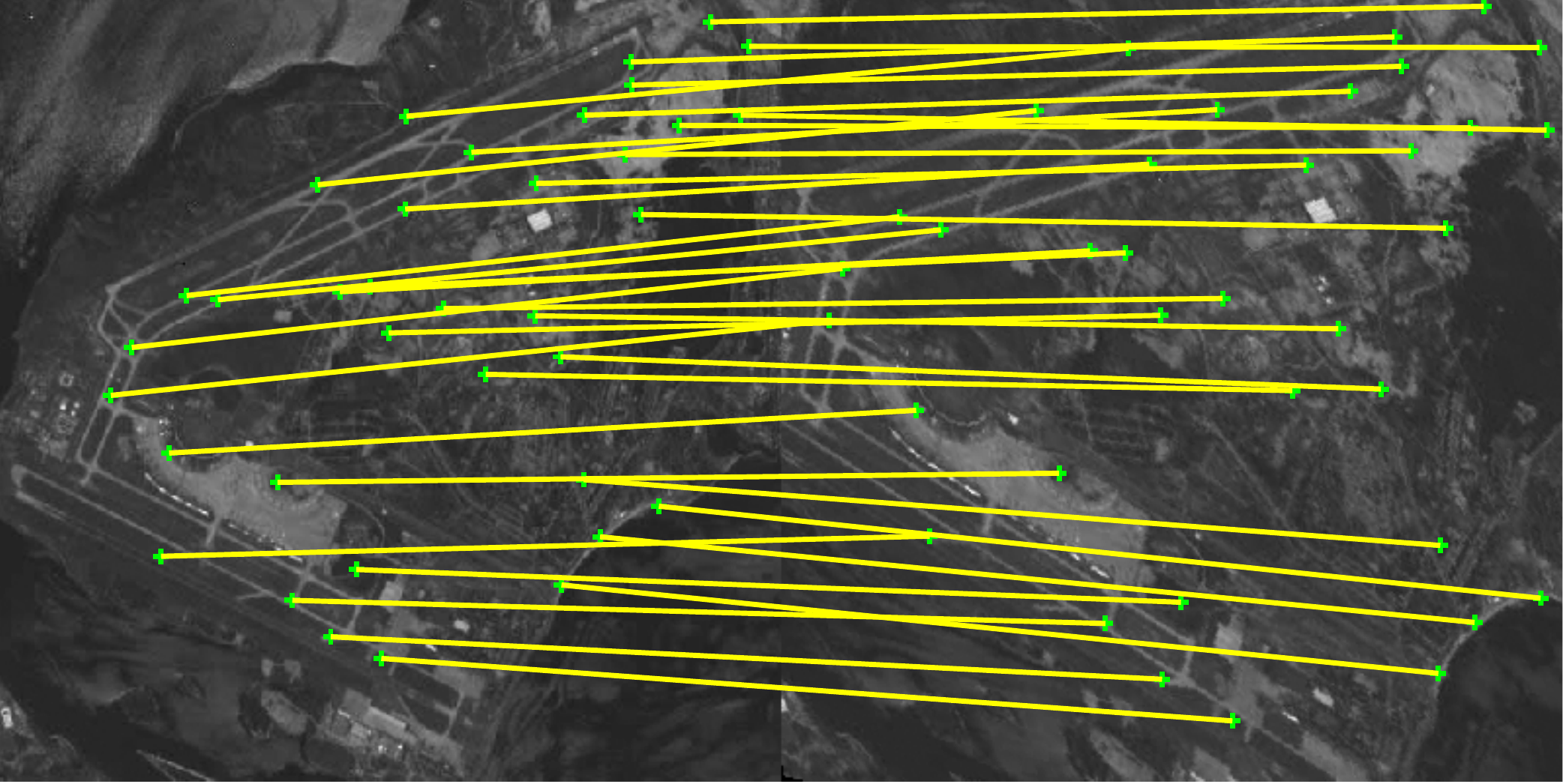}
    \end{minipage}}\\
  \captionstyle{normal}
  \caption{Examples of matching results for ImgSp1-2 with shear deformation factors h=0.1 and v=0.1. (a) SIFT. (b) RANSAC. (c) GTM. (d) VTM. (e) RFVTM.}
  \label{fig-im2}
\end{figure*}

\subsection{Matching Comparisons with Other Algorithms}
In this section, VTM and RFVTM are evaluated and compared with RANSAC \cite{J_MAFichler_1081_CACM} and GTM \cite{J_WAr_2009_IVC}.
RANSAC is a nondeterministic algorithm, which iteratively selects random samples from initial sets to estimate transformation parameters.
GTM builds its KNN graph with the restriction of average distance, and selects the outliers that yields the maximum difference of KNN graphs.
We perform comparisons of these algorithms on three datasets that correspond to representative problems in remote sensing images. KNN graphs in GTM algorithm are generated with an empirically chosen value of $K=5$.
SIFT features are extracted and matched by BBF algorithms \cite{C_JB_PCVPR} as the initial correspondences for all of experiments.
In the example figures of this section, correct matches in residual correspondences are depicted by yellow lines, and false matches are represented by red lines.

\begin{table}[!h]
\centering
  \captionstyle{normal}
  \setlength{\abovecaptionskip}{0pt}
  \setlength{\belowcaptionskip}{10pt}
\caption{THE MATCHING RESULTS FOR 12 SIMULATED IMAGE PAIRS WITH DIFFERENT ROTATION AND SCALE FACTORS}
\begin{lrbox}{\tablebox}
\begin{tabular}{|c|c|c|c|c|c|c|c|c|c|c|c|c|c|}
\hline
\multicolumn{1}{|c|}{No.} &
\multicolumn{1}{c|}{rotate} &
\multicolumn{1}{c|}{scale} &
\multicolumn{3}{c|}{Initial} &
\multicolumn{2}{c|}{RANSAC} &
\multicolumn{2}{c|}{GTM} &
\multicolumn{2}{c|}{VTM} &
\multicolumn{2}{c|}{RFVTM} \\
\cline{4-14}
 & & & $n$ & IC & IF & RC & RF & RC & RF & RC & RF & RC & RF\\
\hline
1 & 30$^\circ$ & 1.5 & 181 & 124 & 57 & 119 & 8 & 43 & 0 & 89 & 0 & 122 & 0\\\hline
2 & 30$^\circ$ & 2.0 & 97 & 58 & 39 & 46 & 11 & 18 & 2 & 39 & 0 & 54 & 0\\\hline
3 & 30$^\circ$ & 3.0 & 33 & 12 & 21 & 10 & 14 & 5 & 0 & 7 & 0 & 10 & 0\\\hline
4 & 60$^\circ$ & 1.5 & 146 & 82 & 64 & 58 & 19 & 20 & 1 & 29 & 0 & 61 & 0\\\hline
5 & 60$^\circ$ & 2.0 & 73 & 46 & 27 & 44 & 7 & 11 & 0 & 11 & 0 & 42 & 0\\\hline
6 & 60$^\circ$ & 3.0 & 45 & 19 & 26 & 16 & 12 & 7 & 3 & 10 & 1 & 14 & 0\\\hline
7 & 90$^\circ$ & 1.5 & 106 & 74 & 32 & 60 & 11 & 36 & 0 & 39 & 0 & 68 & 0\\\hline
8 & 90$^\circ$ & 2.0 & 84 & 51 & 33 & 45 & 13 & 20 & 0 & 27 & 0 & 39 & 0\\\hline
9 & 90$^\circ$ & 3.0 & 37 & 18 & 19 & 16 & 6 & 8 & 0 & 12 & 0 & 15 & 0\\\hline
10 & 120$^\circ$ & 1.5 & 68 & 39 & 29 & 21 & 10 & 13 & 0 & 19 & 0 & 26 & 0\\\hline
11 & 120$^\circ$ & 2.0 & 49 & 30 & 19 & 30 & 7 & 16 & 1 & 17 & 0 & 30 & 0\\\hline
12 & 120$^\circ$ & 3.0 & 25 & 9 & 16 & 6 & 13 & 0 & 5 & 5 & 1 & 7 & 0\\\hline
\end{tabular}
\end{lrbox}
\scalebox{0.75}{\usebox{\tablebox}}
\label{table-rigid}
\end{table}

\begin{table}[!h]
\centering
  \captionstyle{normal}
  \setlength{\abovecaptionskip}{0pt}
  \setlength{\belowcaptionskip}{10pt}
\caption{THE MATCHING RESULTS FOR 8 SIMULATED IMAGE PAIRS WITH DIFFERENT SHEAR FACTORS}
\begin{lrbox}{\tablebox}
\begin{tabular}{|c|c|c|c|c|c|c|c|c|c|c|c|c|c|}
\hline
\multicolumn{1}{|c|}{No.} &
\multicolumn{1}{c|}{horizontal} &
\multicolumn{1}{c|}{vertical} &
\multicolumn{3}{c|}{Initial} &
\multicolumn{2}{c|}{RANSAC} &
\multicolumn{2}{c|}{GTM} &
\multicolumn{2}{c|}{VTM} &
\multicolumn{2}{c|}{RFVTM} \\
\cline{4-14}
 & shear& shear& $n$ & IC & IF & RC & RF & RC & RF & RC & RF & RC & RF\\\hline
1 & 0.0 & 0.1 & 125 & 73 & 52 & 67 & 19 & 11 & 7 & 48 & 2 & 69 & 0\\\hline
2 & 0.0 & 0.2 & 94 & 59 & 35 & 45 & 7 & 8 & 2 & 35 & 0 & 58 & 0\\\hline
3 & 0.0 & 0.3 & 22 & 8 & 14 & 6 & 11 & 0 & 6 & 4 & 1 & 6 & 0\\\hline
4 & 0.1 & 0.1 & 83 & 61 & 22 & 61 & 8 & 4 & 1 & 34 & 0 & 61 & 0\\\hline
5 & 0.1 & 0.2 & 42 & 12 & 30 & 10 & 14 & 0 & 8 & 6 & 0 & 10 & 0\\\hline
6 & 0.1 & 0.3 & 71 & 35 & 36 & 24 & 8 & 8 & 1 & 17 & 0 & 29 & 0\\\hline
7 & 0.2 & 0.2 & 56 & 19 & 37 & 12 & 11 & 5 & 5 & 10 & 0 & 16 & 0\\\hline
8 & 0.3 & 0.3 & 34 & 10 & 24 & 7 & 18 & 0 & 5 & 4 & 2 & 8 & 0\\\hline
\end{tabular}
\end{lrbox}
\scalebox{0.75}{\usebox{\tablebox}}
\label{table-nonrigid}
\end{table}

\begin{figure*}[htb]
\centering
 \setlength{\abovecaptionskip}{0pt}
 \setlength{\belowcaptionskip}{0pt}
 \setlength{\intextsep}{8pt plus 3pt minus 2pt}
  \subfigure[]{
    \label{fig:mini:subfig:a}
    \begin{minipage}[c]{0.4\textwidth}
      \centering
      \includegraphics[width=2.5in]{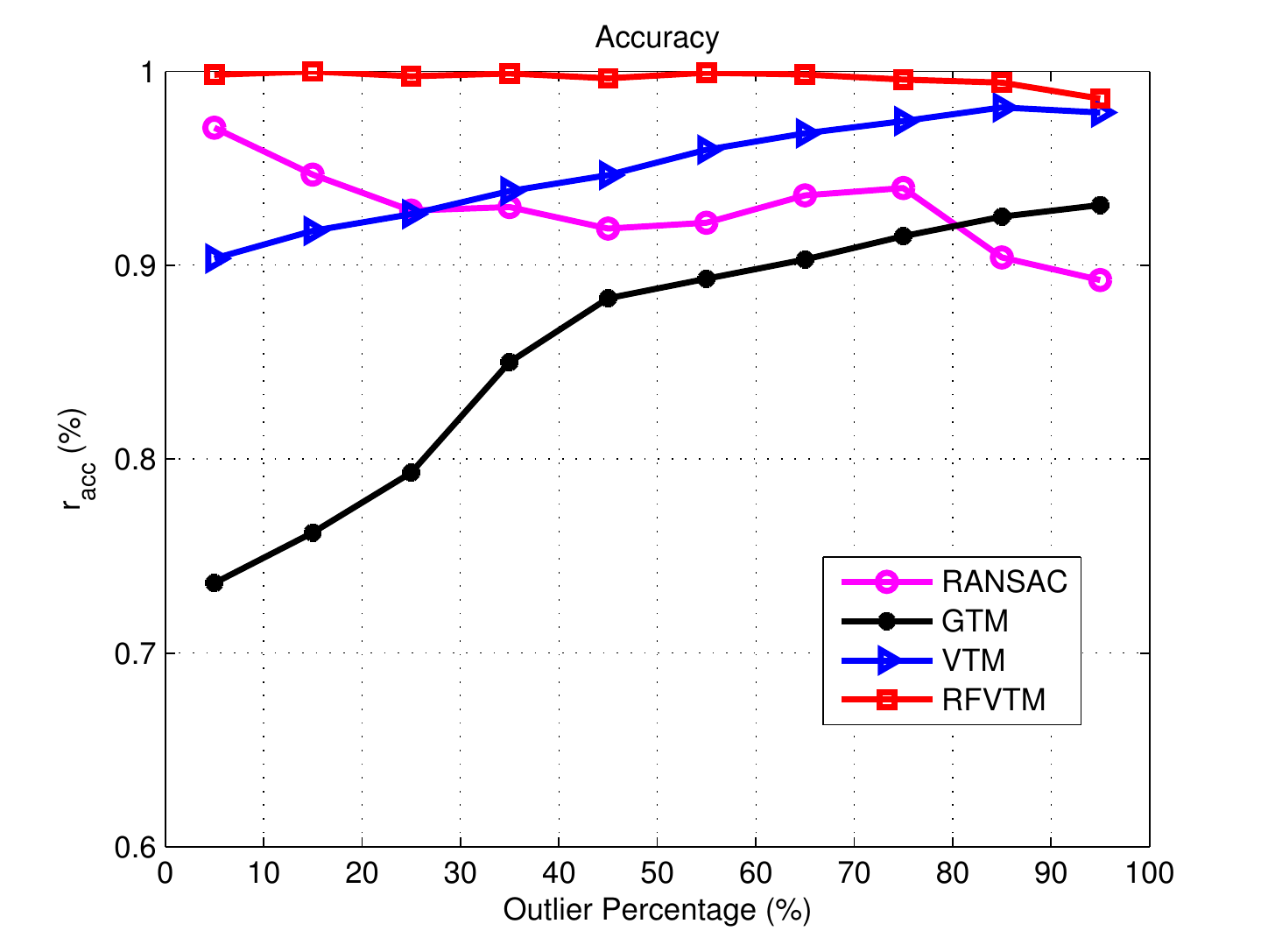}
    \end{minipage}}
  \subfigure[]{
    \label{fig:mini:subfig:b}
    \begin{minipage}[c]{0.4\textwidth}
      \centering
      \includegraphics[width=2.5in]{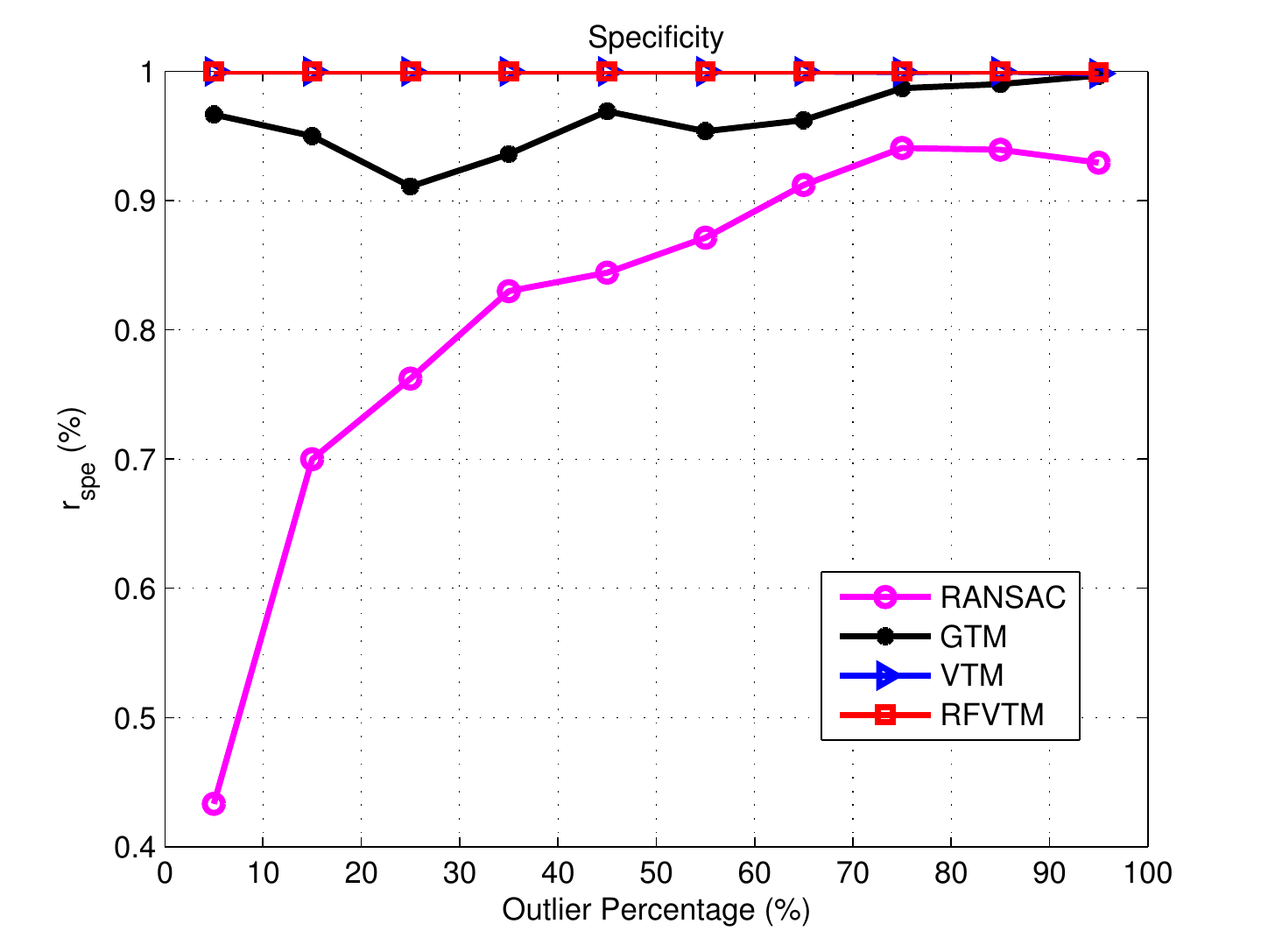}
    \end{minipage}}\\
  \subfigure[]{
    \label{fig:mini:subfig:c}
    \begin{minipage}[c]{0.4\textwidth}
      \centering
      \includegraphics[width=2.5in]{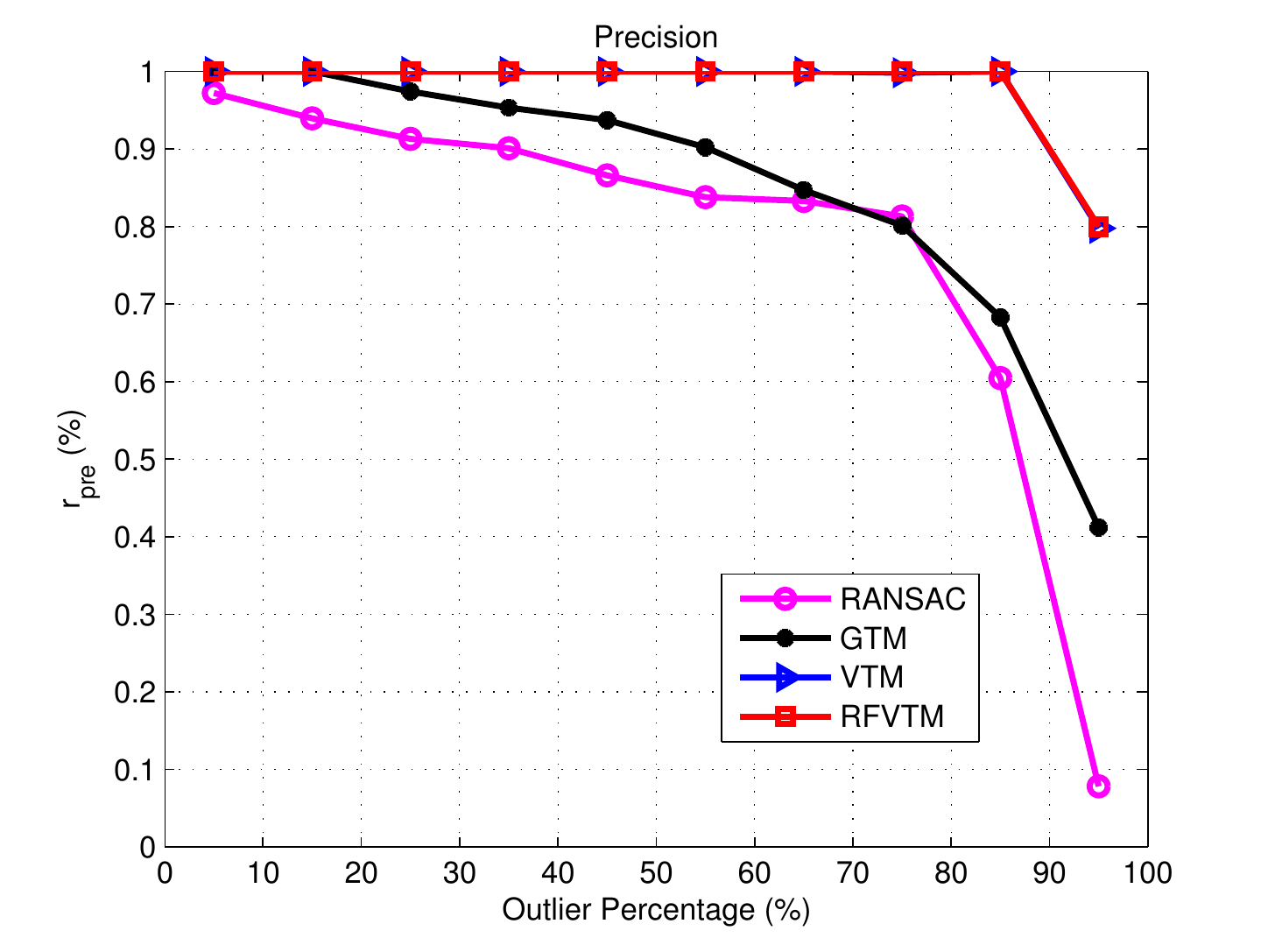}
    \end{minipage}}
  \subfigure[]{
    \label{fig:mini:subfig:a}
    \begin{minipage}[c]{0.4\textwidth}
      \centering
      \includegraphics[width=2.5in]{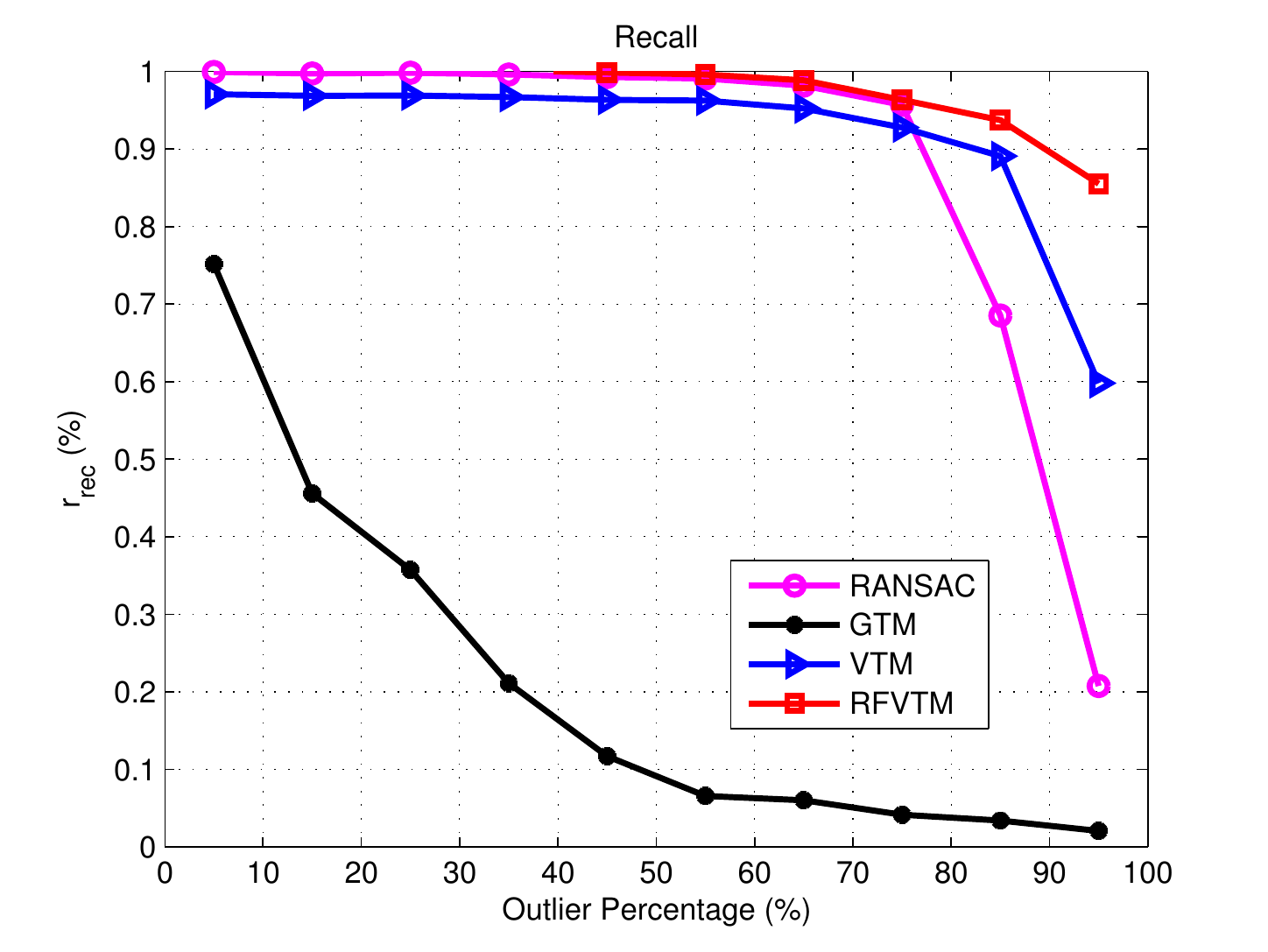}
    \end{minipage}}\\
  \captionstyle{normal}
  \caption{Performance comparison for RANSAC, GTM, VTM, and RFVTM methods under rotation and scale deformations.  (a) accuracy plots. (b) specificity plots. (c) precision plots. (d) recall plots.  }
  \label{fig-plots1_scaleandrotation}
\end{figure*}

\begin{figure*}[htb]
\centering
 \setlength{\abovecaptionskip}{0pt}
 \setlength{\belowcaptionskip}{0pt}
 \setlength{\intextsep}{8pt plus 3pt minus 2pt}
  \subfigure[]{
    \label{fig:mini:subfig:a}
    \begin{minipage}[c]{0.4\textwidth}
      \centering
      \includegraphics[width=2.5in]{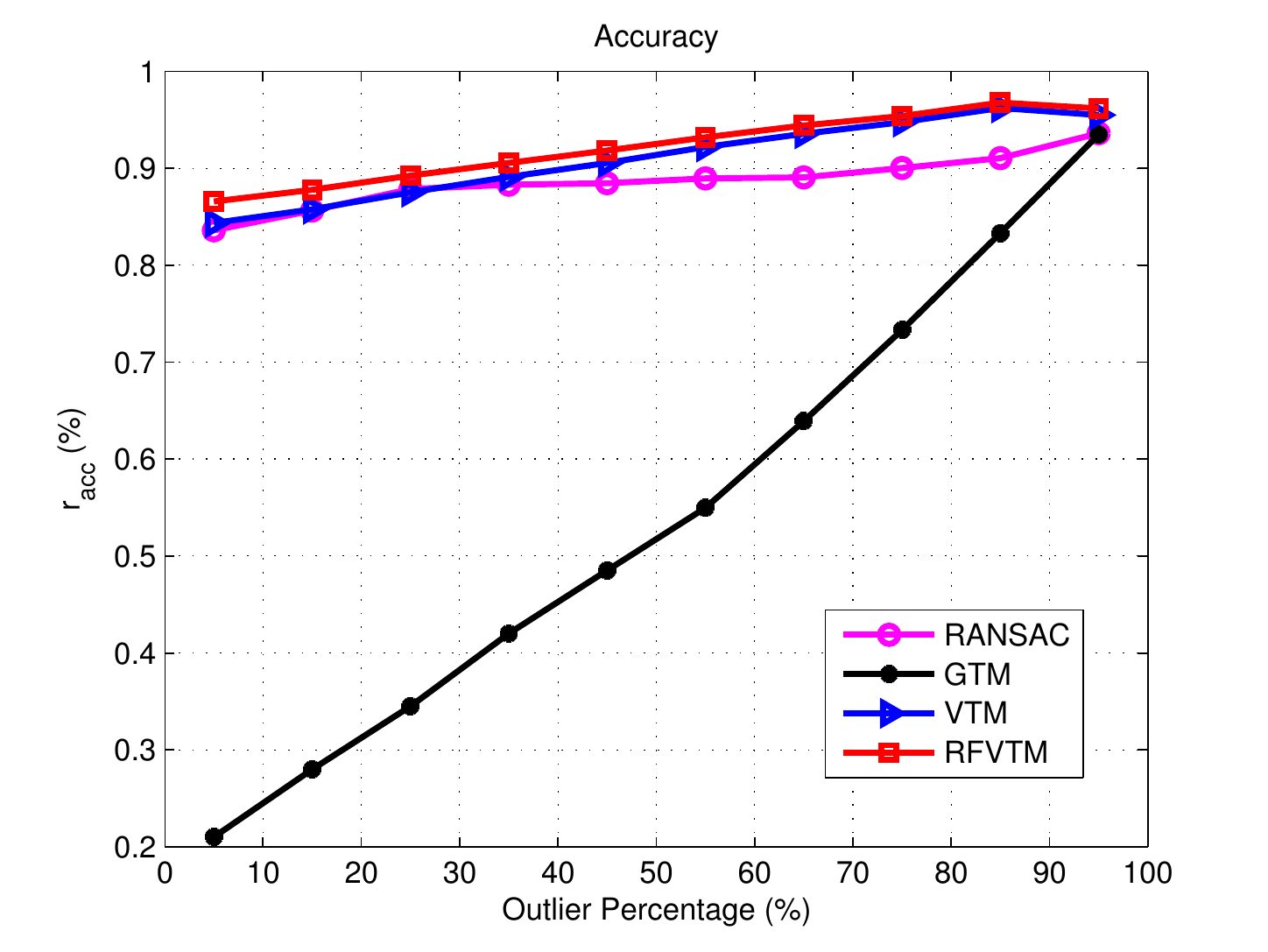}
    \end{minipage}}
  \subfigure[]{
    \label{fig:mini:subfig:b}
    \begin{minipage}[c]{0.4\textwidth}
      \centering
      \includegraphics[width=2.5in]{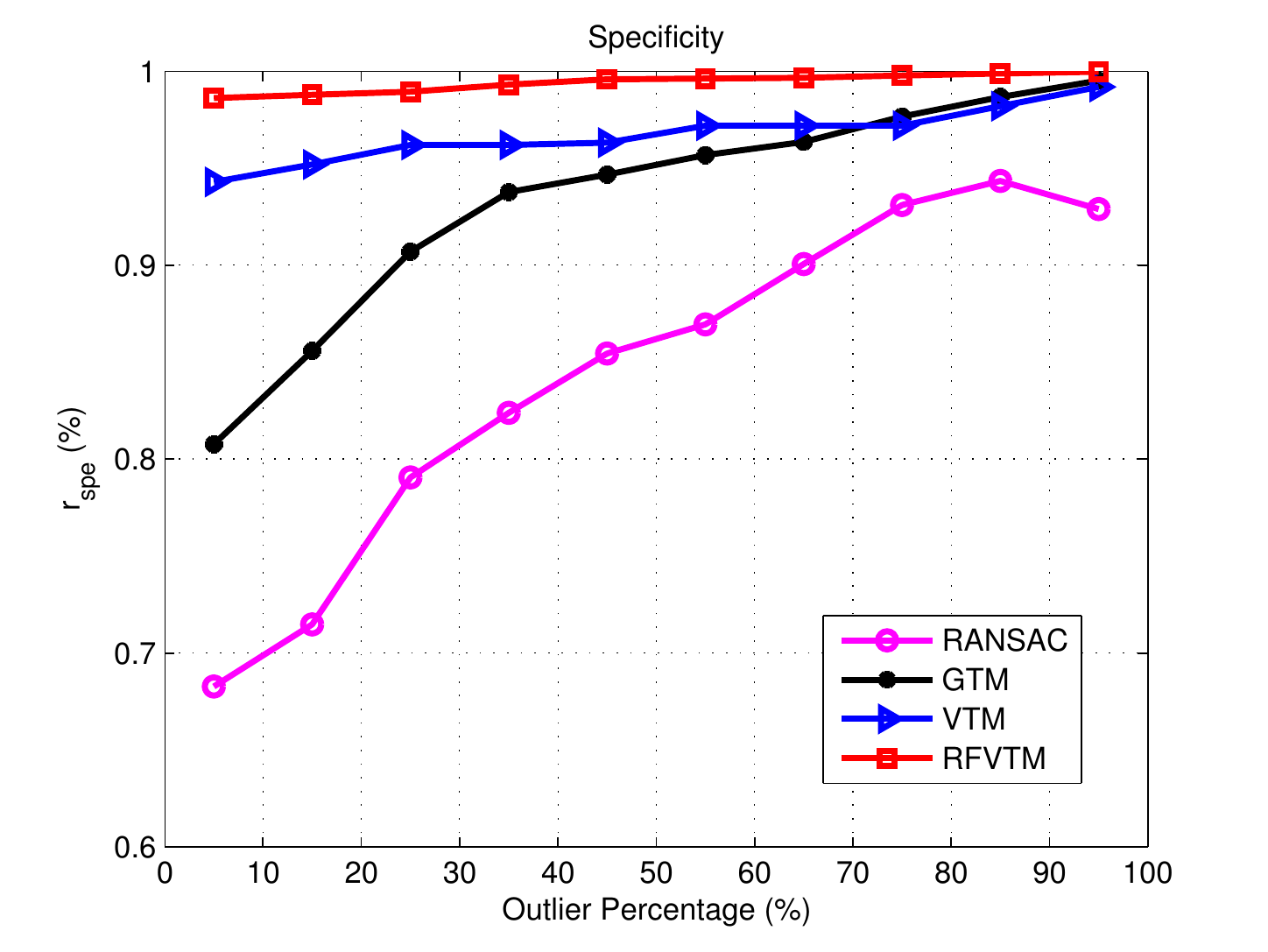}
    \end{minipage}}\\
  \subfigure[]{
    \label{fig:mini:subfig:c}
    \begin{minipage}[c]{0.4\textwidth}
      \centering
      \includegraphics[width=2.5in]{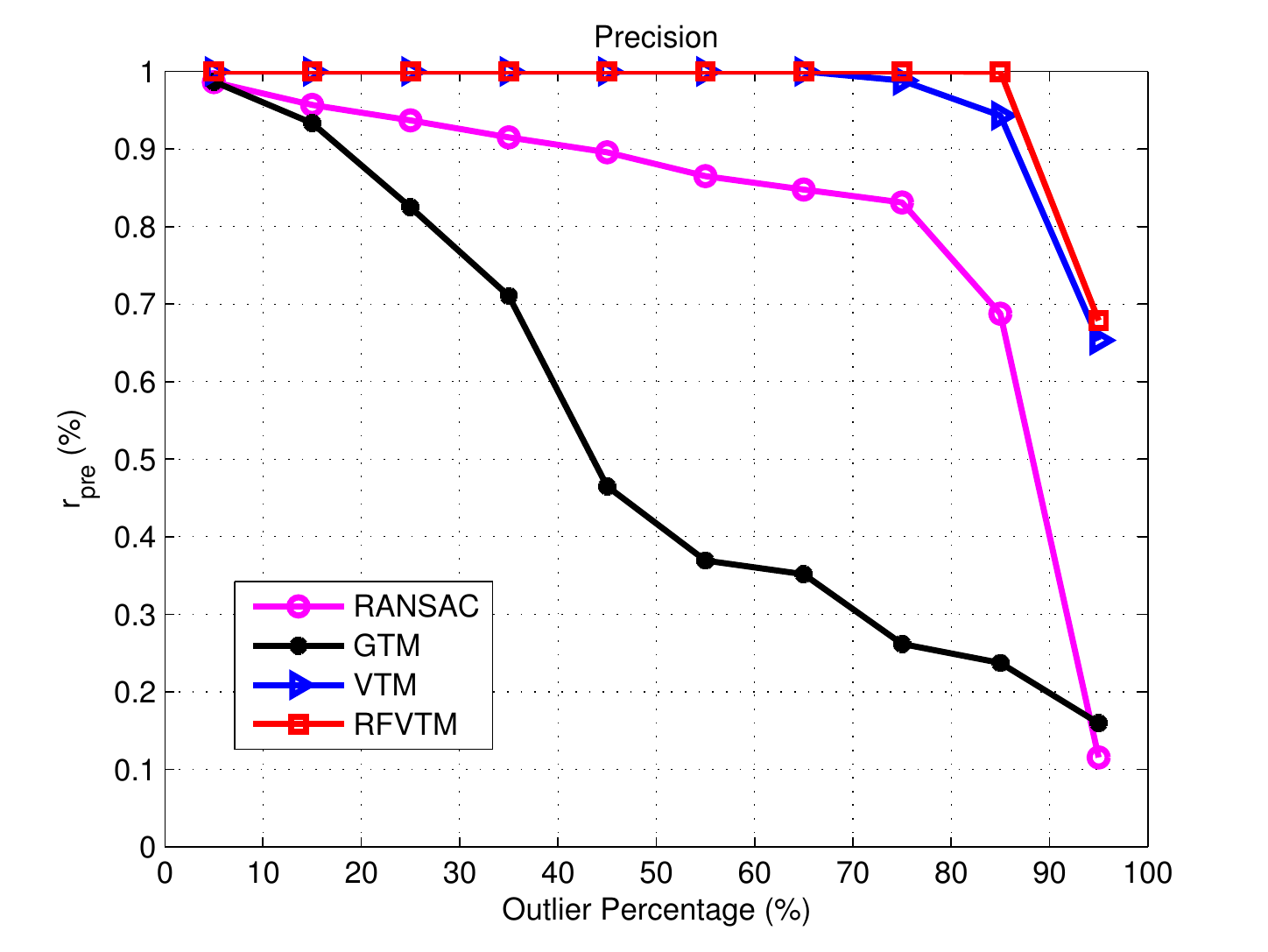}
    \end{minipage}}
  \subfigure[]{
    \label{fig:mini:subfig:a}
    \begin{minipage}[c]{0.4\textwidth}
      \centering
      \includegraphics[width=2.5in]{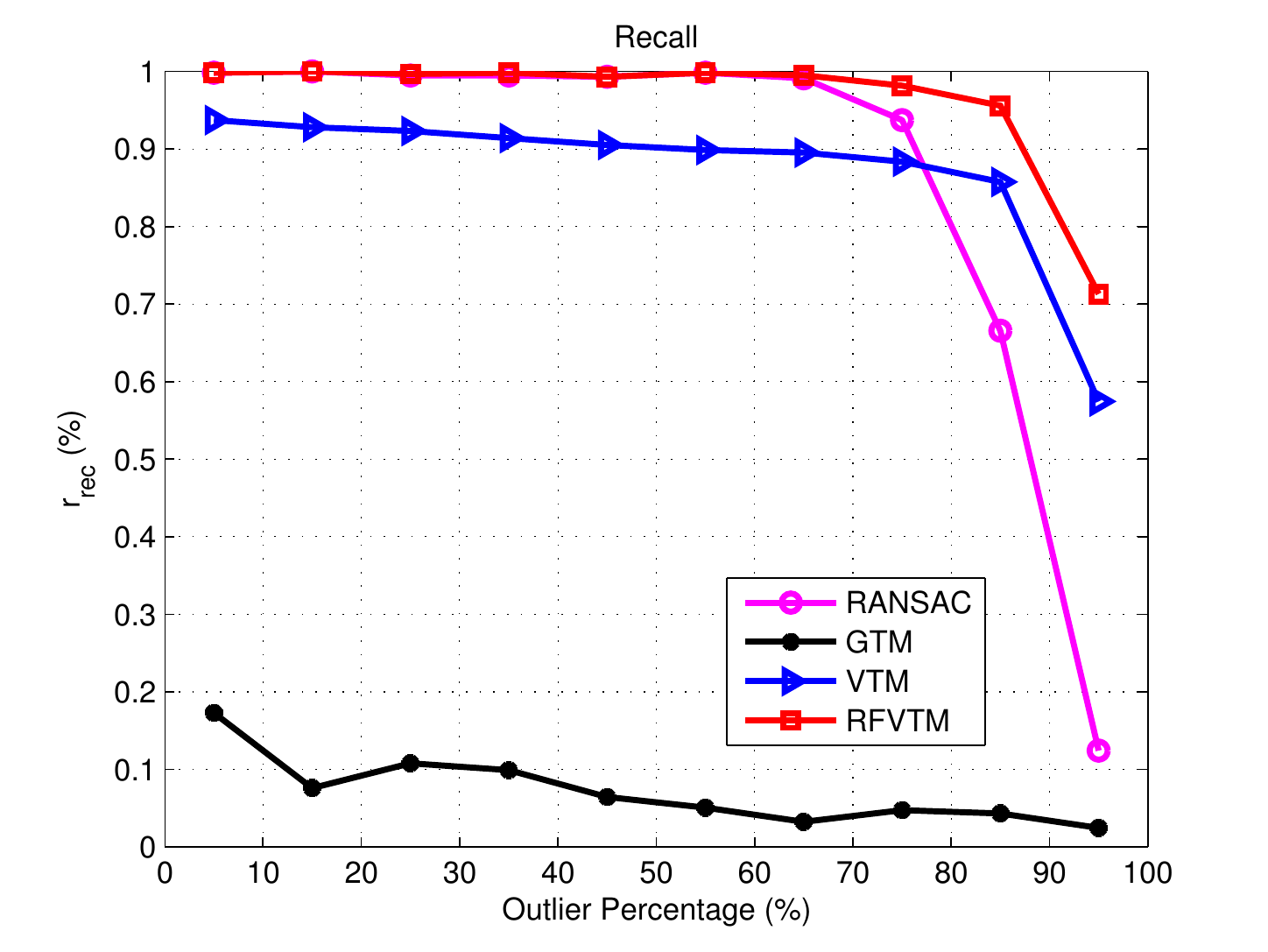}
    \end{minipage}}\\
  \captionstyle{normal}
  \caption{Performance comparison for RANSAC, GTM, VTM, and RFVTM methods under shear deformations. (a) accuracy plots. (b) specificity plots. (c) precision plots. (d) recall plots.  }
  \label{fig-plots2_shear}
\end{figure*}

\begin{itemize}
\item[\emph{1)}] \emph{Experiments on Images with Large Transformations}
\end{itemize}

The first image dataset corresponds to combinations of simulated affine transformation, including 12 simulated image pairs with rigid transformation (i.e., translation, rotation, scale deformations) and 8 simulated image pairs with non-rigid transformation (i.e., shear deformations). All of simulated images are randomly segmented into the same size as the original images with partially overlapping areas.
The results in this section are divided into two parts.
The first part demonstrates comparative matching results of RANSAC, GTM, VTM, and RFVTM for simulated image pairs.
The second part compares the average results of accuracy, specificity, precision, and recall for the four methods through introducing outliers.
Inliers are reserved by hand, and outliers are randomly added into the image pairs of the first image dataset. The percentage of outliers in total number of matches varies from 5\% to 95\% for each image pair.
The same process is repeated 100 times for every pair of images and for every percentage of outliers.

Fig. \ref{fig-im1} and Fig. \ref{fig-im2} demonstrate two visual matching examples of ImgSp1-1 and ImgSp1-2 respectively.
The example of ImgSp1-1 shown in Fig. \ref{fig-im1} consists of the aerial image covering the seacoast of San Diego and the simulated image with rotation of $120^ \circ$  anti-clockwise and scale factor of 2.
Fig. \ref{fig-im2} shows the example of ImgSp1-2, which consists of the aerial image covering the runway of an airport and the sheared image with h=0.1 and v=0.1.
Table \ref{table-rigid} and Table \ref{table-nonrigid} summarize the matching results of the four algorithms for 12 simulated image pairs with rigid transformations and 8 simulated image pairs with shear deformations respectively.
As presented in Table \ref{table-rigid} and \ref{table-nonrigid}, the matching results of VTM and RFVTM outperform RANSAC and GTM for both of rigid and non-rigid transformations. Although RANSAC preserves a large amount of inliers, many outliers still remain in the residual sets.
Compared with other three algorithms, GTM degenerates much more seriously for matching feature points with shear deformations. This is because KNN graph structure of GTM can not keep invariant to shear deformations.

The average performance of the four algorithms with different percentages of outliers are given in Fig. \ref{fig-plots1_scaleandrotation}  and Fig. \ref{fig-plots2_shear}  respectively.
Fig. \ref{fig-plots1_scaleandrotation} shows the average accuracy, specificity, precision, and recall for the 12 simulated image pairs with rotation and scale deformations.
It can be seen that VTM and RFVTM achieve similar specificity and precision values in Fig. \ref{fig-plots1_scaleandrotation} (b) and (c), both outperform RANSAC and GTM. It indicates that fewer false matches are reserved in the residual vertex sets of VTM and RFVTM. RANSAC provides a  recall value close to VTM but degenerates much more seriously when the proportion of outliers increase to 75\%. This is because RANSAC estimates the parameters of the transformation model from a set of correspondences containing outliers and produces reasonable results only within certain proportion of outliers. Since of the recovery strategy integrated in RFVTM, inliers mistakenly supported by outliers in previous iterations can be recovered. It brings the superior of RFVTM to other algorithms in terms of accuracy and recall values respectively in Fig. \ref{fig-plots1_scaleandrotation} (a) and (d).
Fig. \ref{fig-plots2_shear} shows the average accuracy, specificity, precision, and recall for the 8 simulated image pairs with shear deformations.
From the plots, it can be seen that VTM and RFVTM achieve higher precisions than the other algorithms, which validates their abilities to keep inliers in the consistent of shear deformations. Compared with rigid transformation, GTM for shear deformations is much worse than the other three algorithms in terms of precisions and returns poor recalls. This can be explained by the fact that GTM depends on the coherent adjacency relations of corresponding matches, which are not exactly invariant with shear deformations.

\begin{figure*}[htb]
\centering
 \setlength{\abovecaptionskip}{0pt}
 \setlength{\belowcaptionskip}{0pt}
 \setlength{\intextsep}{8pt plus 3pt minus 2pt}
  \subfigure[]{
    \label{fig:mini:subfig:a}
    \begin{minipage}[c]{0.3\textwidth}
      \centering
      \includegraphics[width=2.2in]{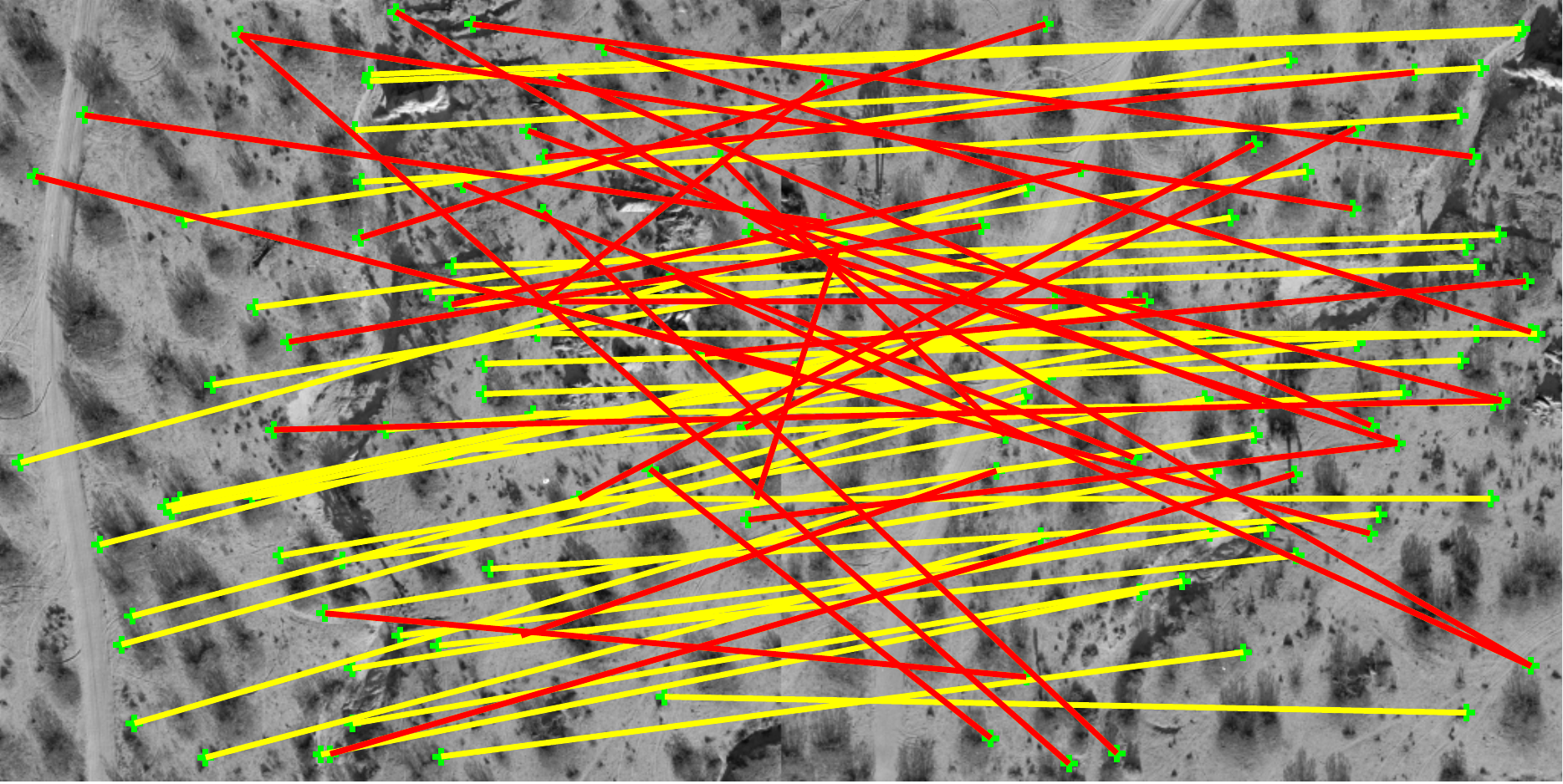}
    \end{minipage}}
  \subfigure[]{
    \label{fig:mini:subfig:b}
    \begin{minipage}[c]{0.3\textwidth}
      \centering
      \includegraphics[width=2.2in]{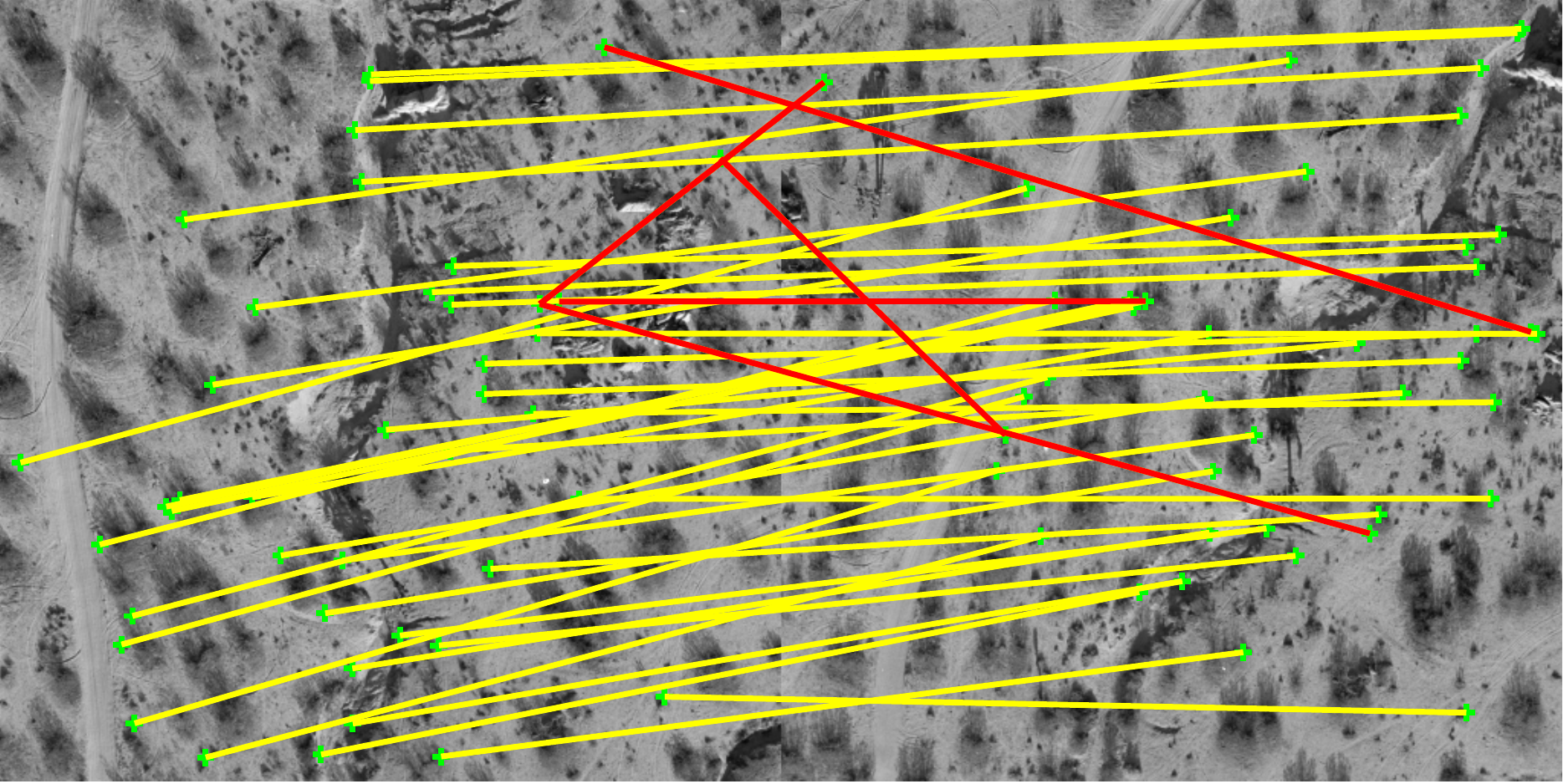}
    \end{minipage}}
  \subfigure[]{
    \label{fig:mini:subfig:c}
    \begin{minipage}[c]{0.3\textwidth}
      \centering
      \includegraphics[width=2.2in]{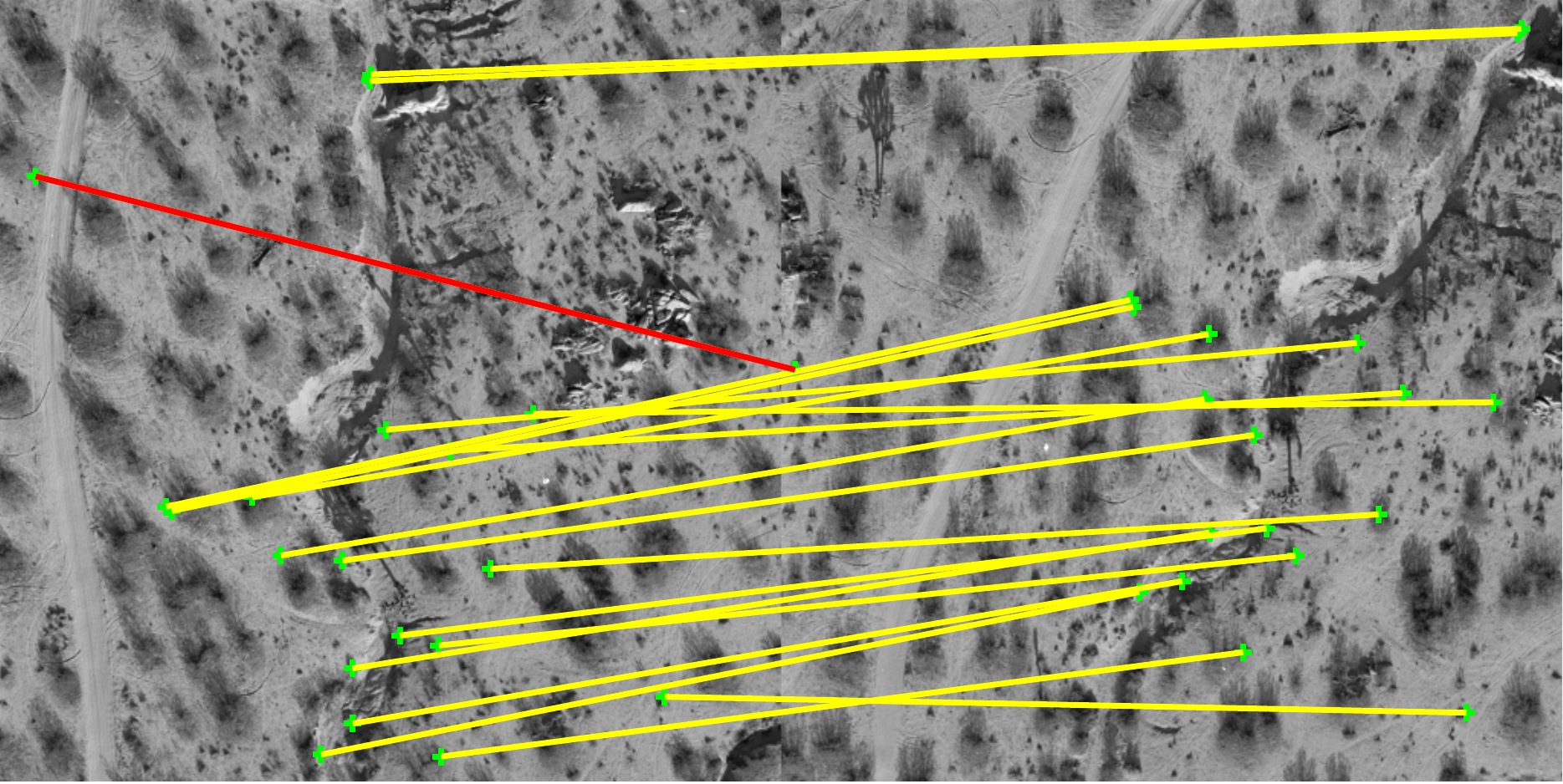}
    \end{minipage}}\\
  \subfigure[]{
    \label{fig:mini:subfig:a}
    \begin{minipage}[c]{0.3\textwidth}
      \centering
      \includegraphics[width=2.2in]{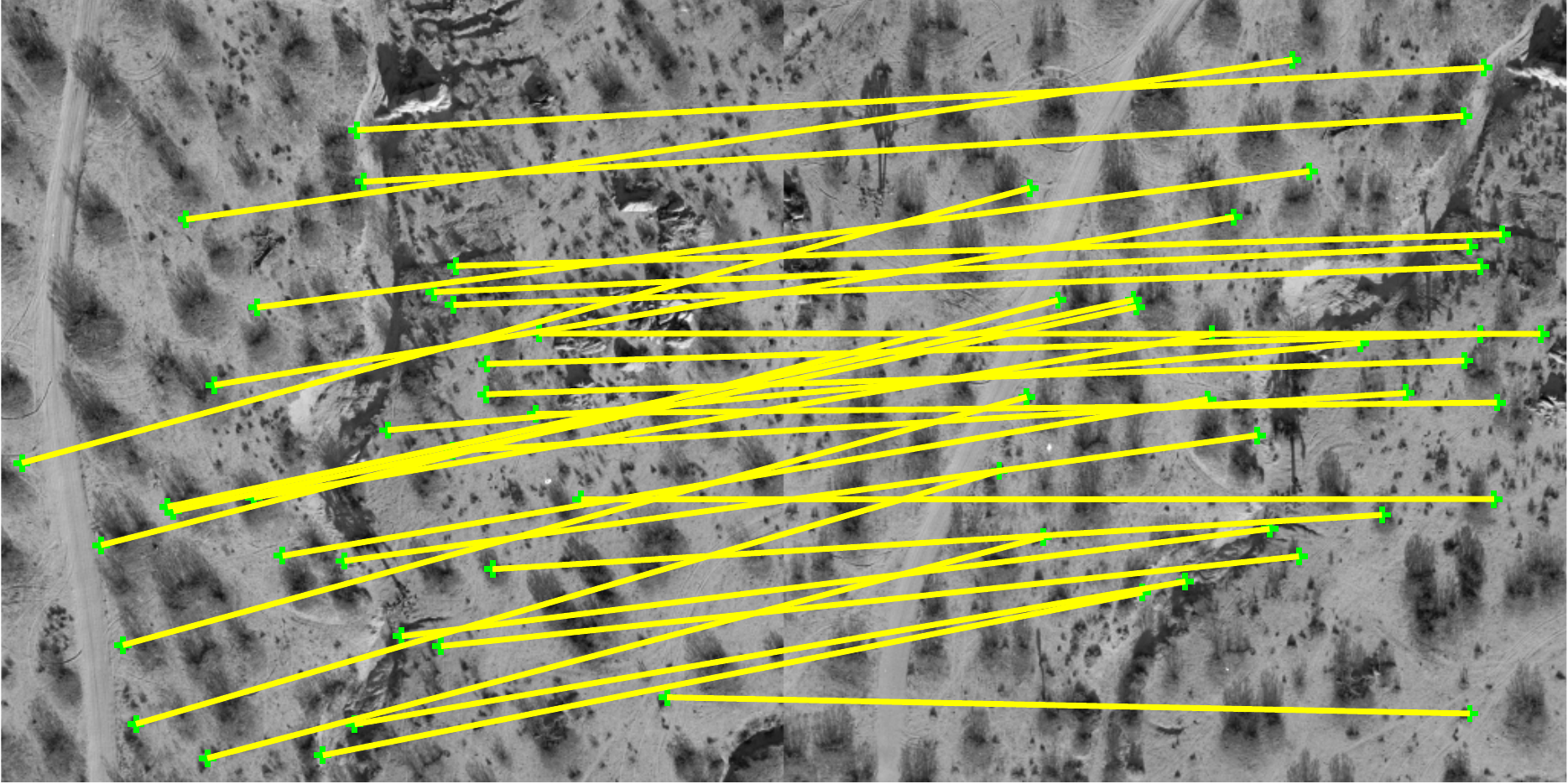}
    \end{minipage}}
  \subfigure[]{
    \label{fig:mini:subfig:a}
    \begin{minipage}[c]{0.3\textwidth}
      \centering
      \includegraphics[width=2.2in]{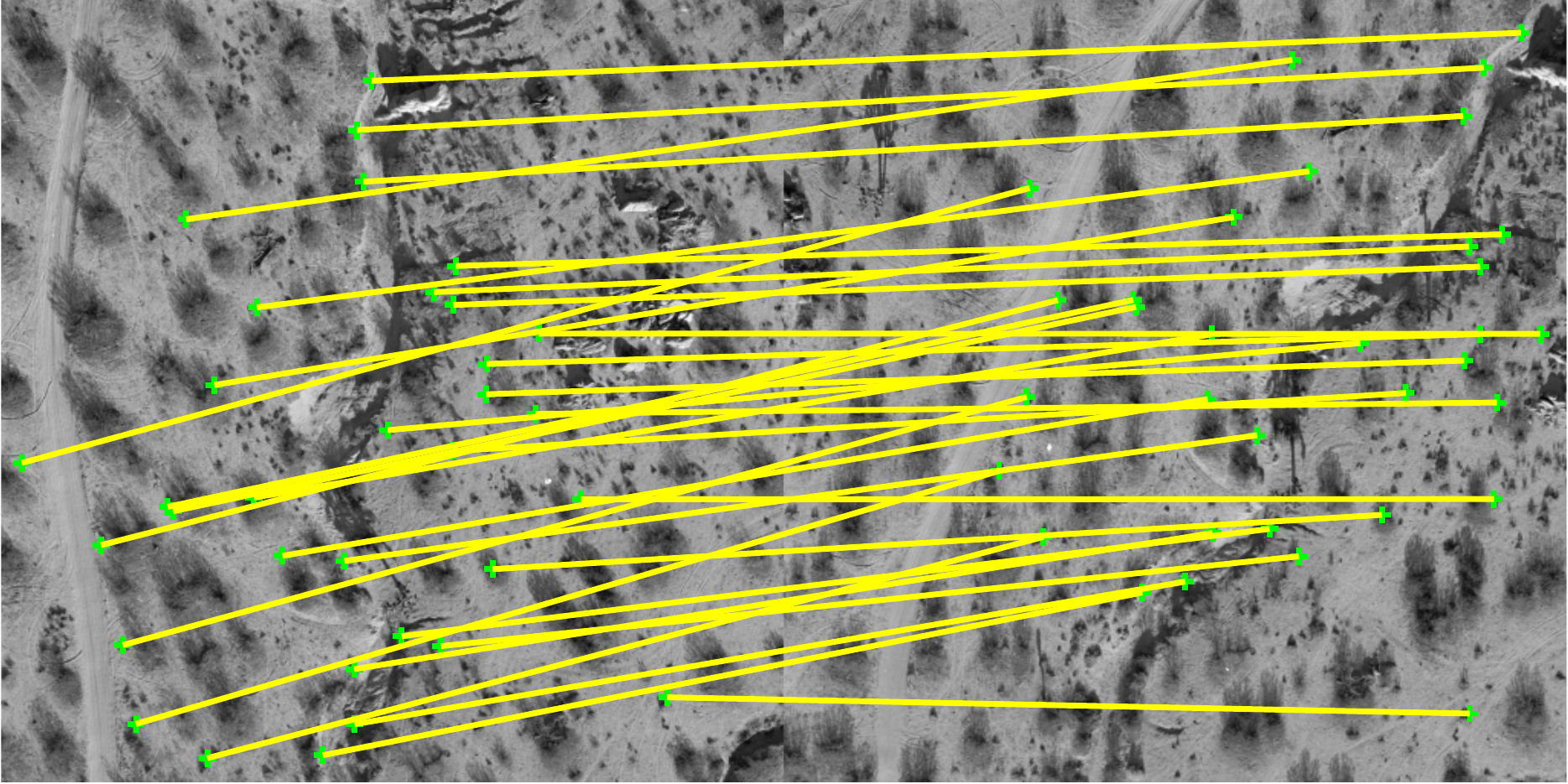}
    \end{minipage}}\\
  \captionstyle{normal}
  \caption{Examples of matching results for ImgSp2-1 with duplicated patterns. (a) SIFT. (b) RANSAC. (c) GTM. (d) VTM. (e) RFVTM.}
  \label{fig-im3}
\end{figure*}

\begin{figure*}[htb]
\centering
 \setlength{\abovecaptionskip}{0pt}
 \setlength{\belowcaptionskip}{0pt}
 \setlength{\intextsep}{8pt plus 3pt minus 2pt}
  \subfigure[]{
    \label{fig:mini:subfig:a}
    \begin{minipage}[c]{0.3\textwidth}
      \centering
      \includegraphics[width=2.2in]{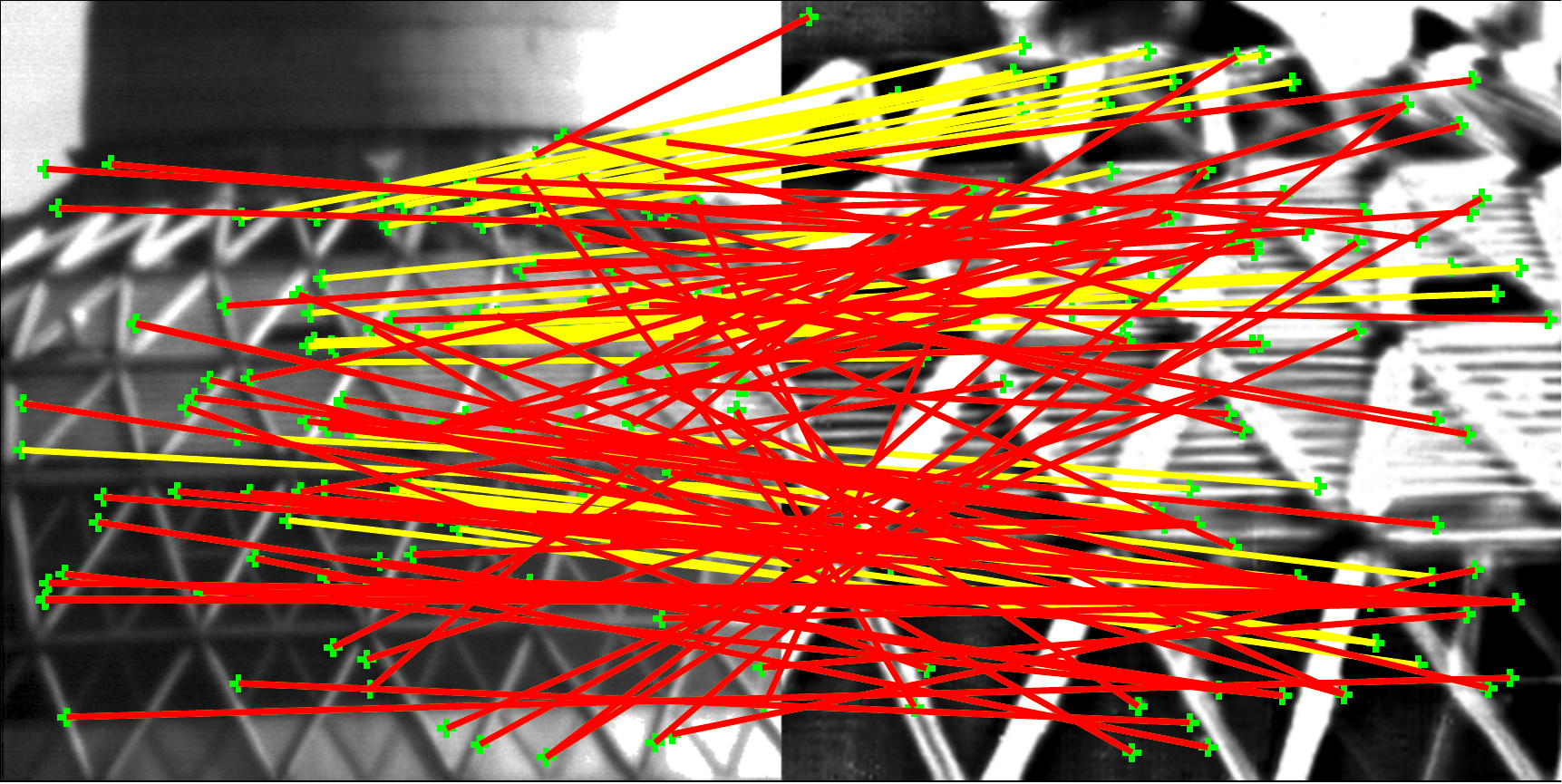}
    \end{minipage}}
  \subfigure[]{
    \label{fig:mini:subfig:b}
    \begin{minipage}[c]{0.3\textwidth}
      \centering
      \includegraphics[width=2.2in]{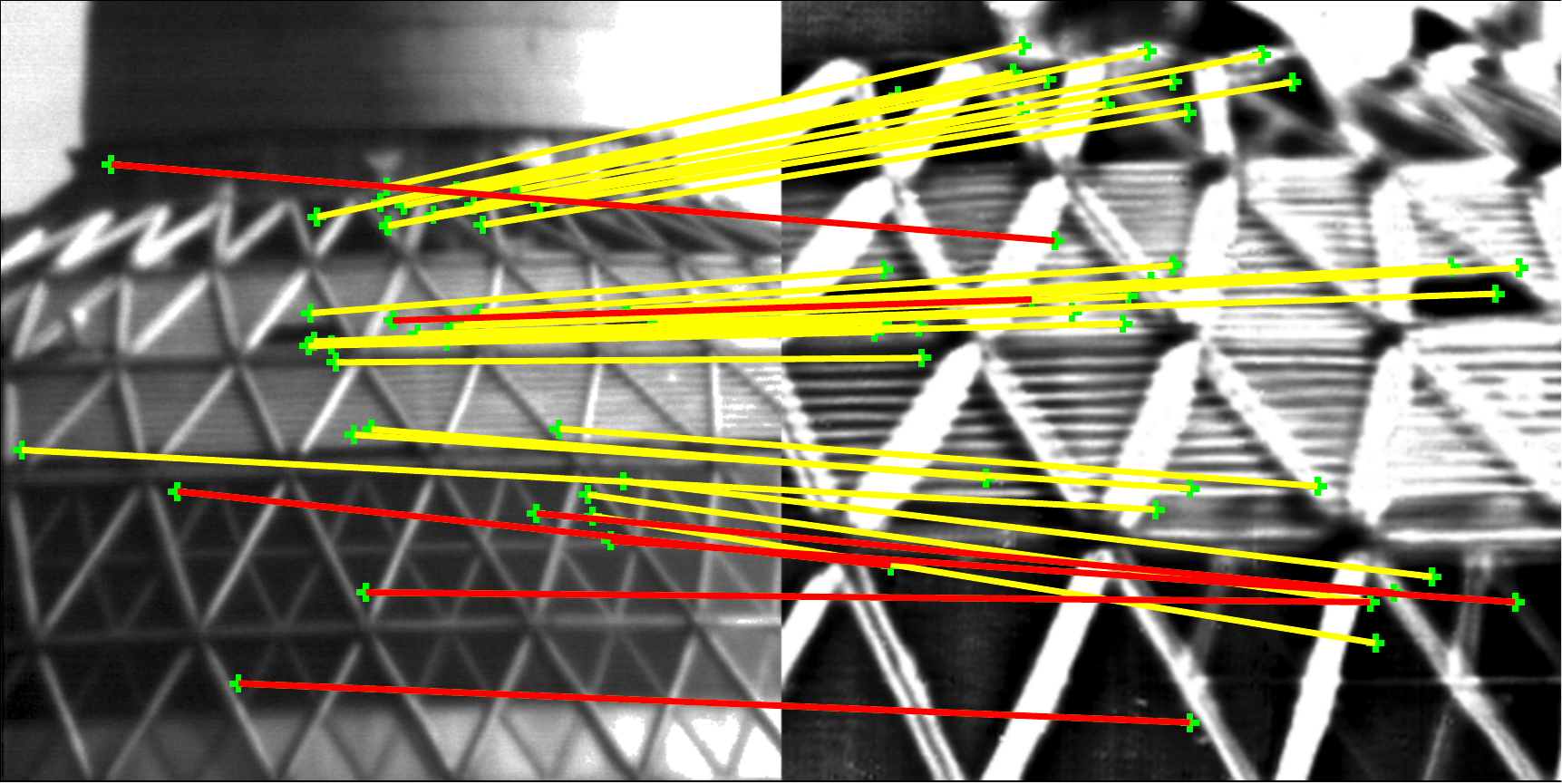}
    \end{minipage}}
  \subfigure[]{
    \label{fig:mini:subfig:c}
    \begin{minipage}[c]{0.3\textwidth}
      \centering
      \includegraphics[width=2.2in]{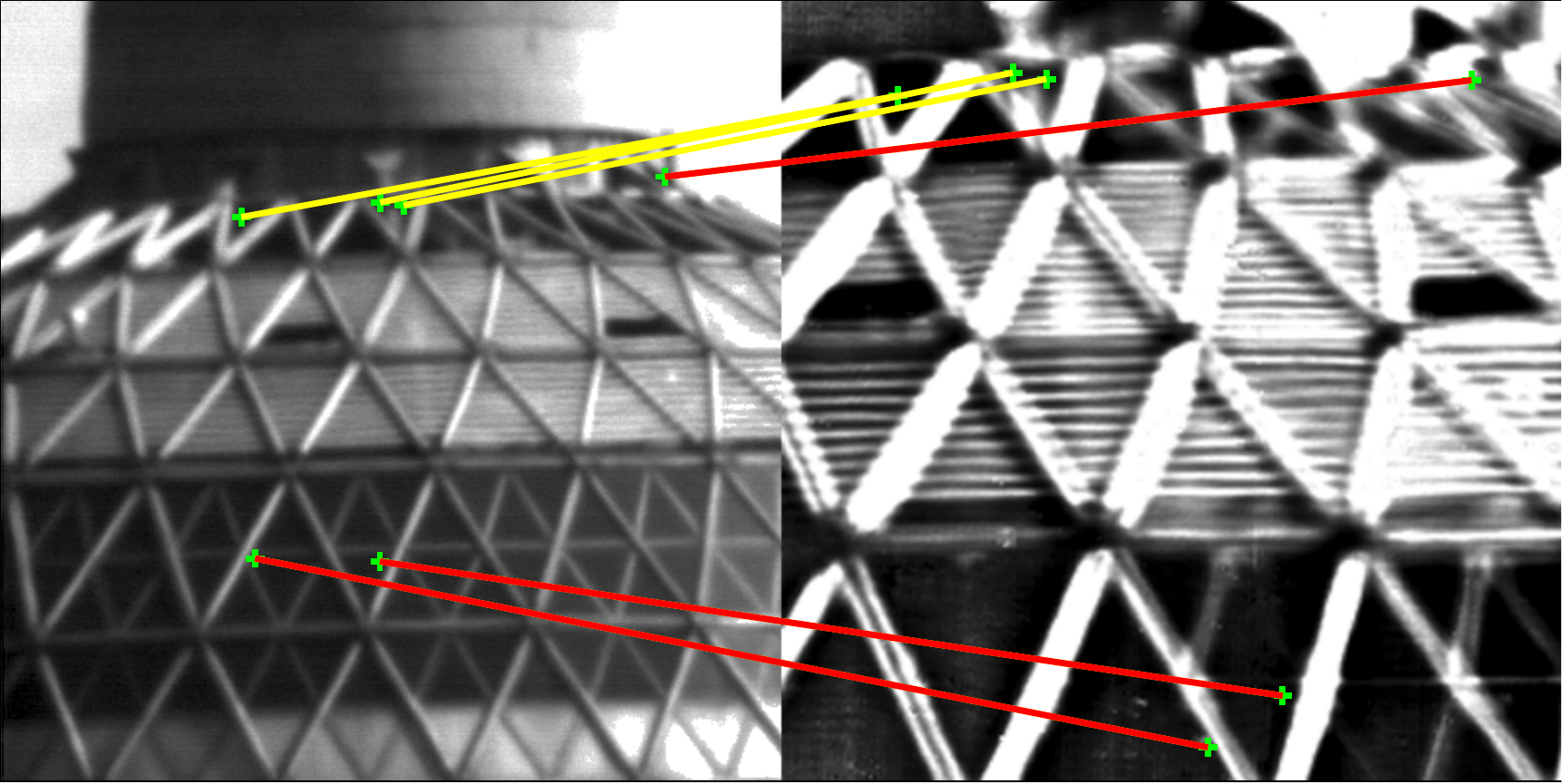}
    \end{minipage}}\\
  \subfigure[]{
    \label{fig:mini:subfig:a}
    \begin{minipage}[c]{0.3\textwidth}
      \centering
      \includegraphics[width=2.2in]{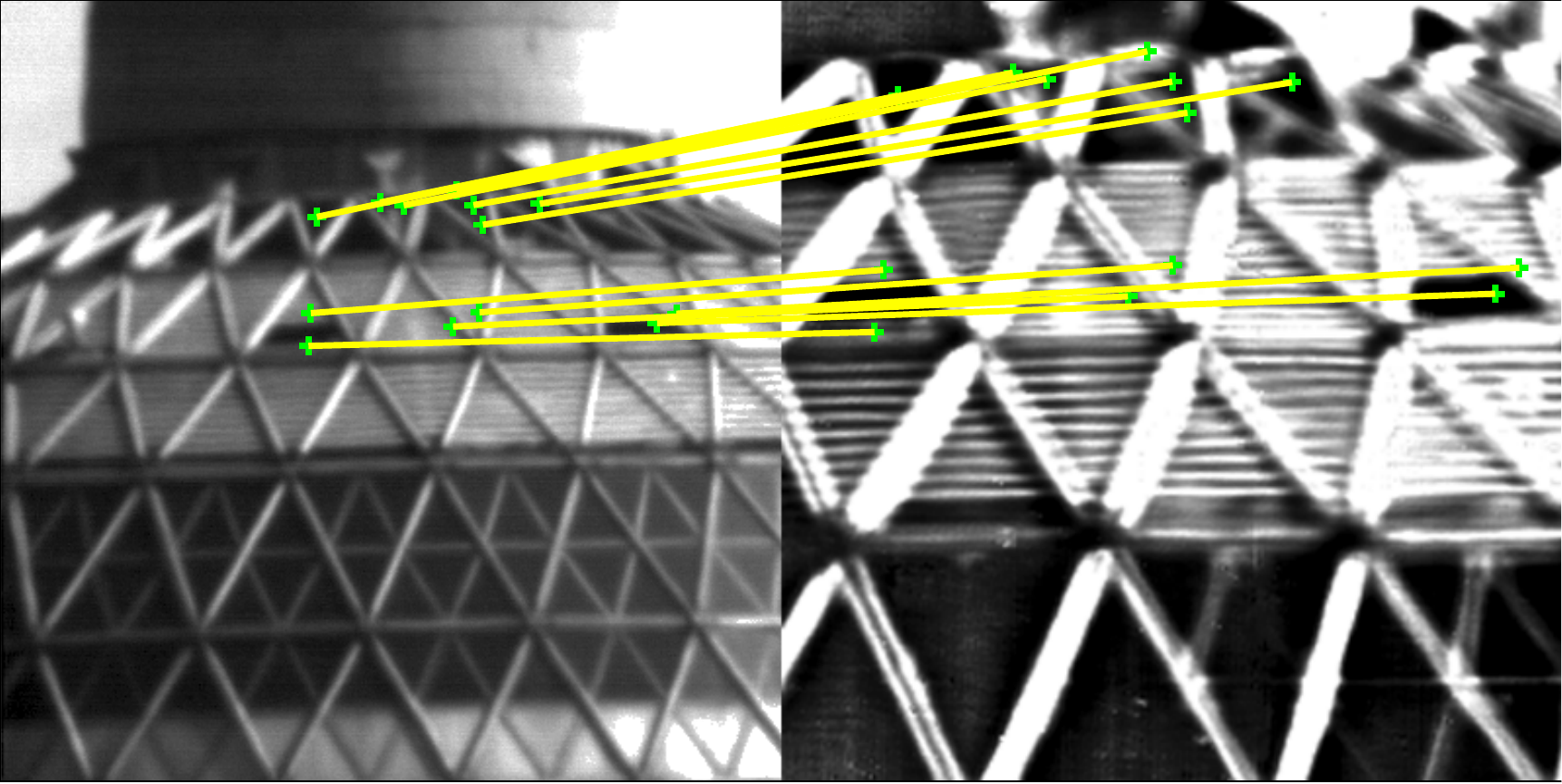}
    \end{minipage}}
  \subfigure[]{
    \label{fig:mini:subfig:a}
    \begin{minipage}[c]{0.3\textwidth}
      \centering
      \includegraphics[width=2.2in]{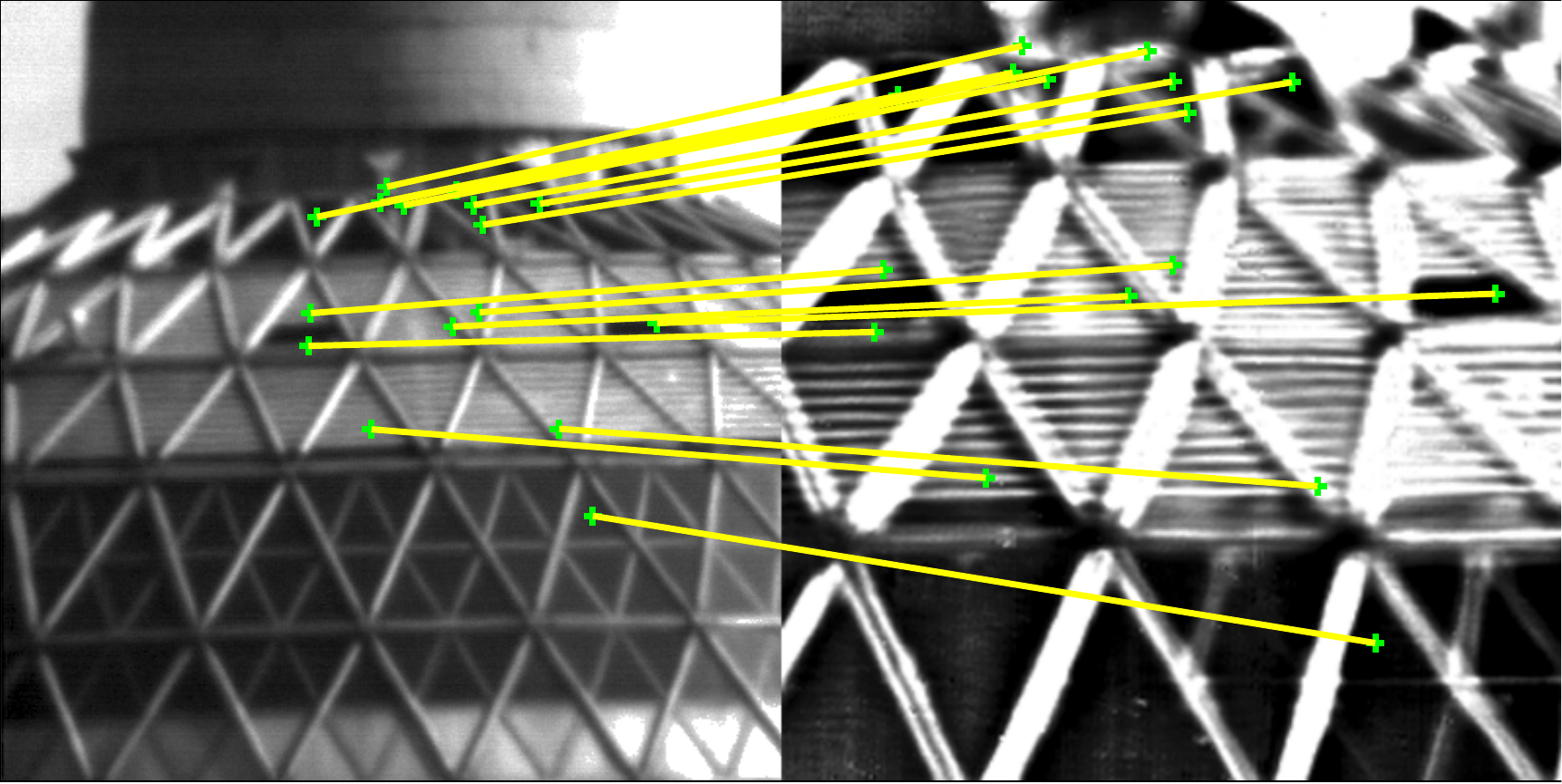}
    \end{minipage}}\\
  \captionstyle{normal}
  \caption{Examples of matching results for ImgSp2-2 with duplicated patterns.  (a) SIFT. (b) RANSAC. (c) GTM. (d) VTM. (e) RFVTM.}
  \label{fig-im4}
\end{figure*}

\begin{figure*}[htb]
\centering
 \setlength{\abovecaptionskip}{0pt}
 \setlength{\belowcaptionskip}{0pt}
 \setlength{\intextsep}{8pt plus 3pt minus 2pt}
  \subfigure[]{
    \label{fig:mini:subfig:a}
    \begin{minipage}[c]{0.4\textwidth}
      \centering
      \includegraphics[width=2.5in]{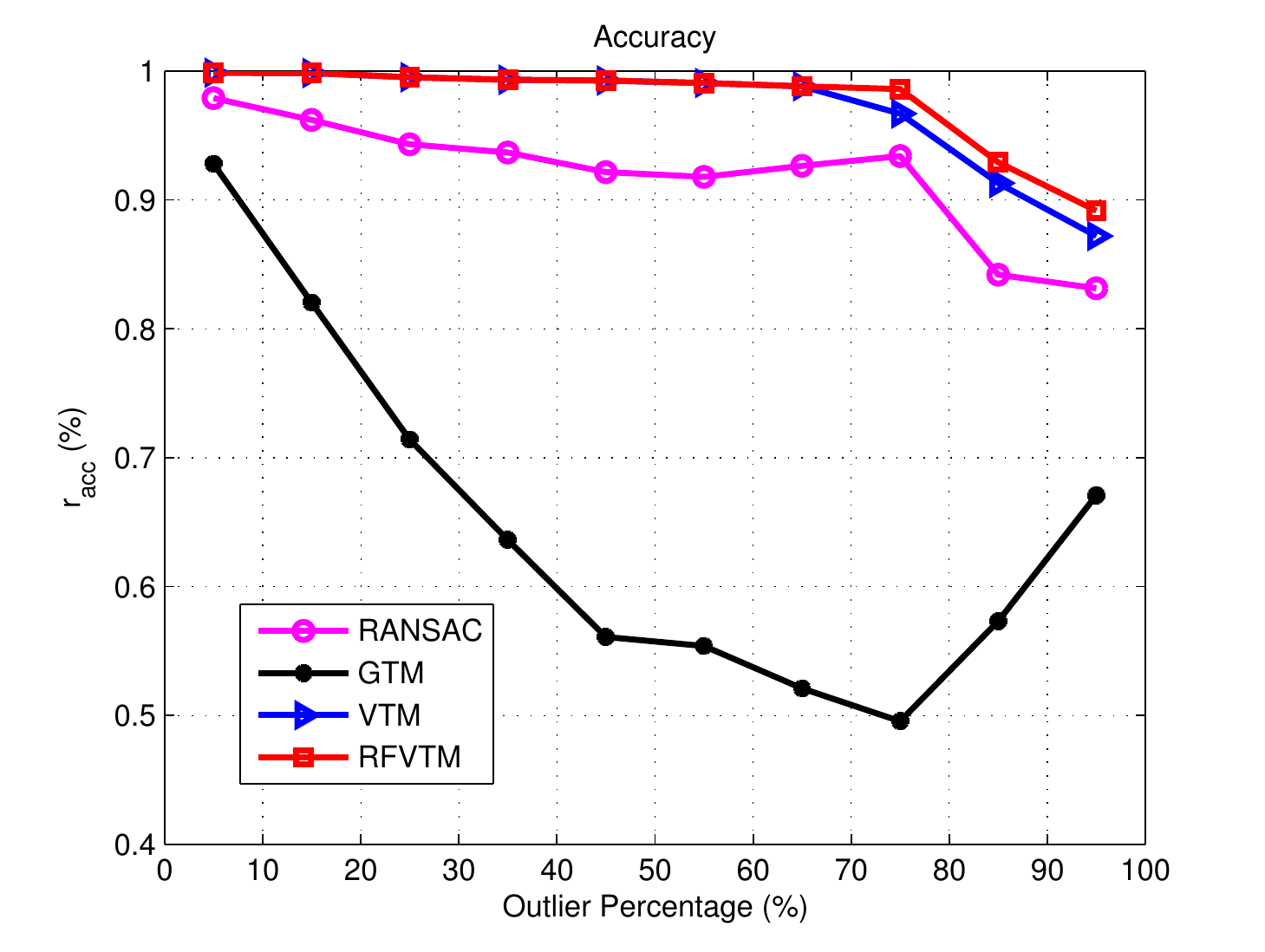}
    \end{minipage}}
  \subfigure[]{
    \label{fig:mini:subfig:b}
    \begin{minipage}[c]{0.4\textwidth}
      \centering
      \includegraphics[width=2.5in]{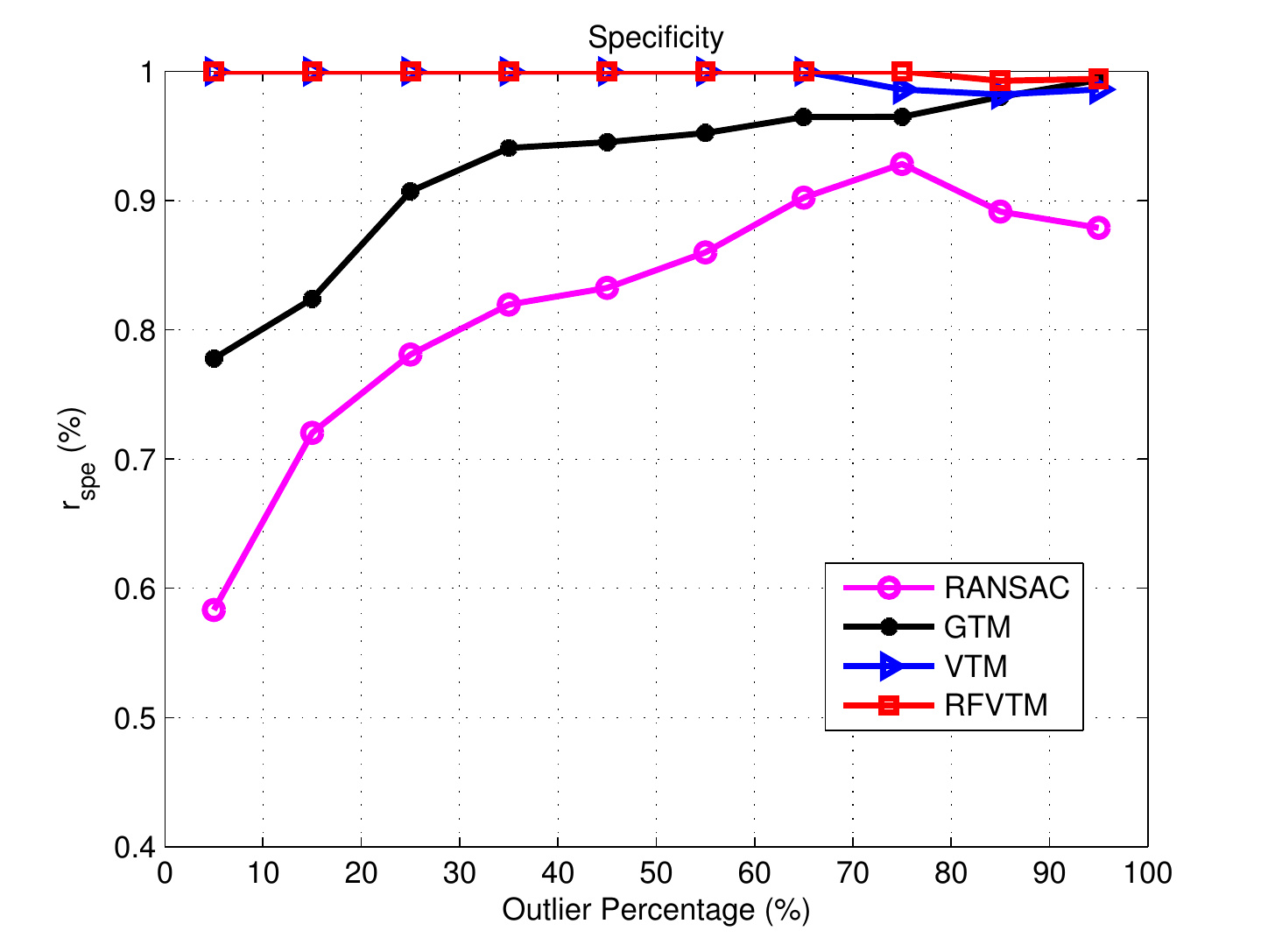}
    \end{minipage}}\\
  \subfigure[]{
    \label{fig:mini:subfig:c}
    \begin{minipage}[c]{0.4\textwidth}
      \centering
      \includegraphics[width=2.5in]{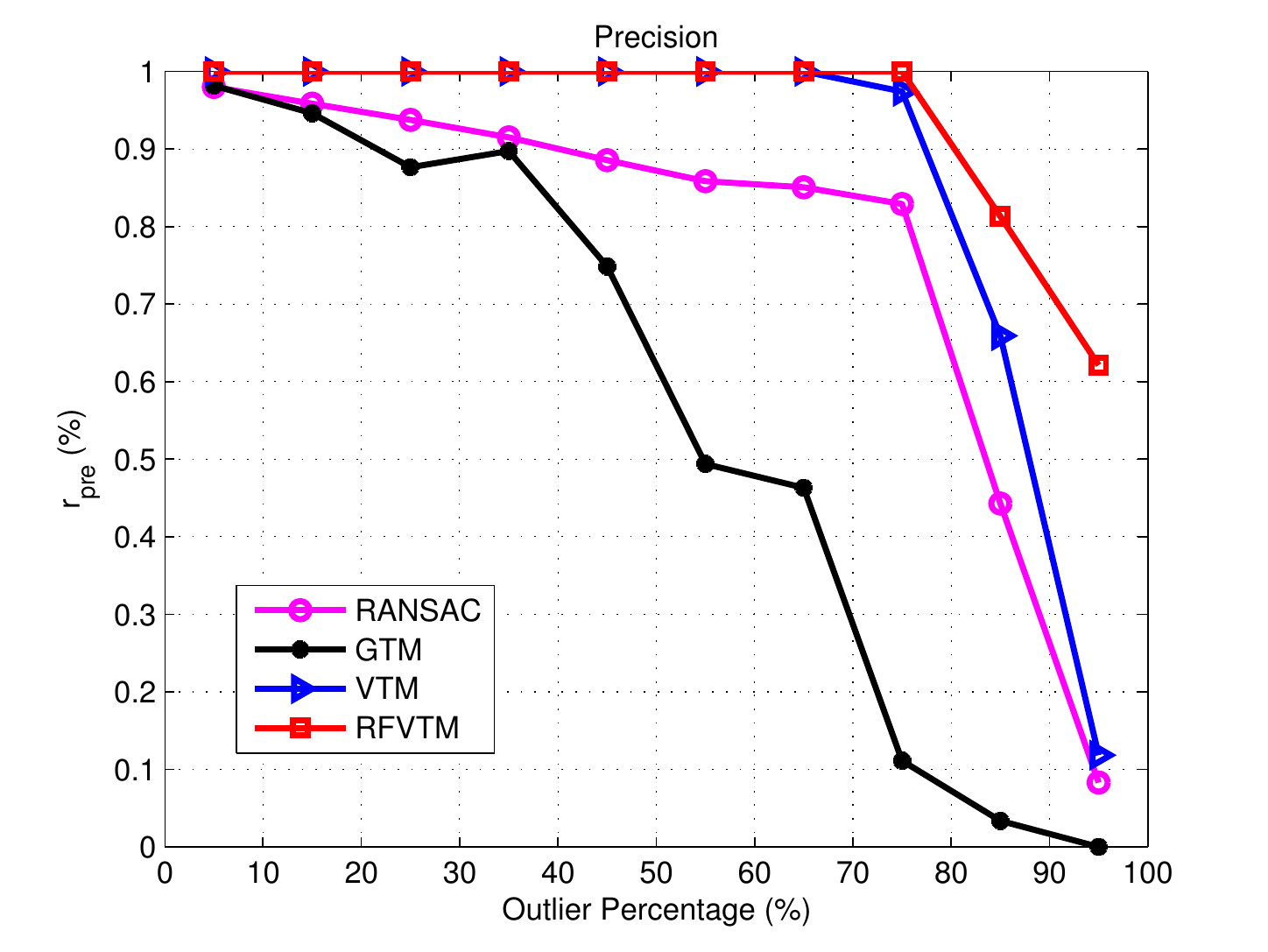}
    \end{minipage}}
  \subfigure[]{
    \label{fig:mini:subfig:a}
    \begin{minipage}[c]{0.4\textwidth}
      \centering
      \includegraphics[width=2.5in]{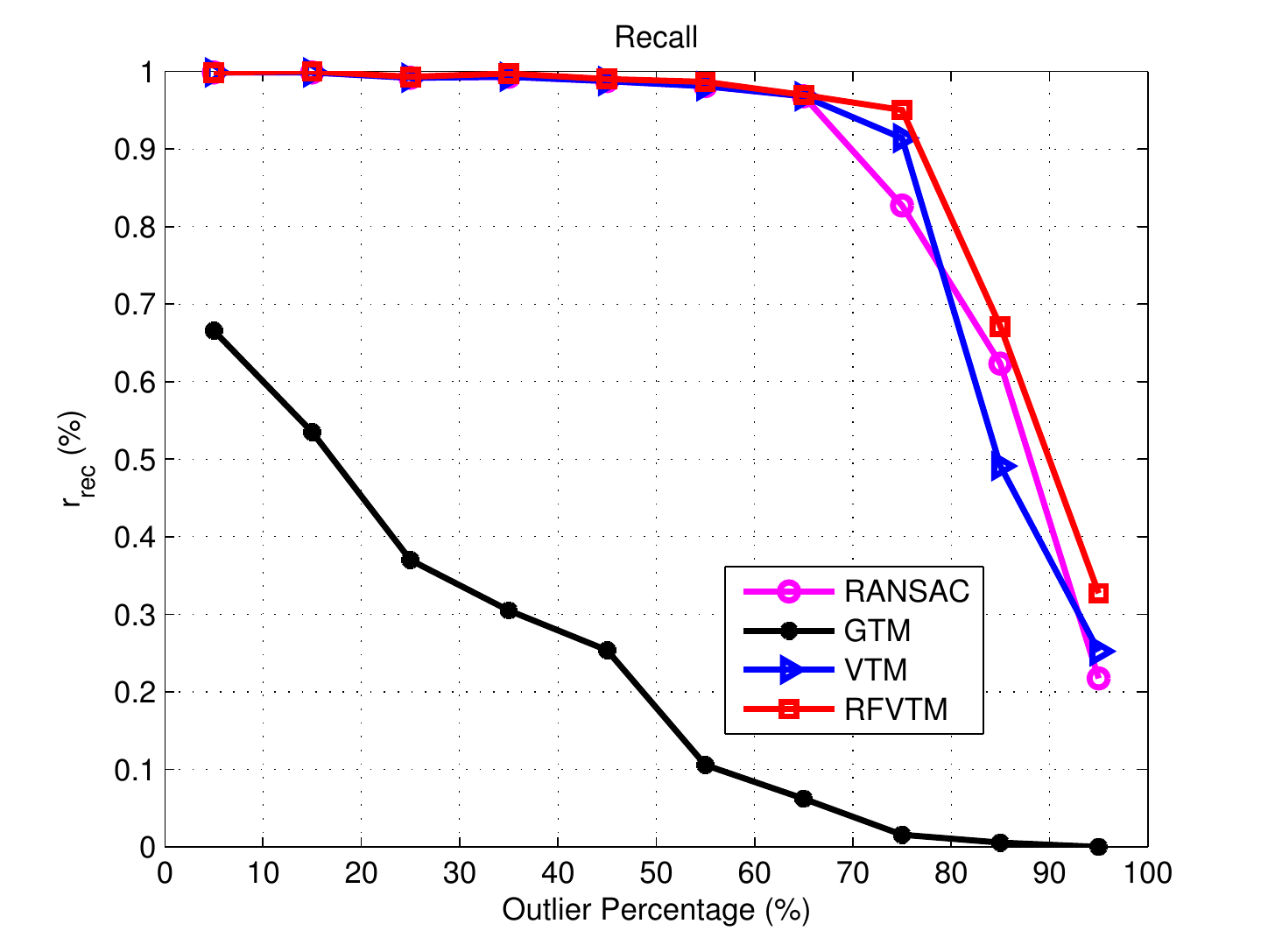}
    \end{minipage}}\\
  \captionstyle{normal}
  \caption{Performance comparison for RANSAC, GTM, VTM, and RFVTM methods under the ambiguities of duplicate patterns. (a) accuracy plots. (b) specificity plots. (c) precision plots. (d) recall plots.  }
  \label{fig-plots3_duplicate}
\end{figure*}

\begin{itemize}
\item[\emph{2)}] \emph{Experiments on Images with Duplicate Patterns}
\end{itemize}

The second image dataset contains twenty image pairs with duplicate patterns.
Fig. \ref{fig-im3} and Fig. \ref{fig-im4} demonstrate two matching samples for ImgSp2-1 and ImgSp2-2 respectively.
The image pair of Fig. \ref{fig-im3} consists of two remote sensing images with the same size of 400$\times$400 from Landsat TM Band 2 over an area with a large amount of similar earth surface.
The image pair of Fig. \ref{fig-im4} consists of VI and SWIR images with the size of 512$\times$512 from the ground test data of Experiment Satellites, which covers a partial area of ``Shanghai Oriental Pearl TV Tower" with many duplicated appearance of the TV tower.
Since similar patterns are duplicated in these images, many multiple mismatches associated with single feature points often exist in the similar but not really corresponding areas, as shown in Fig. \ref{fig-im3} (a) and Fig. \ref{fig-im4} (a).
It can be observed from Fig. \ref{fig-im3} (b)-(d) and Fig. \ref{fig-im4} (b)-(d) that VTM and RFVTM outperform RANSAC and GTM in terms of removing outliers. Despite of preserving as many inliers as RFVTM, RANSAC is incapable to remove a large amount of outliers.

Fig. \ref{fig-plots3_duplicate} provides the average performance of the four algorithms with different outliers. Inliers are manually selected from the initial matching sets for image pairs in the second dataset. Outliers are randomly reintroduced in increasing amounts, and added into each of the initial corresponding sets for the desired outlier proportions from 5\% to 95\%.
From Fig. \ref{fig-plots3_duplicate}, it can be noted that VTM and RFVTM outperform both RANSAC and GTM in terms of precision and recall values. Fig. \ref{fig-plots3_duplicate} (c) and (d) present the highest precision and recall among the four algorithms, and have the ability to keep values above 0.95 even for 75\% of outliers. In GTM, the nearest neighbors of the multiple matches in duplications are falsely identified by the same feature points corresponding to multiple matches. It results that inliers with multiple matches are mistakenly deemed as outliers. Therefore, GTM presents a low recall value but successfully filters out the outliers as long as the proportion of outliers is less than 25\%.

\begin{itemize}
\item[\emph{3)}] \emph{Experiments on Images with Inconsistent Spectral Content}
\end{itemize}

\begin{figure*}[htb]
\centering
 \setlength{\abovecaptionskip}{0pt}
 \setlength{\belowcaptionskip}{0pt}
 \setlength{\intextsep}{8pt plus 3pt minus 2pt}
  \subfigure[]{
    \label{fig:mini:subfig:a}
    \begin{minipage}[c]{0.3\textwidth}
      \centering
      \includegraphics[width=2.2in]{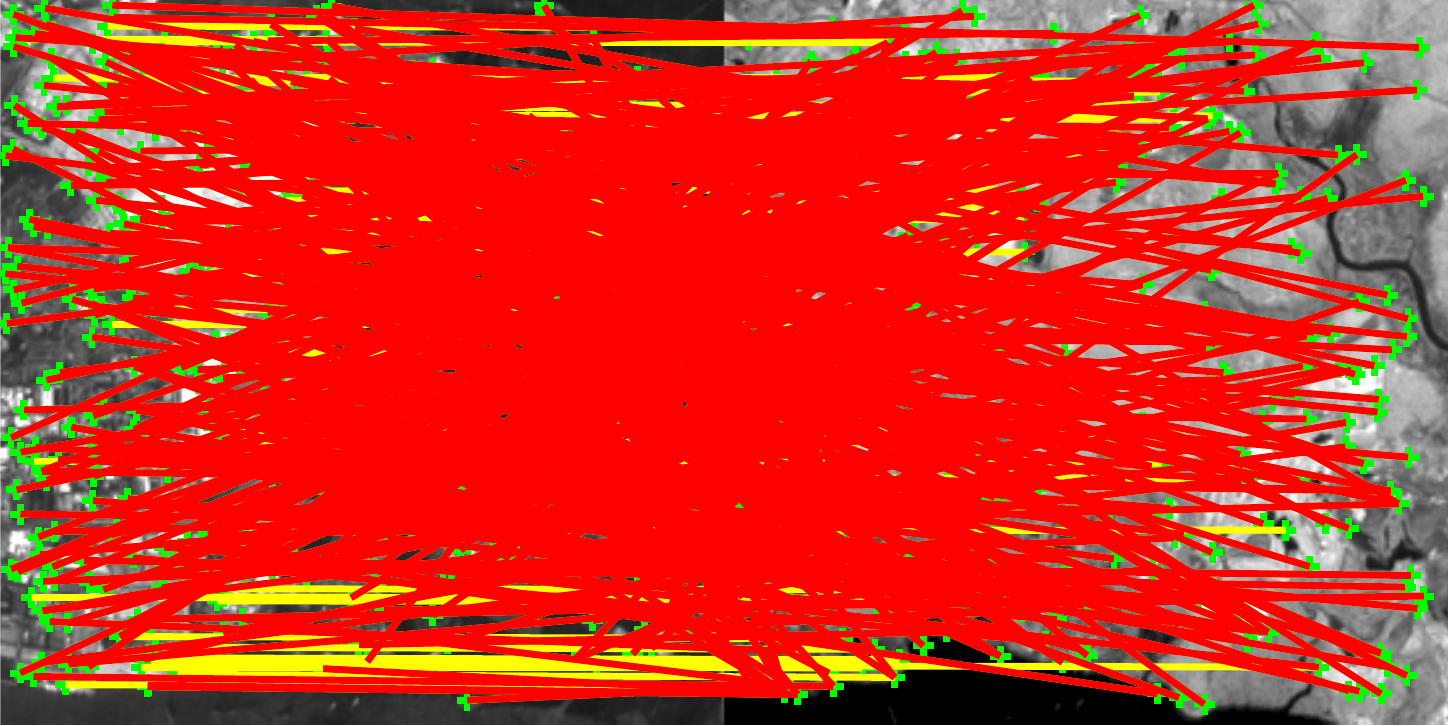}
    \end{minipage}}
  \subfigure[]{
    \label{fig:mini:subfig:b}
    \begin{minipage}[c]{0.3\textwidth}
      \centering
      \includegraphics[width=2.2in]{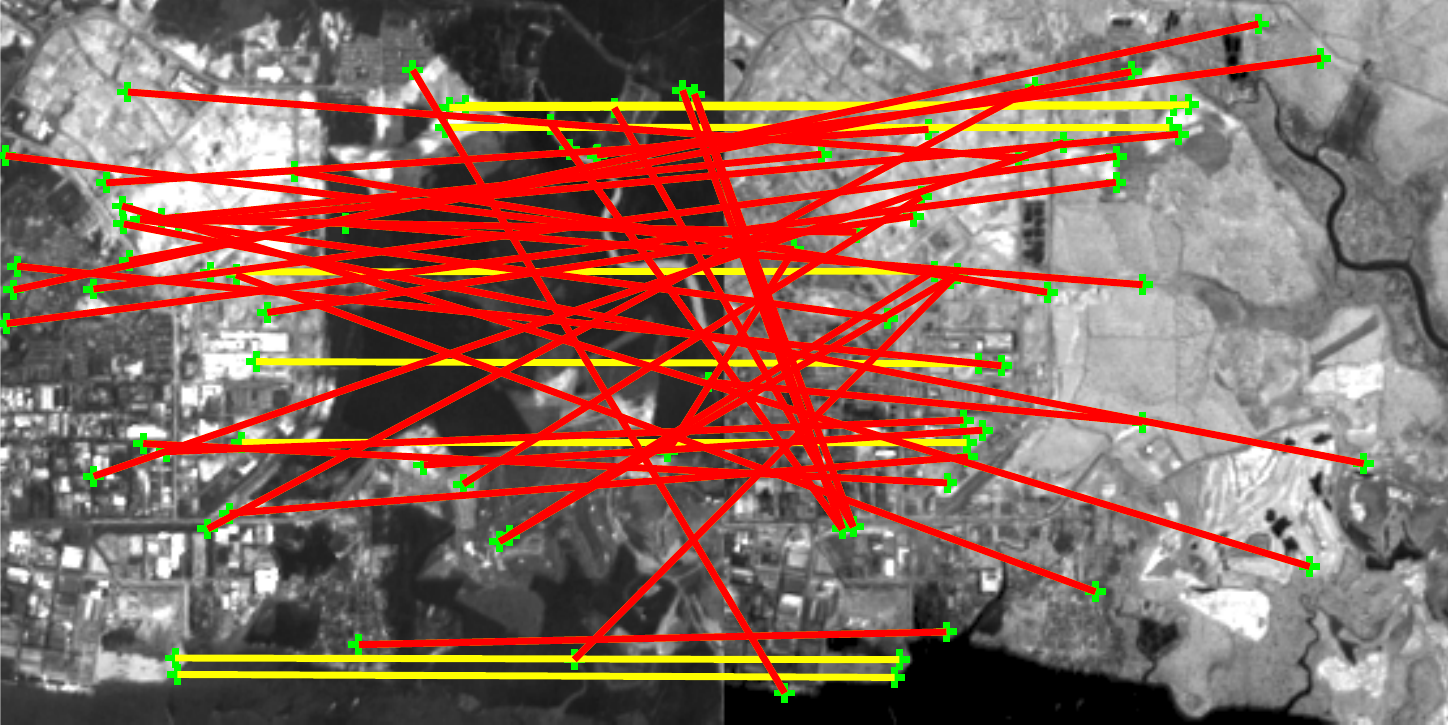}
    \end{minipage}}
  \subfigure[]{
    \label{fig:mini:subfig:c}
    \begin{minipage}[c]{0.3\textwidth}
      \centering
      \includegraphics[width=2.2in]{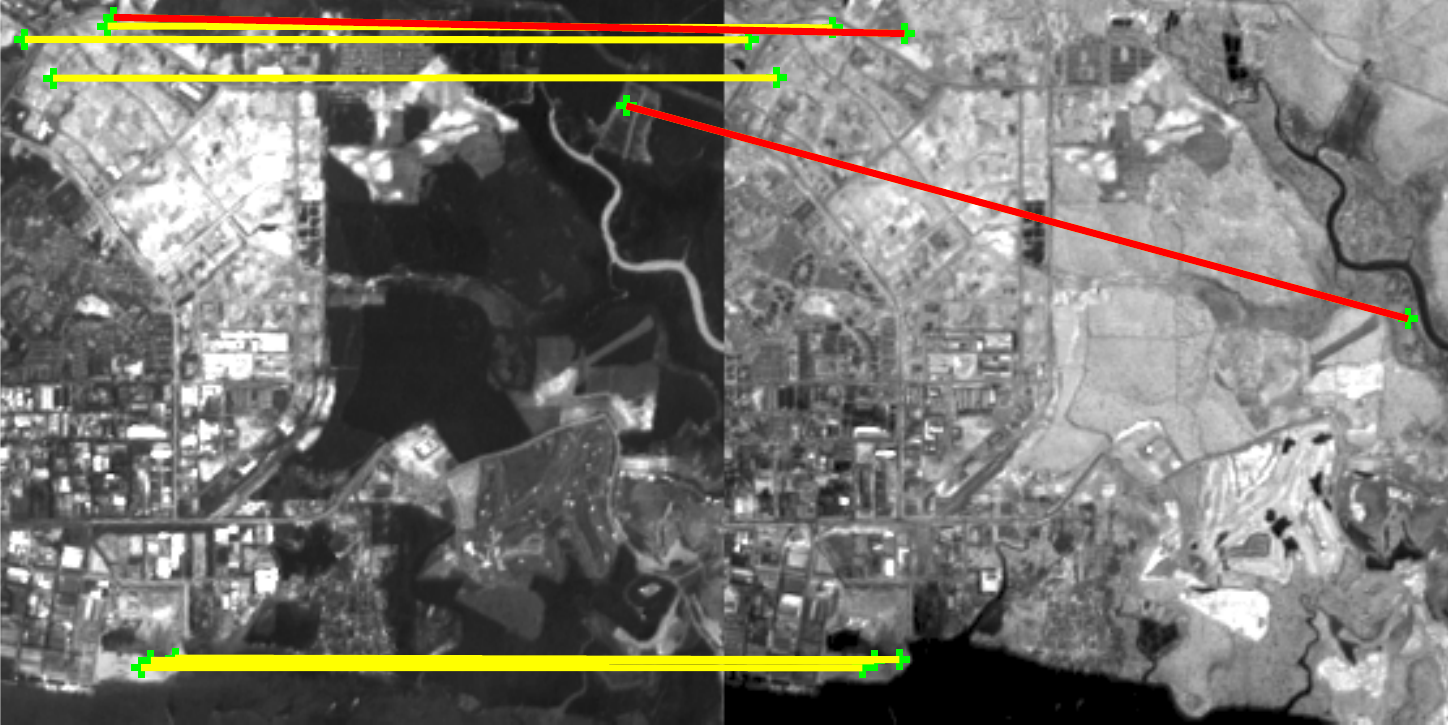}
    \end{minipage}}\\
  \subfigure[]{
    \label{fig:mini:subfig:a}
    \begin{minipage}[c]{0.3\textwidth}
      \centering
      \includegraphics[width=2.2in]{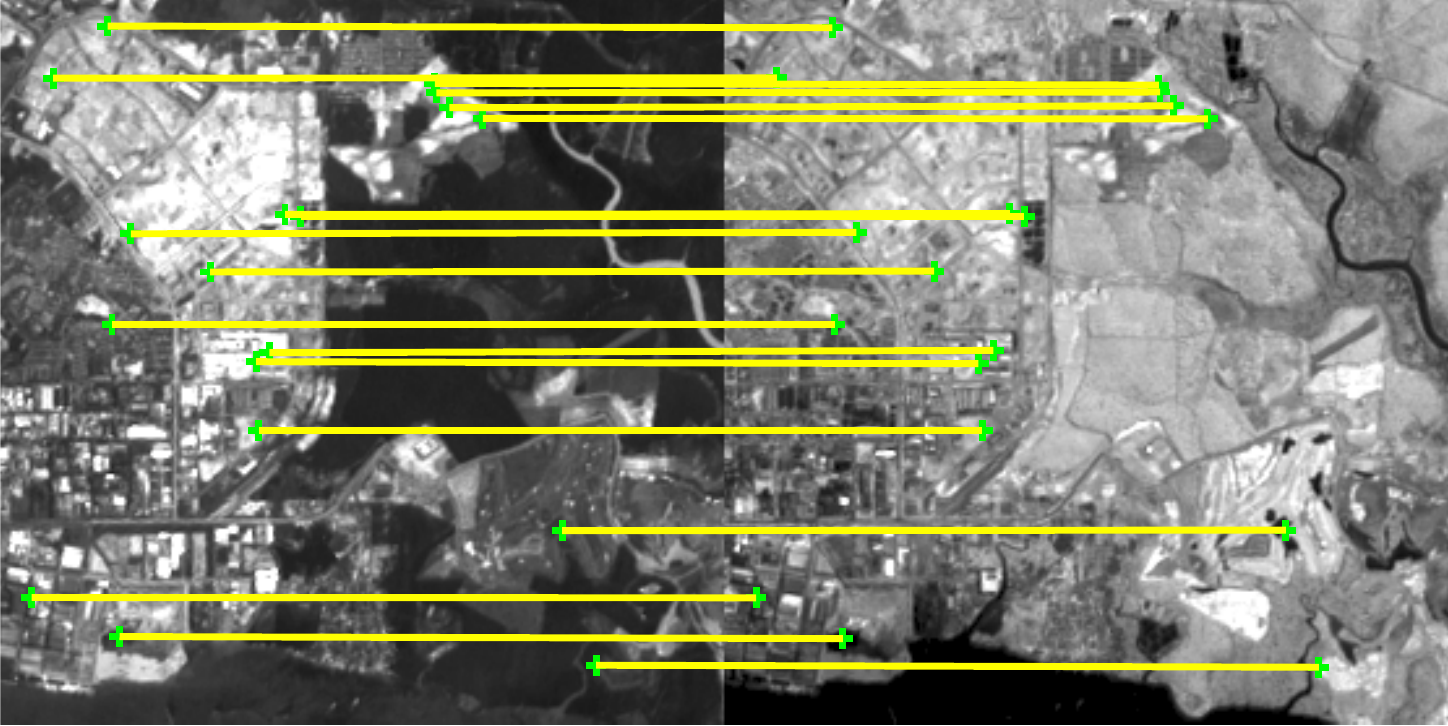}
    \end{minipage}}
  \subfigure[]{
    \label{fig:mini:subfig:a}
    \begin{minipage}[c]{0.3\textwidth}
      \centering
      \includegraphics[width=2.2in]{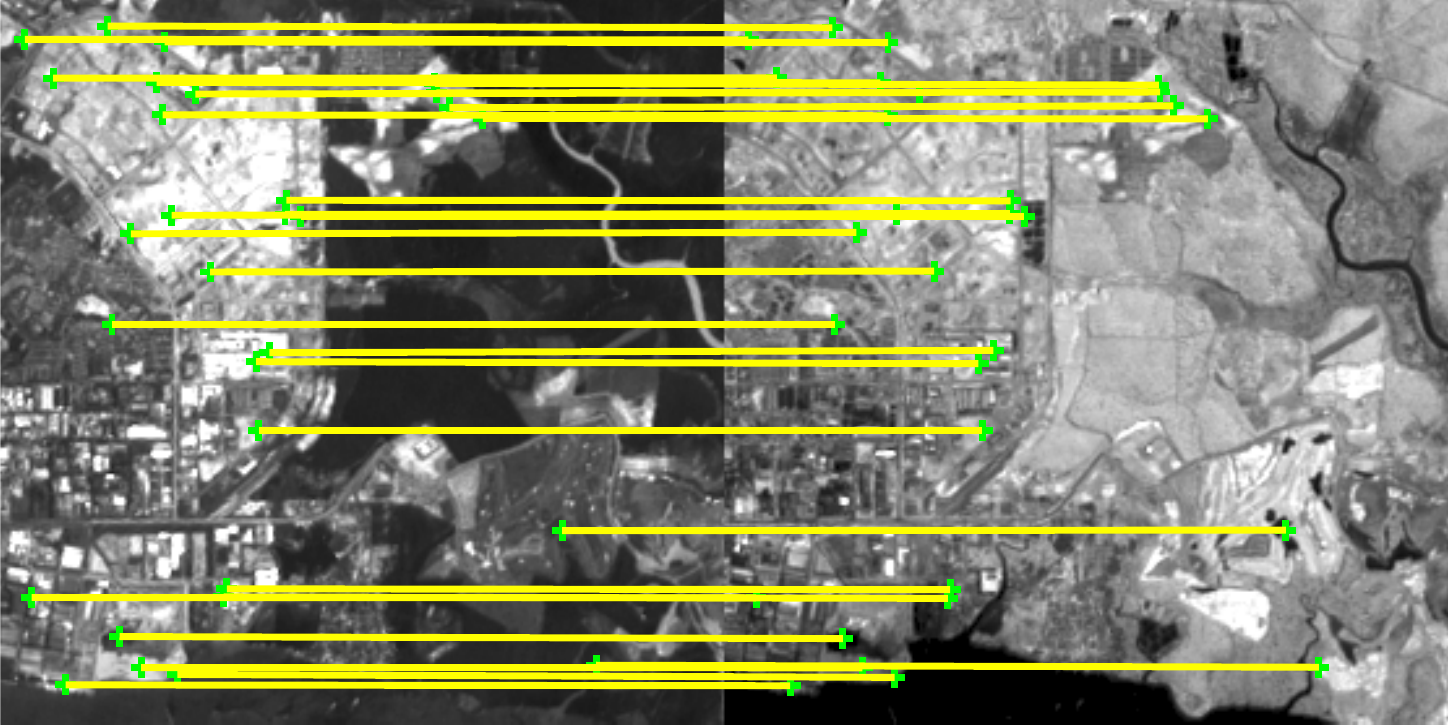}
    \end{minipage}}\\
  \captionstyle{normal}
  \caption{Examples of matching results for ImgSp3-1 with inconsistent spectral content. (a) SIFT. (b) RANSAC. (c) GTM. (d) VTM. (e) RFVTM.}
  \label{fig-im5}
\end{figure*}

\begin{figure*}[htb]
\centering
 \setlength{\abovecaptionskip}{0pt}
 \setlength{\belowcaptionskip}{0pt}
 \setlength{\intextsep}{8pt plus 3pt minus 2pt}
  \subfigure[]{
    \label{fig:mini:subfig:a}
    \begin{minipage}[c]{0.3\textwidth}
      \centering
      \includegraphics[width=2.2in]{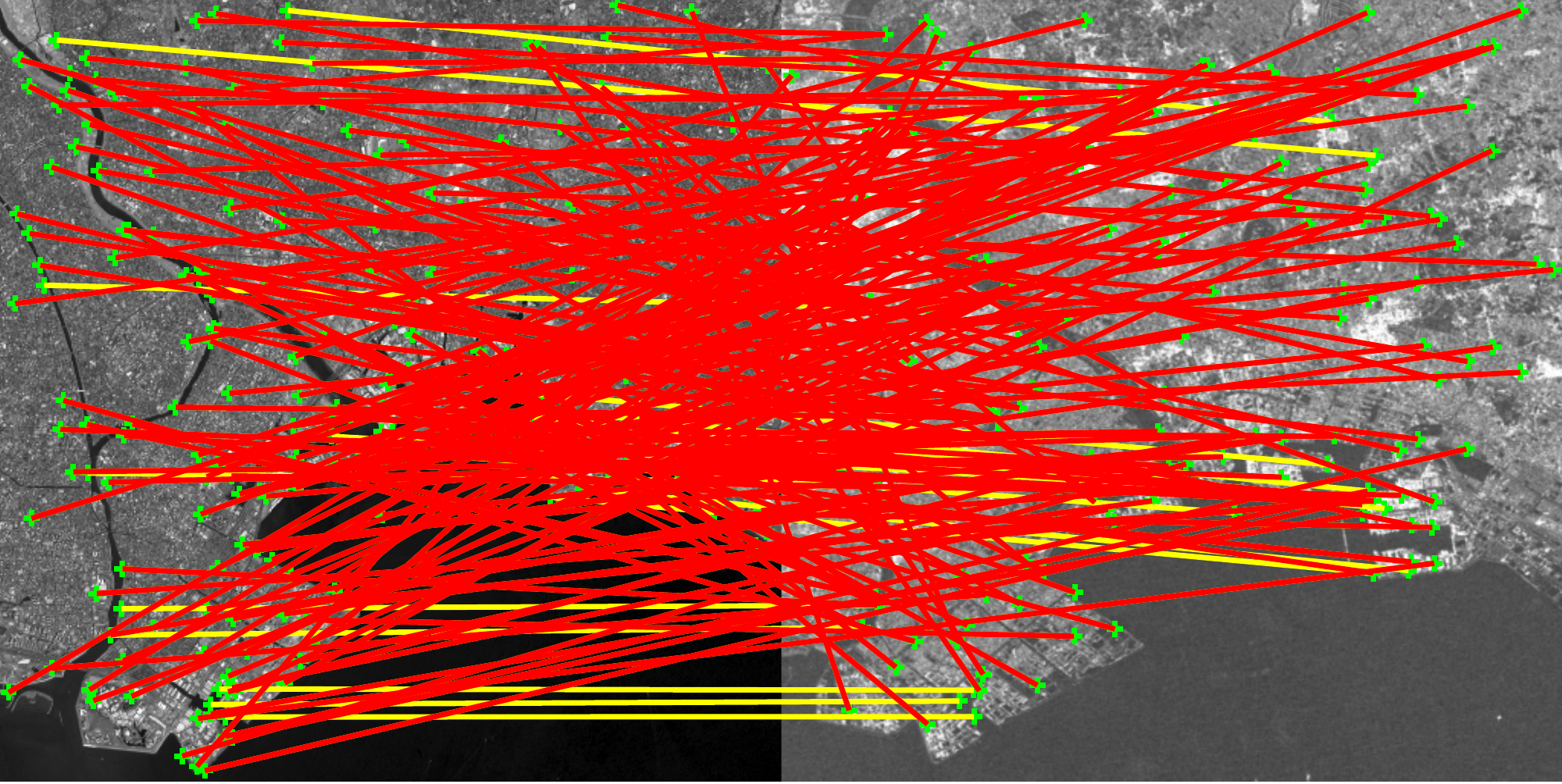}
    \end{minipage}}
  \subfigure[]{
    \label{fig:mini:subfig:b}
    \begin{minipage}[c]{0.3\textwidth}
      \centering
      \includegraphics[width=2.2in]{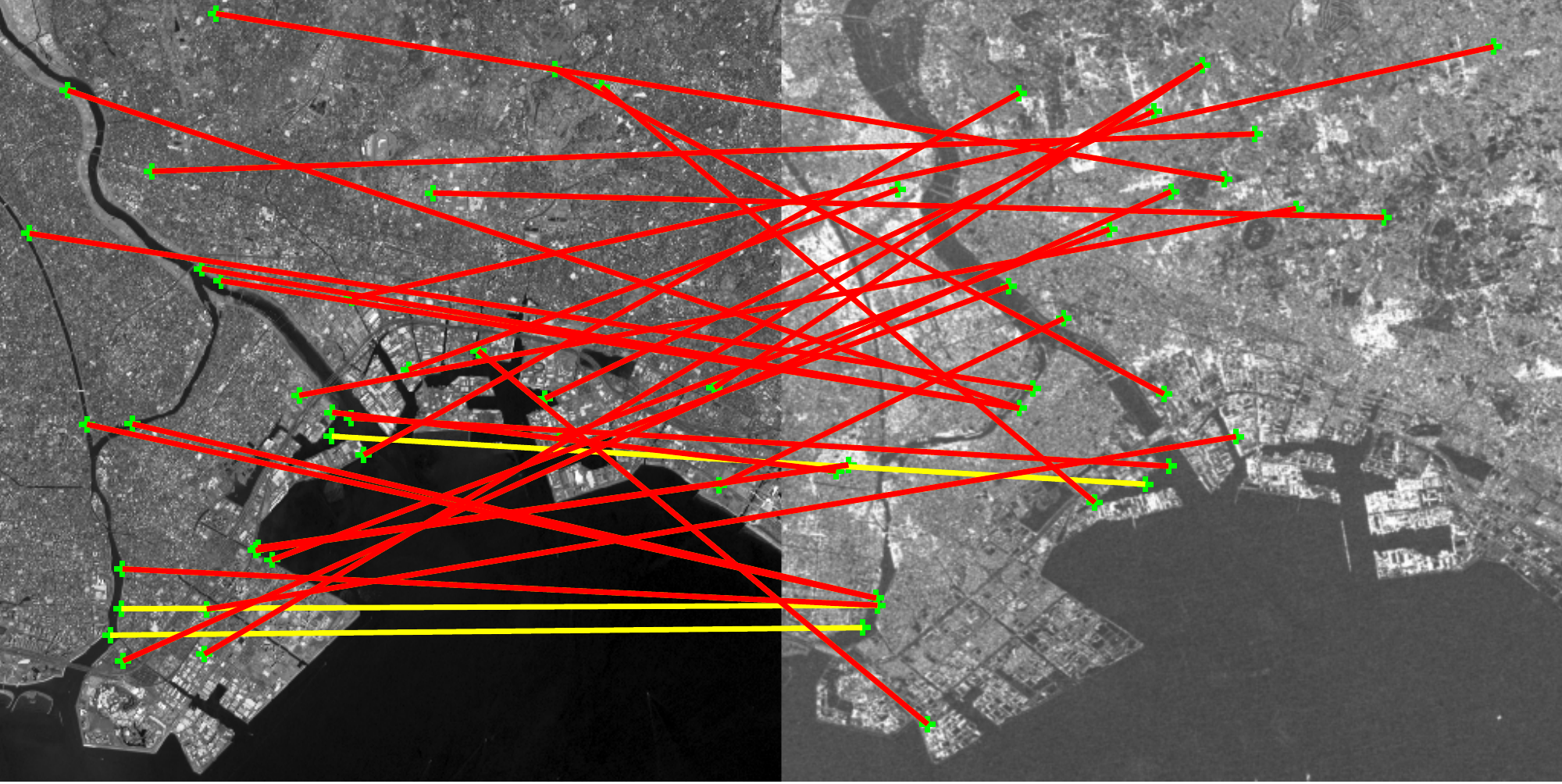}
    \end{minipage}}
  \subfigure[]{
    \label{fig:mini:subfig:c}
    \begin{minipage}[c]{0.3\textwidth}
      \centering
      \includegraphics[width=2.2in]{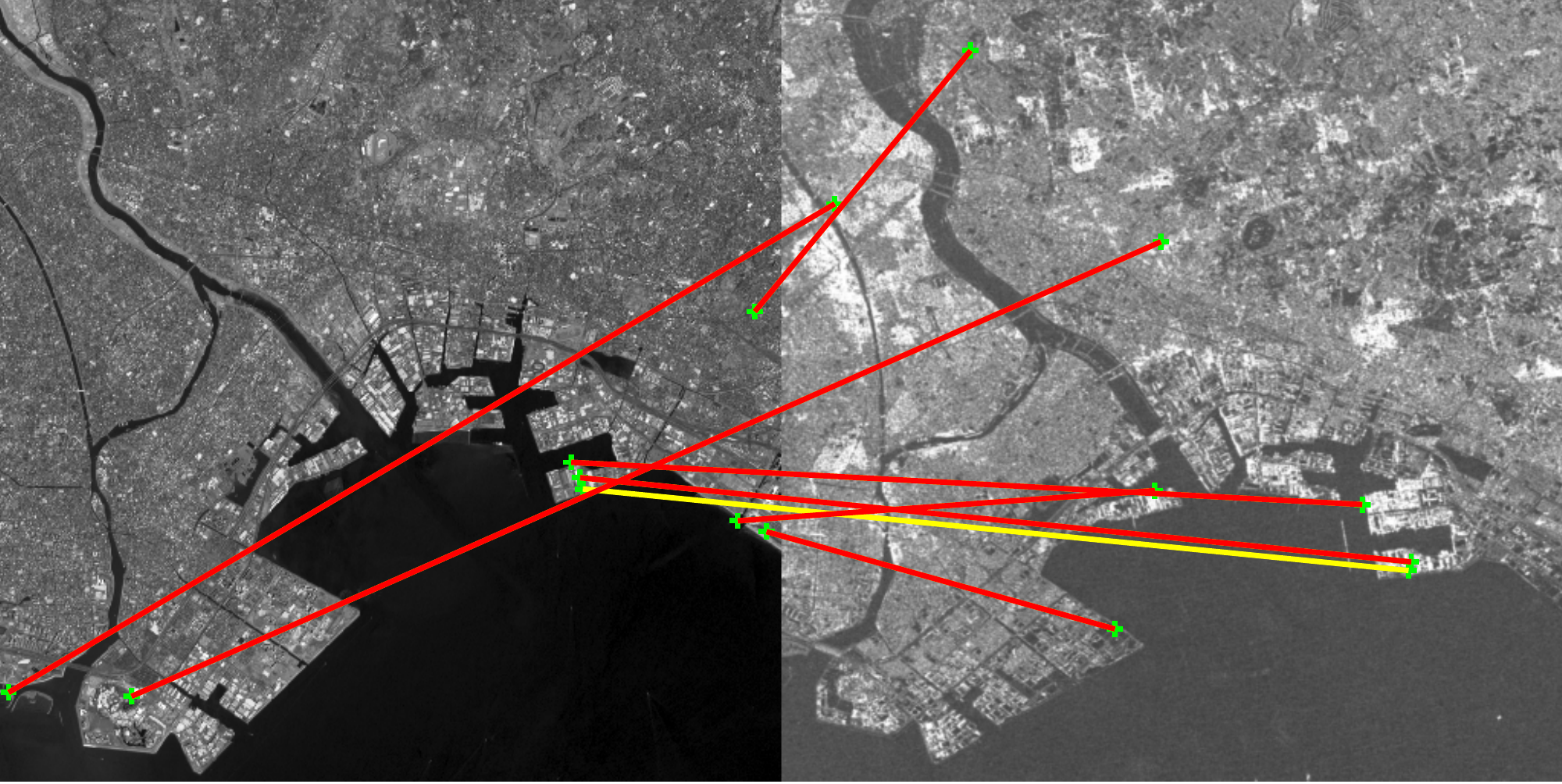}
    \end{minipage}}\\
  \subfigure[]{
    \label{fig:mini:subfig:a}
    \begin{minipage}[c]{0.3\textwidth}
      \centering
      \includegraphics[width=2.2in]{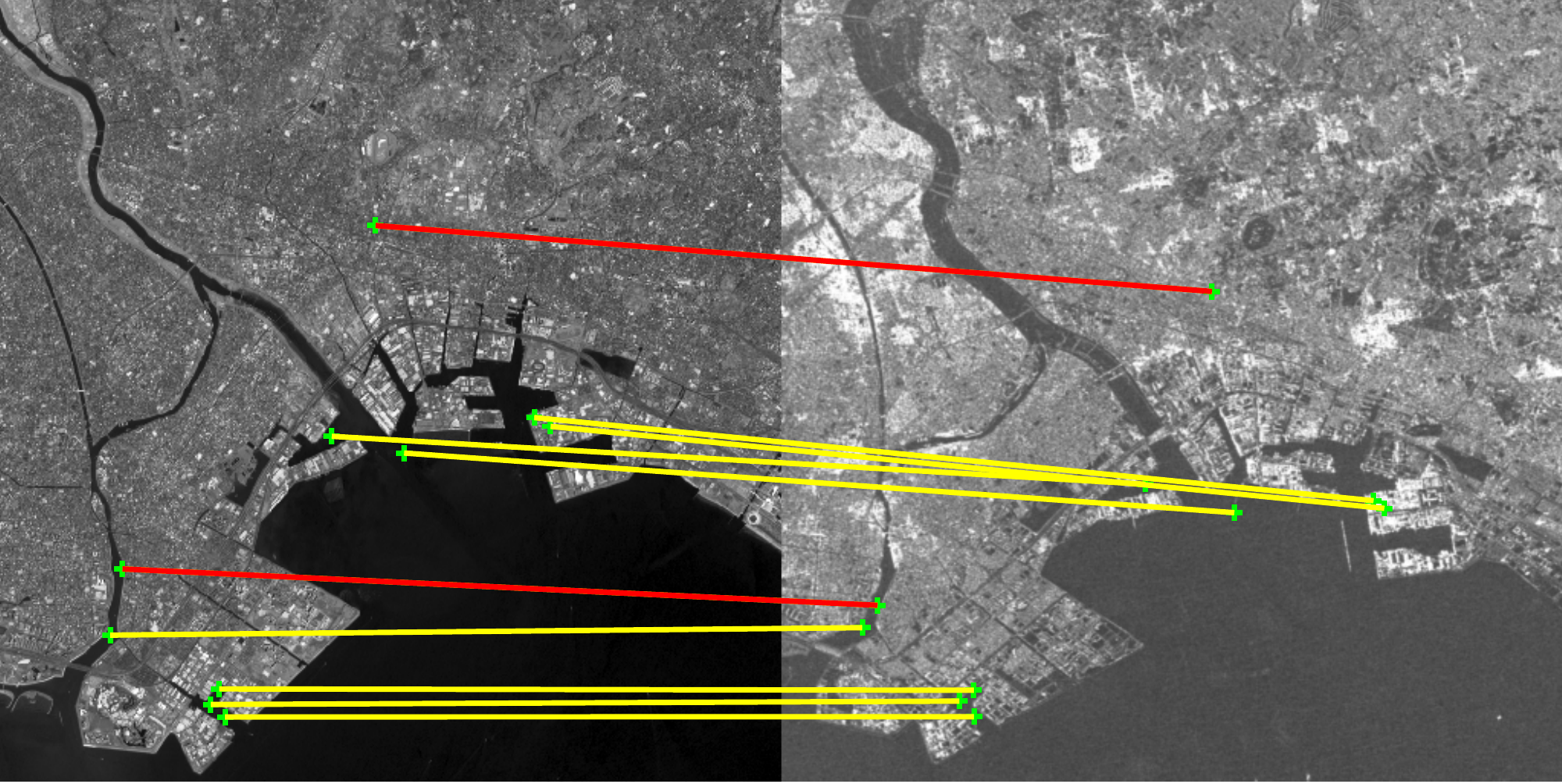}
    \end{minipage}}
  \subfigure[]{
    \label{fig:mini:subfig:a}
    \begin{minipage}[c]{0.3\textwidth}
      \centering
      \includegraphics[width=2.2in]{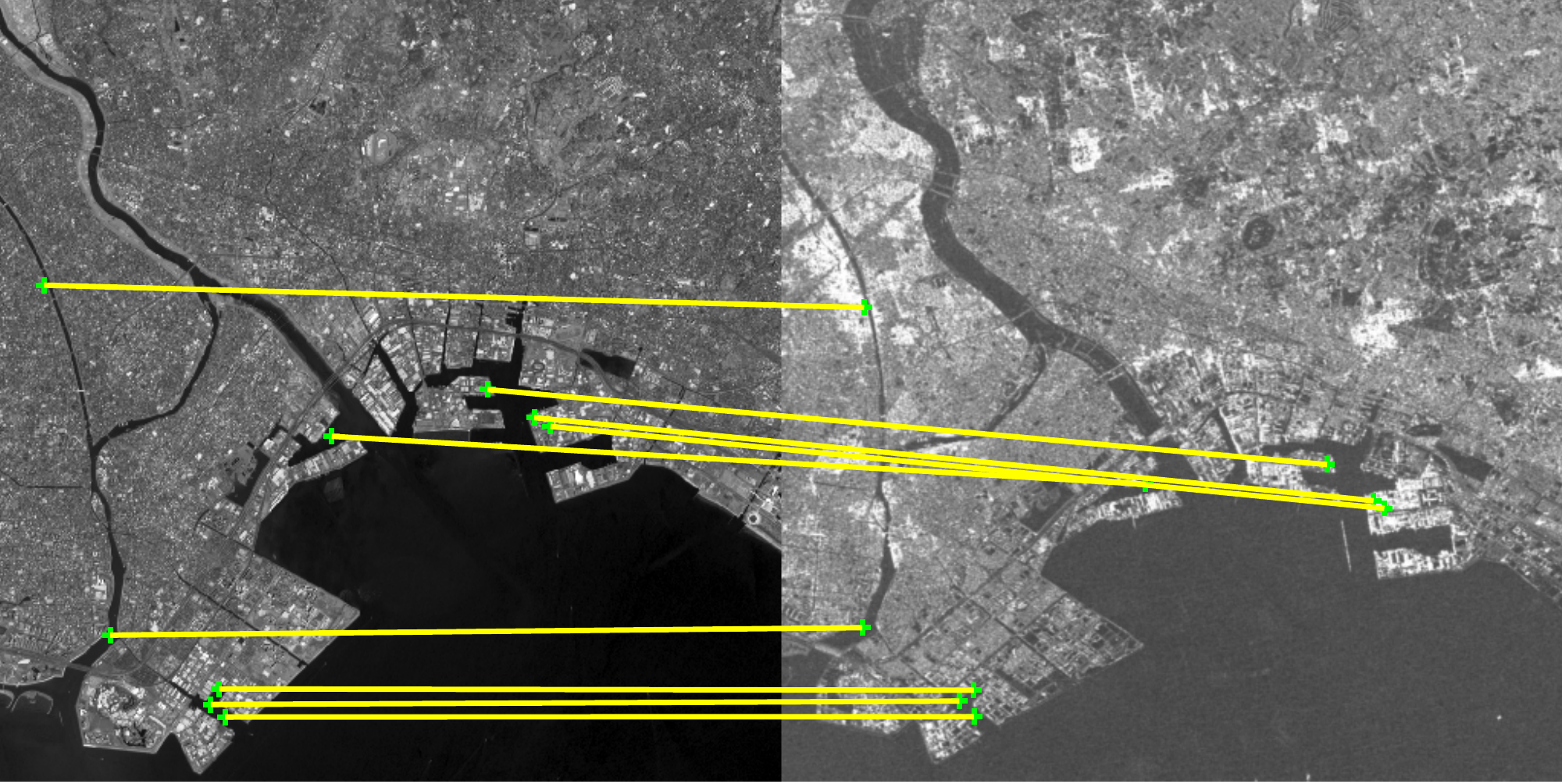}
    \end{minipage}}\\
  \captionstyle{normal}
  \caption{Examples of matching results for ImgSp3-2 with inconsistent spectral content.  (a) SIFT. (b) RANSAC. (c) GTM. (d) VTM. (e) RFVTM.}
  \label{fig-im6}
\end{figure*}

The third image dataset contains ten image pairs with inconsistent spectral content, involving remote sensing optical and synthetic aperture radar (SAR) images taken at multispectral and multimodal situations.
Fig. \ref{fig-im5} and Fig. \ref{fig-im6} give two matching examples for ImgSp3-1 and ImgSp3-2 respectively.
The image pair of Fig. \ref{fig-im5} with the same size of 278$\times$278 are selected from the multispectral imagery SPOT HRV from band XS1 (0.50-0.59m) and band XS3 (0.79-0.89m) over the same area. Most appearances are significantly different between the two images of the same scene, e.g., the river appears bright in XS1, but dark in XS3.
The image pair of  Fig. \ref{fig-im6} consists of two images with the size of 512$\times$512 taken by the sensor of ASTER L1B band 1 and PALSAR fine mode, covering the bay of Tokyo, Japan.
The substantial disparity according to the visual appearance can be observed between the optical and SAR images. The SAR image from PALSAR is inevitably contaminated
by the speckle noise and scatter signals from the earth surface, which bring more challenges to feature point extraction and matching.
The intensities of the same objects in both of examples are significant different, so that a large amount of outliers are involved in the initial correspondences shown in Fig. \ref{fig-im5} (a) and Fig. \ref{fig-im6} (a).
It can be observed that RANSAC reserves much more outlier than other three algorithms in the both of two cases.
As shown in Fig. \ref{fig-im5} (b) and Fig. \ref{fig-im6} (b), few outliers still exist with the same K Nearest Neighbor structures after GTM matching.
RFVTM outperforms than other three algorithms in terms of removing outliers and preserving inliers. As shown in Fig. \ref{fig-im5} (c)-(d) and Fig. \ref{fig-im6} (c)-(d), few outliers that close to the correct locations cannot be removed by VTM, but can be filtered out by RFVTM.

\begin{figure*}[htb]
\centering
 \setlength{\abovecaptionskip}{0pt}
 \setlength{\belowcaptionskip}{0pt}
 \setlength{\intextsep}{8pt plus 3pt minus 2pt}
  \subfigure[]{
    \label{fig:mini:subfig:a}
    \begin{minipage}[c]{0.4\textwidth}
      \centering
      \includegraphics[width=2.5in]{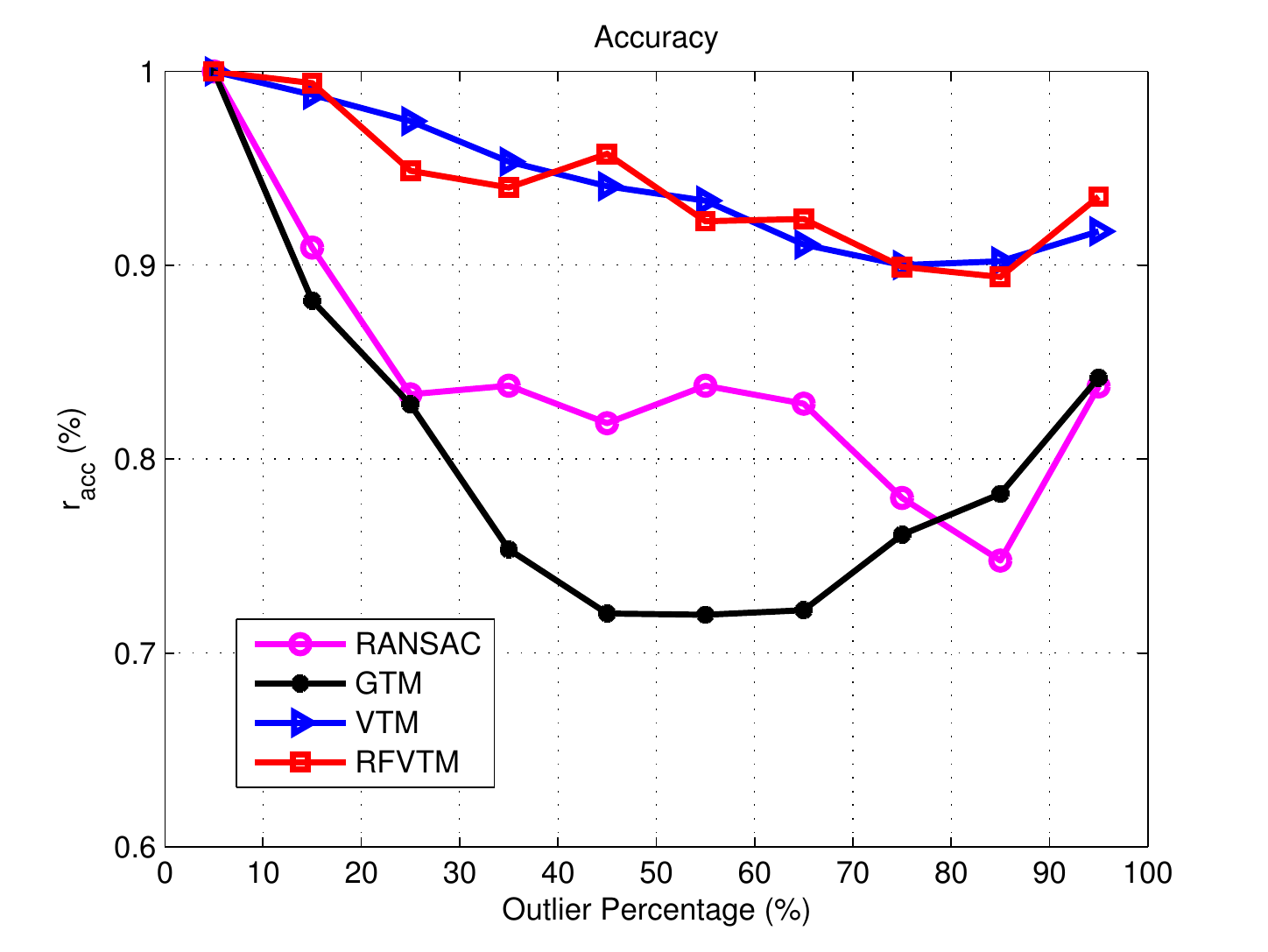}
    \end{minipage}}
  \subfigure[]{
    \label{fig:mini:subfig:b}
    \begin{minipage}[c]{0.4\textwidth}
      \centering
      \includegraphics[width=2.5in]{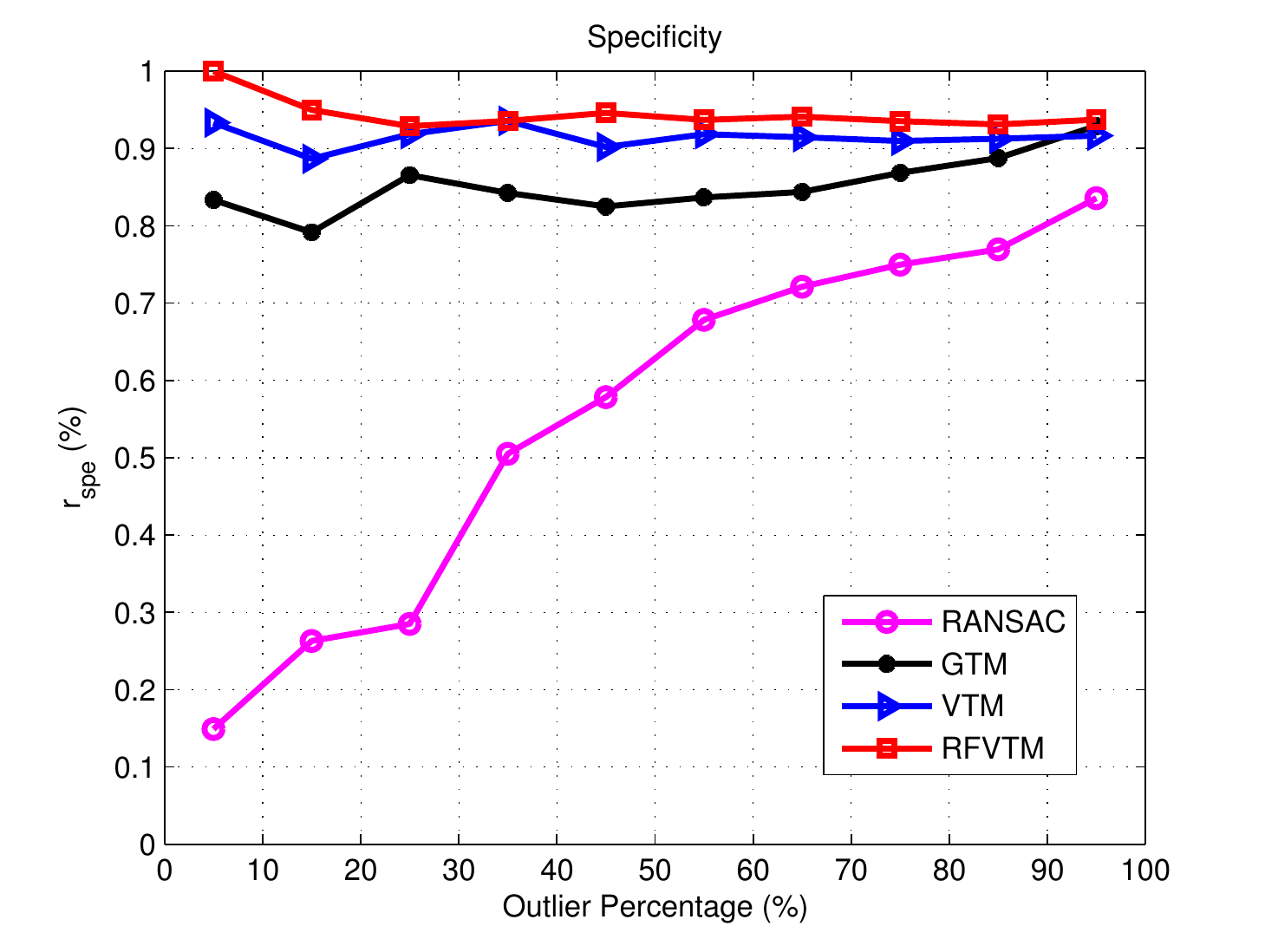}
    \end{minipage}}\\
  \subfigure[]{
    \label{fig:mini:subfig:c}
    \begin{minipage}[c]{0.4\textwidth}
      \centering
      \includegraphics[width=2.5in]{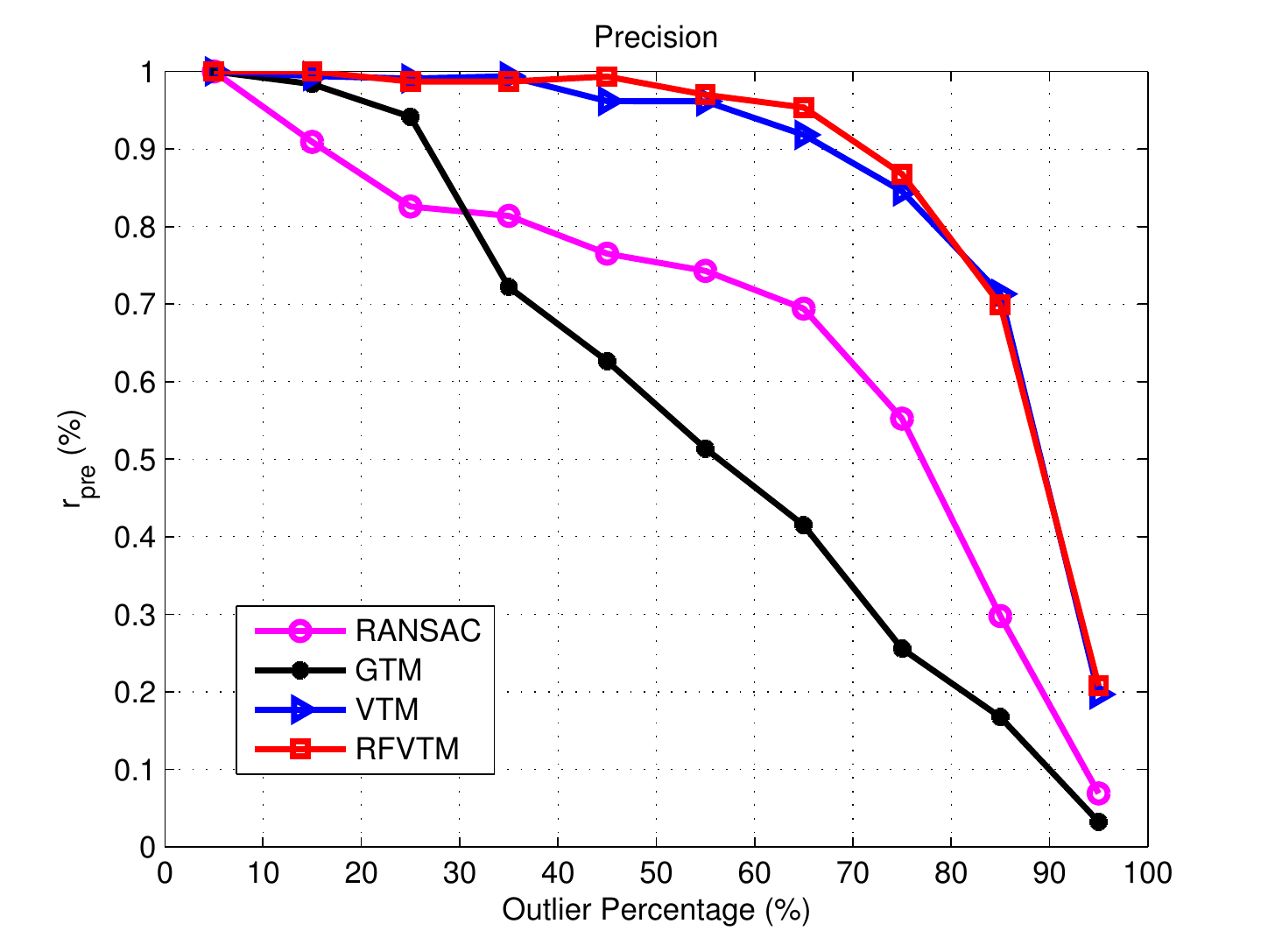}
    \end{minipage}}
  \subfigure[]{
    \label{fig:mini:subfig:a}
    \begin{minipage}[c]{0.4\textwidth}
      \centering
      \includegraphics[width=2.5in]{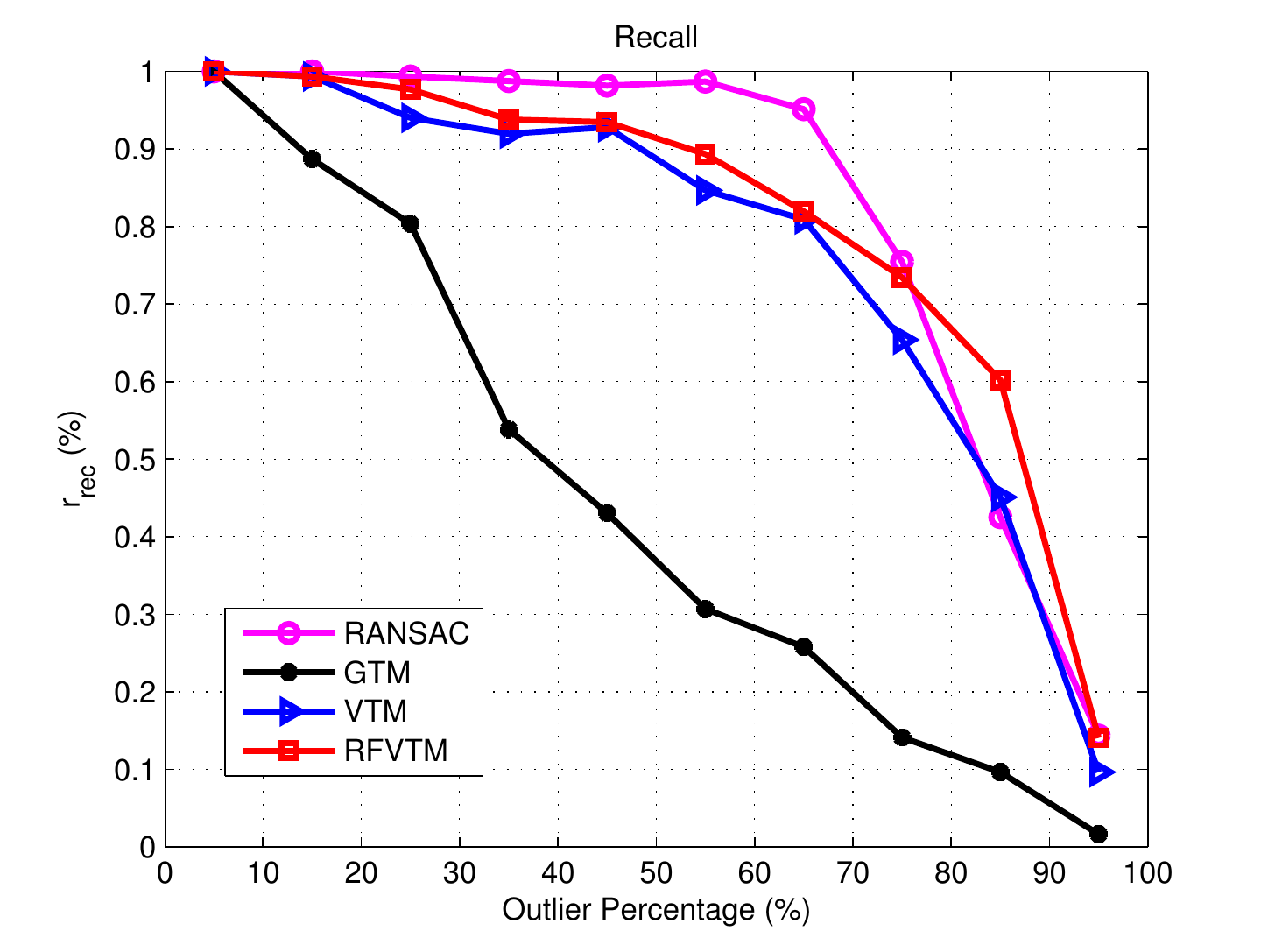}
    \end{minipage}}\\
  \captionstyle{normal}
  \caption{Performance comparison for RANSAC, GTM, VTM, and RFVTM methods under the ambiguities of inconsistent spectral content. (a) accuracy plots. (b) specificity plots. (c) precision plots. (d) recall plots.  }
  \label{fig-plots4_inconsistent}
\end{figure*}

Fig. \ref{fig-plots4_inconsistent} shows the mean accuracy, specificity, precision, and recall of the four algorithms for the images with inconsistent spectral content.
As the same treatment in the previous experiments, the inliers are reserved and outliers are added to the inliers in increasing amounts. The percentage of outliers is changed from 5\% to 95\%.
Fig. \ref{fig-plots4_inconsistent} (b) shows that, with less than 30\% outliers, RANSAC incorrectly identifies more than half of them as correct matches.
It can be seen from Fig. \ref{fig-plots4_inconsistent} (c) and (d) that the precisions of VTM and RFVTM are significantly higher than GTM when the proportion of outliers is more than 30\%.
This is because the local KNN adopted by GTM has a weaker structures when there are fewer inliers in the initial sets \cite{J_MI_2012_ITGRS}. Besides that, when there are too many outliers existing in the K Nearest Neighborhoods, these inliers would be arbitrarily removed during GTM iterations.
RANSAC provides high recall values for low proportions of outliers but its recall falls dramatically when the proportion of outliers is above 60\%.

\subsection{Sensitivity Evaluation on VTM and RFVTM}
\begin{itemize}
\item[\emph{1)}] \emph{Sensitivity With Respect to the Number of Inliers}
\end{itemize}

\begin{figure*}[htb]
\centering
 \setlength{\abovecaptionskip}{0pt}
 \setlength{\belowcaptionskip}{0pt}
 \setlength{\intextsep}{8pt plus 3pt minus 2pt}
  \subfigure[]{
    \label{fig:mini:subfig:a}
    \begin{minipage}[c]{0.4\textwidth}
      \centering
      \includegraphics[width=2.5in]{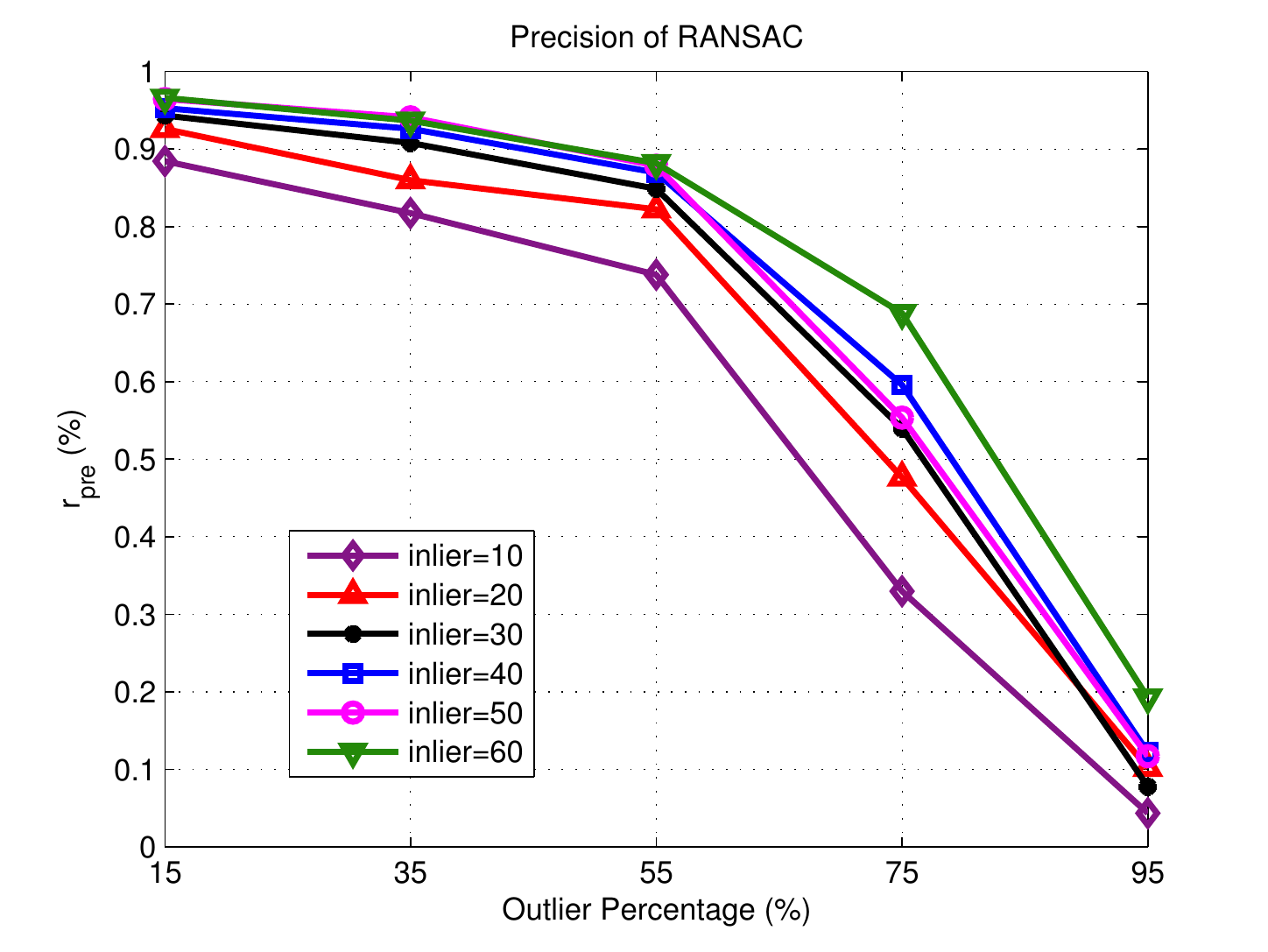}
    \end{minipage}}
  \subfigure[]{
    \label{fig:mini:subfig:b}
    \begin{minipage}[c]{0.4\textwidth}
      \centering
      \includegraphics[width=2.5in]{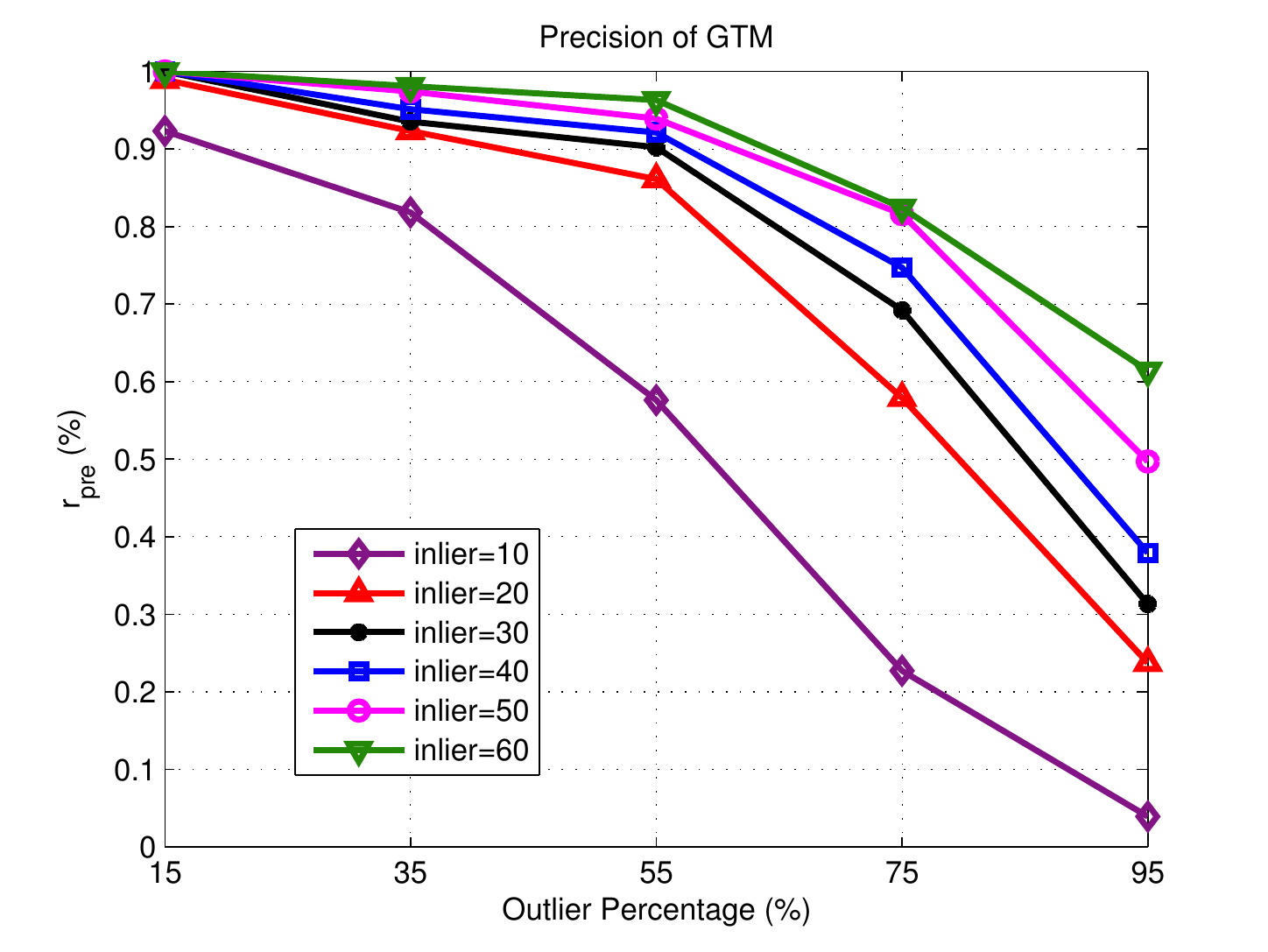}
    \end{minipage}}\\
  \subfigure[]{
    \label{fig:mini:subfig:c}
    \begin{minipage}[c]{0.4\textwidth}
      \centering
      \includegraphics[width=2.5in]{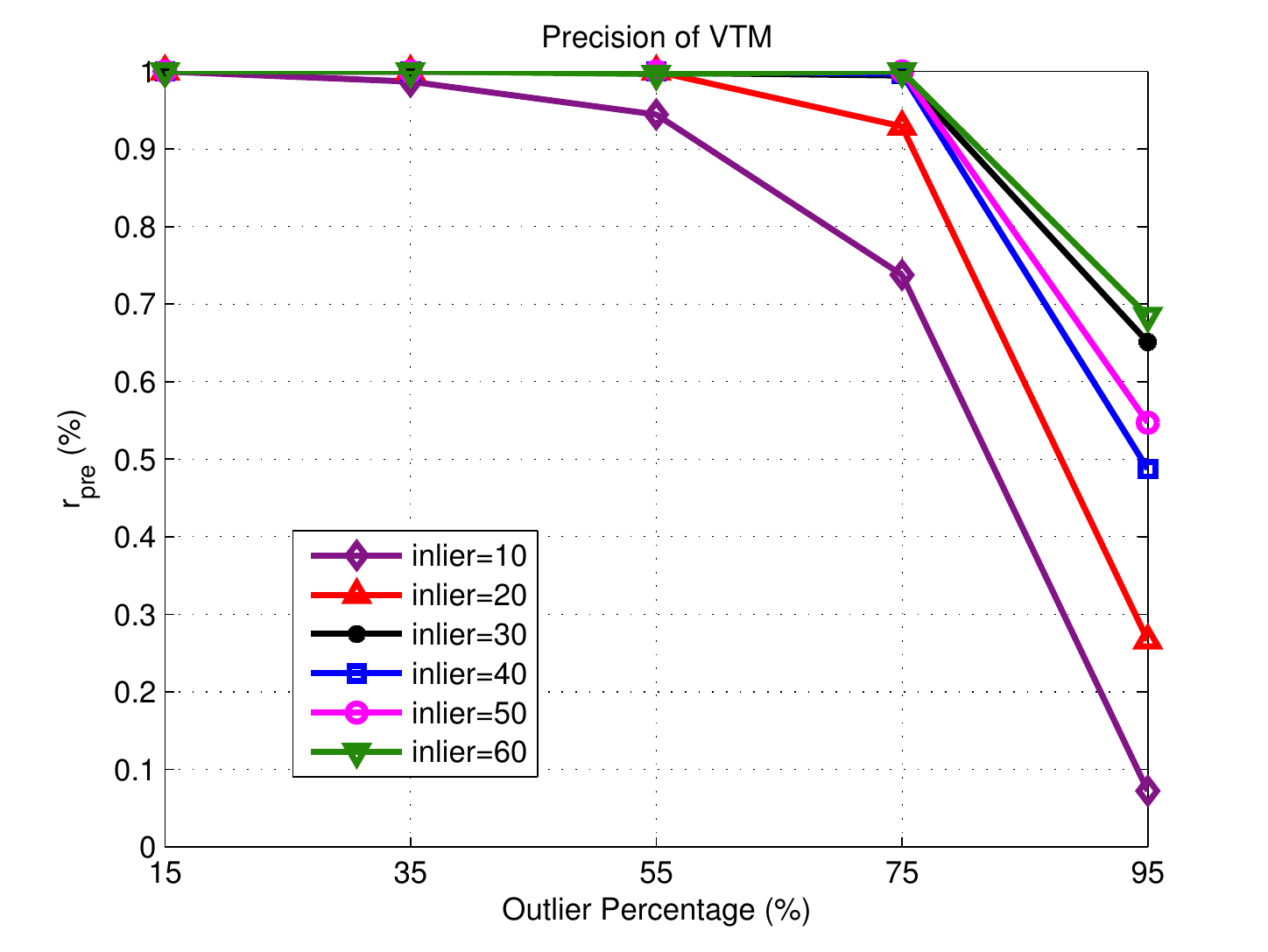}
    \end{minipage}}
  \subfigure[]{
    \label{fig:mini:subfig:a}
    \begin{minipage}[c]{0.4\textwidth}
      \centering
      \includegraphics[width=2.5in]{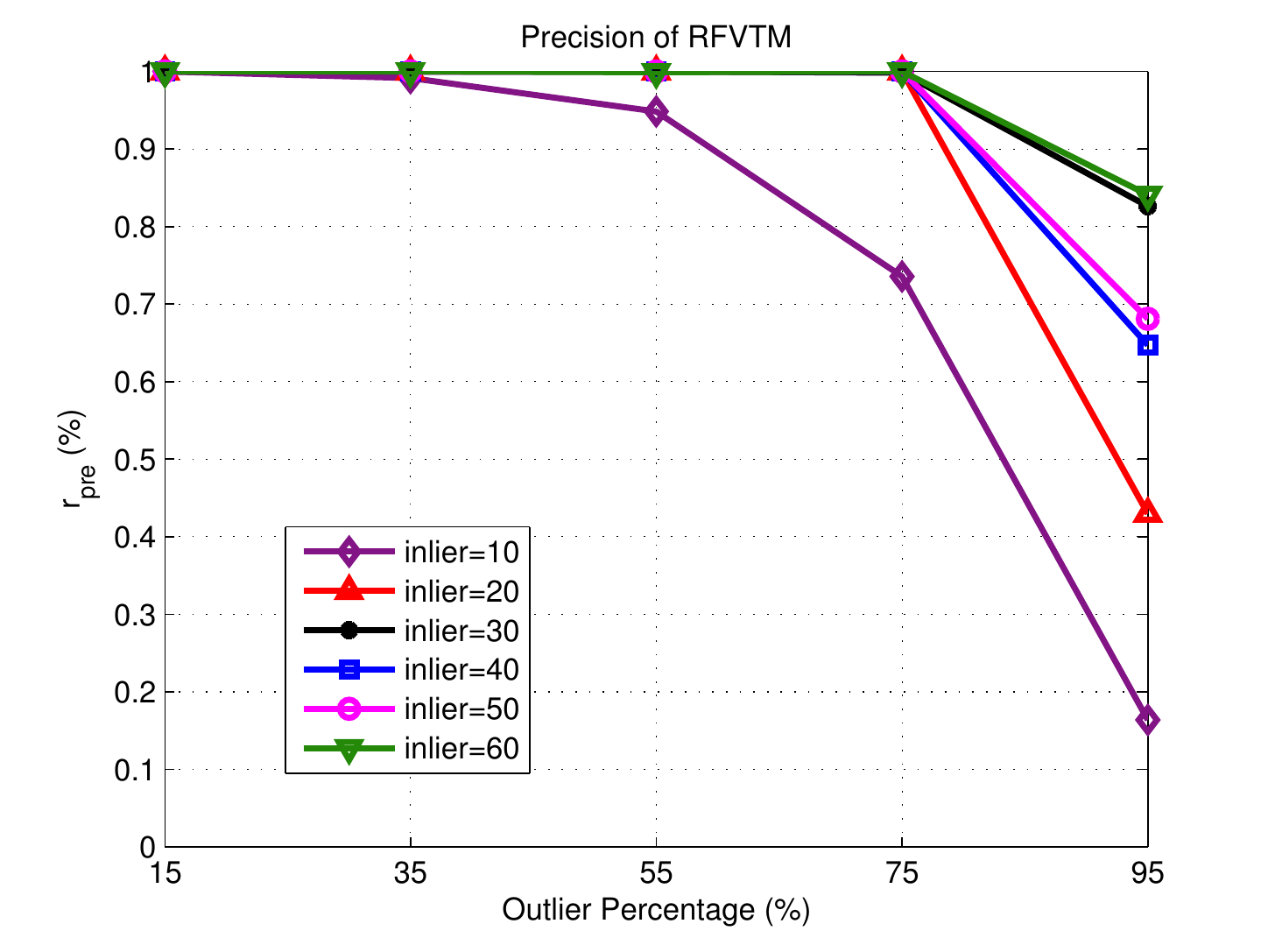}
    \end{minipage}}\\
  \captionstyle{normal}
  \caption{Performance comparison of average precision for RANSAC, GTM, VTM, and RFVTM methods for different number of inliers. (a) RANSAC. (b) GTM with K=5. (c) VTM. (d) RFVTM}
  \label{fig-plots_inliers}
\end{figure*}

The four algorithms achieve outlier rejections according to the spatial relations between feature points.
In this part, the sensitivities with respect to the number of inliers are discussed in brief.
Thirty test image pairs are selected from the three representative image datasets respectively. The inliers in each image pair are increased from 10 to 60. The percentage of added outliers varies from 15\% to 95\% in increments of 20\%.
Fig. \ref{fig-plots_inliers} demonstrates the comparison of statistical precision for RANSAC, GTM, VTM, and RFVTM.
As depicted in Fig. \ref{fig-plots_inliers} (a)-(d), the precisions of all of four algorithms increase substantially with the increase of inliers. When the number of inliers is less than 20, all algorithms are inferior to the cases with more than 20 inliers in terms of precisions. However, comparing to GTM and RANSAC, VTM and RFVTM still perform more robustly with few inliers.
It can be explained by the fact that the graphs of GTM relying on the nearest neighborhood have vulnerable local structures consisted of K-Nearest-Neighbor of points (K=5).
The existing inliers only validate graph structures of their K-Nearest-Neighbors.
When there are only fewer inliers, the graph structures would become less credible.
In contrast to the local structures of GTM, the graphs of the proposed vertex trichotomy descriptor are composed by all of feature points in the form of three trichotomy sets. Each of graph structures is validated by overall inliers in the initial sets.
Therefore, the performance of VTM and RFVTM is less affected by the number of inliers than GTM.
Also, the precisions of GTM and RANSAC drop much more significantly than those of VTM and RFVTM when outliers are greater than 55\%.

Table \ref{table-time} presents the average execution time for each algorithm with 20, 40, and 60 inliers.
The running time of RANSAC is close to GTM, VTM, and RFVTM when the percentage of outlier is below 35\%, but increases much faster when the percentage of outliers is above 55\%.
The number of iterations in RANSAC relies on the probability to randomly select three correct matches from the initial sets, so that the average running time of RANSAC increases significantly with a high percentage of outliers.
Compared to VTM and RFVTM, GTM is more computationally expensive than VTM and RFVTM. This is because GTM requires to reconstruct the KNN graphs with the changed KNN feature points at each iteration.

\begin{table*}[htb]
\centering
  \captionstyle{normal}
  \setlength{\abovecaptionskip}{0pt}
  \setlength{\belowcaptionskip}{10pt}
\caption{THE AVERAGE EXECUTION TIME FOR THE FOUR ALGORITHMS WITH DIFFERENT NUMBERS OF INLIERS}
\begin{lrbox}{\tablebox}
\begin{tabular}{|c|c|c|c|c|c|c|c|c|c|c|c|c|}
\hline
\multirow{1}{*}{Outlier} &
\multicolumn{4}{c|}{Time with 20 inliers (sec)} &
\multicolumn{4}{c|}{Time with 40 inliers (sec)} &
\multicolumn{4}{c|}{Time with 60 inliers (sec)} \\
\cline{2-13}
$(\%)$ & RANSAC	& GTM &	VTM & RFVTM & RANSAC & GTM & VTM & RFVTM & RANSAC & GTM & VTM & RFVTM\\
\hline
15 & 0.032 & 0.039 & 0.023 & 0.035 & 0.086 & 0.094 & 0.061 & 0.812 & 0.116 & 0.187 & 0.119 & 0.152\\
\hline
35 & 0.079 & 0.048 & 0.030 & 0.042 & 0.267 & 0.206 & 0.106 & 0.198 & 0.643 & 0.462 & 0.204 & 0.315\\
\hline
55 & 0.344 & 0.193 & 0.052 & 0.067 & 1.230 & 0.381 & 0.192 & 0.265 & 4.866 & 1.081 & 0.576 & 0.732\\
\hline
75 & 2.803 & 0.486 & 0.143 & 0.219 & 6.094 & 2.705 & 0.874 & 1.590 & 10.639 & 3.764 & 1.407 & 2.186\\
\hline
95 & 10.747 & 6.941 & 3.752 & 4.851 & 24.110 & 14.863 & 8.305 & 10.741 & 42.004 & 21.263 & 11.018 & 15.190\\
\hline
\end{tabular}
\end{lrbox}
\scalebox{1}{\usebox{\tablebox}}
\label{table-time}
\end{table*}

\begin{itemize}
\item[\emph{2)}] \emph{Sensitivity to Subdivision for Initial Correspondence}
\end{itemize}

Section IV-C propose an optimization for vertex trichotomy descriptor to subdivide the initial vertices of size $n$ into $m$ groups and implement VTM and RFVTM with $n/m$ correspondences for $m$ groups respectively.
In this section, we select ten image pairs from dataset to present a brief discussion with the performance of subdivision.
Table \ref{table-sub} shows the performance of VTM and RFVTM with three different subdivision, i.e, $m=1$, $m=5$, and $m=10$.
``IC" and ``IF" represent the initial correct correspondences and initial false correspondences respectively.
The following can be observed from Table \ref{table-sub}:
1) Compared with non-subdivided cases (m=1), the time complexities of VTM and RFVTM with subdivisions are reduced. The more subdivisions, the lower time complexities are achieved.
2) The matching results of VTM with subdivisions are inferior to the corresponding cases without subdivisions in terms of ``RC" and ``RF", especially when the numbers of initial inliers are small. It can be explained by the fact that the number of inliers in each group decreases with subdivisions, and the matching performance will be degenerated with the decrease of inliers as we discussed in the above section.
3) With the same subdivisions, RFVTM suffers a less performance loss than VTM, since that the mistakenly removed inliers in each subgroup have the probability to be recovered by the recovered and filtering strategy.

\begin{table*}[htb]
\centering
  \captionstyle{normal}
  \setlength{\abovecaptionskip}{0pt}
  \setlength{\belowcaptionskip}{10pt}
\caption{PERFORMANCE OF VTM AND RFVTM WITH DIFFERENT SUBDIVISIONS}
\begin{lrbox}{\tablebox}
\begin{tabular}{|c|c|c|c|c|c|c|c|c|c|c|c|c|c|c|c|c|c|c|c|c|c|c|c|c|}
\hline
\multirow{3}{*}{No.} &
\multicolumn{3}{c|}{\multirow{2}{*}{Initial}} &
\multicolumn{9}{c|}{VTM} &
\multicolumn{9}{c|}{RFVTM} \\
\cline{5-22}&
  \multicolumn{3}{c|}{} & \multicolumn{3}{c|}{$m=1$} & \multicolumn{3}{c|}{$m=5$} & \multicolumn{3}{c|}{$m=10$} & \multicolumn{3}{c|}{$m=1$} & \multicolumn{3}{c|}{$m=5$} & \multicolumn{3}{c|}{$m=10$}  \\
\cline{2-22} &$n$ &IC &IF	
&RC	&RF	&time(s)	&RC	&RF	&time(s)    &RC	&RF	&time(s)	
&RC	&RF	&time(s)	&RC	&RF	&time(s)	&RC	&RF	&time(s)\\
\hline
1 &37 &22	&15	&22	&0	&0.031	&22	&0	&0.014	&17	&0	&0.006	&22	&0	&0.045	&22	&0	&0.027	&20	&0	&0.011\\
\hline
2 &76 &27	&49	&20	&4	&0.108	&16	&6	&0.085	&9	&4	&0.064	&26	&0	&0.179	&21	&4	&0.104	&13	&3	&0.073\\
\hline
3 &139	&35	&104	&18	&0	&0.675	&13	&2	&0.204	&8	&2	&0.187	&29	&0	&0.896	&24	&0	&0.395	&19	&0	&0.117\\
\hline
4 &111	&42	&69	&42	&0	&0.214	&42	&0	&0.127	&42	&0	&0.083	&42	&0	&0.284	&42	&0	&0.143	&42	&0	&0.104\\
\hline
5 &155	&48	&107	&48	&2	&0.891	&48	&2	&0.343	&48	&6	&0.212	&48	&0	&1.247	&48	&0	&0.650	&48	&2	&0.416\\
\hline
6 &95	&66	&29	&66	&0	&0.183	&66	&0	&0.104	&66	&0	&0.094	&66	&0	&0.253	&66	&0	&0.146	&66	&0	&0.102\\
\hline
7 &150	&72	&78	&72	&0	&0.656	&72	&0	&0.284	&70	&0	&0.173	&72	&0	&0.829	&72	&0	&0.324	&72	&0	&0.206\\
\hline
8 &155	&94	&61	&94	&0	&0.809	&94	&0	&0.397	&94	&0	&0.215	&94	&0	&1.213	&94	&0	&0.562	&94	&0	&0.311\\
\hline
9 &197	&102	&95	&102	&0	&1.295	&97	&0	&0.520	&84	&0	&0.386	&102	&0	&1.574	&102	&0	&0.821	&102	&0	&0.609\\
\hline
10 &327	&117	&210	&93	&0	&7.422	&86	&5	&2.393	&66	&8	&1.715	&110	&0	&8.643	&92	&0	&3.442	&87	&0	&2.875\\
\hline
\end{tabular}
\end{lrbox}
\scalebox{0.85}{\usebox{\tablebox}}
\label{table-sub}
\end{table*}

\section{ CONCLUSION}\label{sec:precoder}
In this paper, a graph-based feature point matching approach called RFVTM is presented to establish reliable and sufficient matches for remote sensing images. The global spatial relations are described by vertex trichotomy descriptor, which is invariant for images with affine transformations including rigid and non-rigid deformations. The candidate outliers are determined by comparing the disparity of corresponding vertex trichotomy descriptors. On the basis of VTM and two restoration restrictions, the recovery and filtering strategy is designed to reserve sufficient inliers and reject stubborn outliers. The matching results in the experiments have indicated the superiority of RFVTM.
Future work includes incorporating intensity features into graph matching, as well as
reliable image segmentations applied in image registrations for multispectral/multimodal images.

\bibliographystyle{IEEEtran}
\bibliography{MyReference}

\begin{thebibliography}{10}
\providecommand{\url}[1]{#1}
\csname url@samestyle\endcsname
\providecommand{\newblock}{\relax}
\providecommand{\bibinfo}[2]{#2}
\providecommand{\BIBentrySTDinterwordspacing}{\spaceskip=0pt\relax}
\providecommand{\BIBentryALTinterwordstretchfactor}{4}
\providecommand{\BIBentryALTinterwordspacing}{\spaceskip=\fontdimen2\font plus
\BIBentryALTinterwordstretchfactor\fontdimen3\font minus
  \fontdimen4\font\relax}
\providecommand{\BIBforeignlanguage}[2]{{%
\expandafter\ifx\csname l@#1\endcsname\relax
\typeout{** WARNING: IEEEtran.bst: No hyphenation pattern has been}%
\typeout{** loaded for the language `#1'. Using the pattern for}%
\typeout{** the default language instead.}%
\else
\language=\csname l@#1\endcsname
\fi
#2}}
\providecommand{\BIBdecl}{\relax}
\BIBdecl

\bibitem{J_BZitova_2003_IVC}
{B. Zitoza and B. Flusser}, ``Image registration methods: a survey,''
  \emph{Image Vis. Comput.}, vol.~21, no.~11, pp. 977--1000, Jul. 2003.

\bibitem{J_GY_2014_ITCSVT}
{G. Yammine, E. Wige, F. Simmet, D. Niederkorn, and A. Kaup}, ``Novel
  similarity-invariant line descriptor and matching algorithm for global motion
  estimation,'' \emph{IEEE Trans. Circuits Syst. Video Technol.}, vol.~24,
  no.~8, pp. 1323--1335, Aug. 2014.

\bibitem{R1}
{C. Gomez}, ``{Historical 3D topographic reconstruction of the Iwaki Volcano
  using Structure from Motion from Uncalibrated Aerial Photographs},'' 2012.

\bibitem{R2}
C.~Gomez, ``{Digital photogrammetry and GIS-based analysis of the
  bio-geomorphological evolution of {Sakurajima Volcano}, diachronic analysis
  from 1947 to 2006},'' \emph{Journal of Volcanology and Geothermal Research},
  vol. 280, pp. 1--13, 2014.

\bibitem{R3}
M.~R. James and S.~Robson, ``Sequential digital elevation models of active lava
  flows from ground-based stereo time-lapse imagery,'' \emph{ISPRS Journal of
  Photogrammetry and Remote Sensing}, vol.~97, pp. 160--170, 2014.

\bibitem{R4}
{J. I. Farquharson, M. R. James, and H. Tuffen}, ``{Examining rhyolite lava
  flow dynamics through photo-based 3D reconstructions of the 2011-2012 lava
  flowfield at Cord\'{o}n-Caulle, Chile},'' \emph{Journal of Volcanology and
  Geothermal Research}, vol. 304, pp. 336--348, 2015.

\bibitem{R5}
{L. K. Krosley, P. T. Shaffner, E. Oerter, T. Ortiz}, ``Digital ground-based
  photogrammetry for measuring discontinuity orientations in steep rock
  exposures,'' in \emph{Proceedings of the 41st U.S. Symposium of Rock
  Mechanics}, 2006, pp. 1--13.

\bibitem{R6}
{M. Sturzenegger and D. Stead}, ``Close-range terrestrial digital
  photogrammetry and terrestrial laser scanning for discontinuity
  characterization on rock cuts,'' \emph{Engineering Geology}, vol. 106, no.~3,
  pp. 163--182, 2009.

\bibitem{R7}
{S. P. Bemis, S. Micklethwaite, D. Turner, M. R. James, S. Akciz, S. T. Thiele,
  and H. A. Bangash}, ``{Ground-based and UAV-Based photogrammetry: a
  multi-scale, highresolution mapping tool for structural geology and
  paleoseismology},'' \emph{Journal of Structural Geology}, vol.~69, pp.
  163--178, 2014.

\bibitem{R8}
{S. K. Nouwakpo, M. R. James, M. A. Weltz, C. H. Huang, I. Chagas, and L.
  Lima}, ``Evaluation of structure from motion for soil microtopography
  measurement,'' \emph{The Photogrammetric Record}, vol.~29, no. 147, pp.
  297--316, 2014.

\bibitem{J_ZXLiu_2012_ITGRS}
{Z. X. Liu, J. B. An and Y. Jing}, ``A simple and robust feature point matching
  algorithm based on restricted spatial order constraints for aerial image
  registration,'' \emph{IEEE Trans. Geosci. Remote Sens.}, vol.~50, no.~5, pp.
  514--527, May. 2012.

\bibitem{J_ZLS_2014_ITGRS}
{Z. L. Song, S. G. Zhou, and J. H. Guan}, ``A novel image registration
  algorithm for remote sensing under affine transformation,'' \emph{IEEE Trans.
  Geosci. Remote Sens.}, vol.~52, no.~8, pp. 4895--4912, Aug. 2014.

\bibitem{J_MZ_2015_IGRSL}
{M. Zhao, B. W. An, Y. P. Wu, B. Y. Chen and S. L. Sun}, ``A robust delaunay
  triangulation matching for multispectral/multidate remote sensing image
  registration,'' \emph{IEEE Geosci. Remote Sens Lett.}, vol.~12, no.~4, pp.
  711--715, Apr. 2015.

\bibitem{J_MI_2012_ITGRS}
{M. Izadi and P. Saeedi}, ``Robust weighted graph transformation matching for
  rigid and nonrigid image registration,'' \emph{IEEE Trans. Geosci. Remote
  Sens.}, vol.~21, no.~10, pp. 4369--4382, Oct. 2012.

\bibitem{J_WAr_2009_IVC}
{W. Aguilar, Y. Frauel, F. Escolano, M. E. Martinez-Perez, A. Espinosa-Romero,
  and M. A. Lozano}, ``A robust graph transformation matching for non-rigid
  registration,'' \emph{Image Vis. Comput.}, vol.~27, no.~7, pp. 897--910, Jun.
  2009.

\bibitem{R10}
{F. Maes, A. Collignon, D. Vandermeulen, G. Marchal, and P. Suetens},
  ``Multimodality image registration by maximization of mutual information,''
  \emph{IEEE Trans. Medical Imaging}, vol.~16, no.~2, pp. 187--198, Apr. 1997.

\bibitem{R12}
{R. J. Althof, M. G. J. Wind, and J. T. Dobbins}, ``A rapid and automatic image
  registration algorithm with subpixel accuracy,'' \emph{IEEE Trans. Pattern
  Anal. Mach. Intell.}, vol.~16, no.~3, pp. 308--316, Jun. 1997.

\bibitem{R9}
{J. L. Moigne, W. J. Campbell, and R. F. Cromp}, ``An automated parallel image
  registration technique based on the correlation of wavelet features,''
  \emph{IEEE Trans. Geosci. Remote Sens.}, vol.~40, no.~8, pp. 1849--1864, Aug.
  2002.

\bibitem{R11}
{J. M. Murphy, J. L. Moigne, and D. J. Harding}, ``Automatic image registration
  of multimodal remotely sensed data with global shearlet features,''
  \emph{IEEE Trans. Geosci. Remote Sens.}, vol.~54, no.~3, pp. 1685--1704, Mar.
  2016.

\bibitem{J_DGL_2004_IJCV}
{D. G. Lowe}, ``Distinctive image features from scale-invariant keypoints,''
  \emph{Int. J. Comput. Vis.}, vol.~60, no.~2, pp. 91--110, Nov. 2004.

\bibitem{J_KM_2005_ITPAMI}
{K. Mikolajzyk and C. Schmid}, ``A performance evaluation of local
  descriptors,'' \emph{IEEE Trans. Pattern Anal. Mach. Intell.}, vol.~27,
  no.~10, pp. 1615--1630, Oct. 2005.

\bibitem{J_FD_2015_ITGRS}
{F. Dellinger, J. Delon, Y. Gousseau, J. Michel, and F. Tupin}, ``{SAR-SIFT: A
  SIFT-Like algorithm for SAR images},'' \emph{IEEE Trans. Geosci. Remote
  Sens.}, vol.~53, no.~1, pp. 453--465, Jan. 2015.

\bibitem{J_SHW_2012_IGRSL}
{S. H. Wang, H. J. You and K. Fu}, ``{BFSIFT}: A novel method to find feature
  matches for sar image registration,'' \emph{IEEE Geosci. Remote Sens Lett.},
  vol.~9, no.~4, pp. 649--653, Oct. 2012.

\bibitem{J_AS_2015_ITGRS}
{A. Sedaghat and H. Ebadi}, ``Remote sensing image matching based on adaptive
  binning {SIFT} descriptor,'' \emph{IEEE Trans. Geosci. Remote Sens.},
  vol.~53, no.~10, pp. 5283--5293, Oct. 2015.

\bibitem{KASIFT}
{X.Z. Liu, Z. Tian, Q. Lua, L. Yang, and C. Y. Chai}, ``{A new affine invariant
  descriptor framework in shearlets domain for SAR image multiscale
  registration},'' \emph{AEU-International Journal of Electronics and
  Communications}, vol.~67, no.~9, pp. 743--753, 2013.

\bibitem{ASIFT}
{J. M. Morel and G. Yu}, ``{ASIFT}: A new framework for fully affine invariant
  image comparison,'' \emph{SIAM Journal on Imaging Sciences}, vol.~2, no.~2,
  pp. 438--469, 2009.

\bibitem{C_HB_PECCV}
{H. Bay, T. Tuytelaars, and L. V. Gool}, ``{SURF}: Speeded up robust
  features,'' in \emph{Proc. Eur. Conf. Comput. Vis.}, 2006, pp. 404--417.

\bibitem{J_AW_2007_ITGRS}
{A. Wong and D. A. Clausi}, ``{ARRSI}: Automatic registration of remote-sensing
  images,'' \emph{IEEE Trans. Geosci. Remote Sens.}, vol.~45, no.~5, pp.
  1483--1493, May. 2007.

\bibitem{J_MAFichler_1081_CACM}
{M. A. Fichler and R. C. Bolles}, ``Random sample consensus: a paradigm for
  model fitting with applications to image analysis and automated
  cartography,'' \emph{Commun. ACM}, vol.~24, no.~6, pp. 381--395, Jun. 1981.

\bibitem{J_PB_1992_ITPAMI}
{P. Besl and N. Mckay}, ``A method for registration of 3-d shapes,'' \emph{IEEE
  Trans. Pattern Anal. Mach. Intell.}, vol.~14, no.~2, pp. 239--256, Feb. 1992.

\bibitem{J_HC_2003_CVIU}
{H. Chui and A. Rangarajan}, ``A new point matching algorithm for nonrigid
  registration,'' \emph{Comput. Vis. Image Understanding}, vol.~89, no. 2-3,
  pp. 114--141, Feb. 2003.

\bibitem{J_AM_2010_ITPAMI}
{A. Myronenko and X. B. Song}, ``Point set registration: Coherent point
  drift,'' \emph{IEEE Trans. Pattern Anal. Mach. Intell.}, vol.~32, no.~12, pp.
  2262--2275, Dec. 2010.

\bibitem{J_BJ_2011_ITPAMI}
{B. Jian and B. C. Vemuri}, ``Robust point set registration using gaussian
  mixture models,'' \emph{IEEE Trans. Pattern Anal. Mach. Intell.}, vol.~33,
  no.~8, pp. 1633--1645, Aug. 2011.

\bibitem{J_SB_2002_ITPAMI}
{S. Belongie, J. Malik, and J. Puzicha}, ``Shape matching and object
  recognition using shape contexts,'' \emph{IEEE Trans. Pattern Anal. Mach.
  Intell.}, vol.~24, no.~24, pp. 509--522, Apr. 2002.

\bibitem{J_YZ_2006_ITPAMI}
{Y. Zheng and D. Doermann}, ``Robust point matching for nonrigid shapes by
  preserving local neighborhood structures,'' \emph{IEEE Trans. Pattern Anal.
  Mach. Intell.}, vol.~28, no.~4, pp. 643--649, Apr. 2006.

\bibitem{J_MZ_2013_IGRSL}
{M. Zhao, B. W. An, Y. P. Wu, and C. Q. Lin}, ``{Bi-SOGC}: A graph matching
  approach based on bilateral {KNN} spatial orders around geometric centers for
  remote sensing image registration,'' \emph{IEEE Geosci. Remote Sens Lett.},
  vol.~10, no.~6, pp. 1429--1434, Nov. 2013.

\bibitem{J_MZ_2014_JIMW}
{M. Zhao, B. W. An, T. Z. Wang, Y. Y. Xu, C. Q. Lin, and S. L. Sun}, ``{A graph
  matching algorithm based on filtering strategy of Bi-directional K Nearest
  Neighbors},'' \emph{J. Infrared Millim. Waves}, vol.~33, no.~1, pp. 78--89,
  Feb. 2014.

\bibitem{C_YM_PWCSE}
Y.~Ma and Q.~Chen, ``{Fast Delaunay Triangulation and Voronoi diagram
  generation on the sphere},'' in \emph{Proc. WCSE}, 2010, pp. 187--190.

\bibitem{J_SU_1991_ITPAMI}
S.~Umeyama, ``Least-squares estimation of transformation parameters between two
  point patterns,'' \emph{IEEE Trans. Pattern Anal. Mach. Intell.}, vol.~13,
  no.~4, pp. 376--380, Apr. 1991.

\bibitem{J_GJB_2009_CVIU}
{G. J. Burghouts and J. M. Geusebroek}, ``Performance evaluation of local
  colour invariants,'' \emph{Compute. Vis. Image Understanding}, vol. 113,
  no.~1, pp. 48--62, Jan. 2009.

\bibitem{C_JB_PCVPR}
J.~Beis and D.~G. Lowe, ``Shape indexing using approximate nearest-neighbour
  search in highdimensional spaces,'' in \emph{Proc. Computer Vision and
  Pattern Recognition}, vol.~2, 1997, pp. 1000--1006.

\end{thebibliography}

\begin{IEEEbiography}[{\includegraphics[width=1in,height=1.25in,clip,keepaspectratio]{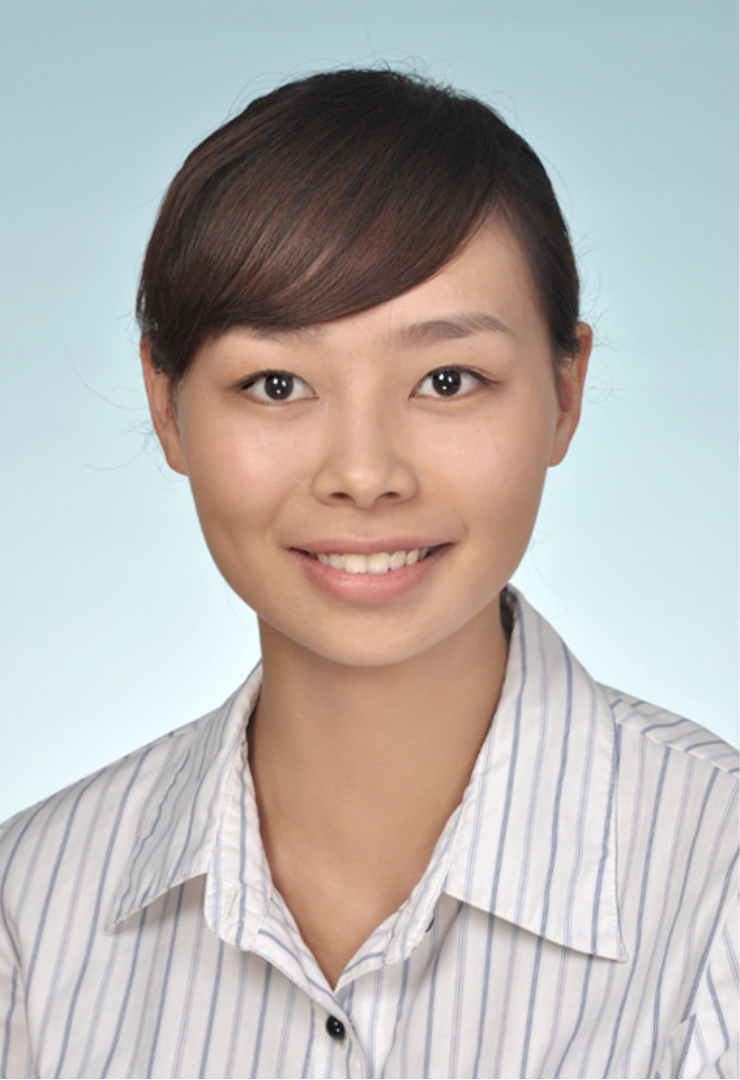}}]
{Ming Zhao}
received the B.S. degree in telecommunication engineering from Wuhan University, Wuhan, China, in July 2007,
the Ph.D. degree in physical electronics with Shanghai Institute of Technical Physics, the institute of Chinese Academy of Sciences (CAS),
Shanghai, China, in June 2012.

She is currently an associate professor with Shanghai Maritime University, China.
From 2015 to 2016, she was a Visiting Scholar with Friedrich-Alexander University Erlangen-N$\ddot{u}$rnberg, Germany.
Her research interests include remote sensing image processing, pattern recognition, and computer vision.
\end{IEEEbiography}

\begin{IEEEbiography}[{\includegraphics[width=1in,height=1.25in,clip,keepaspectratio]{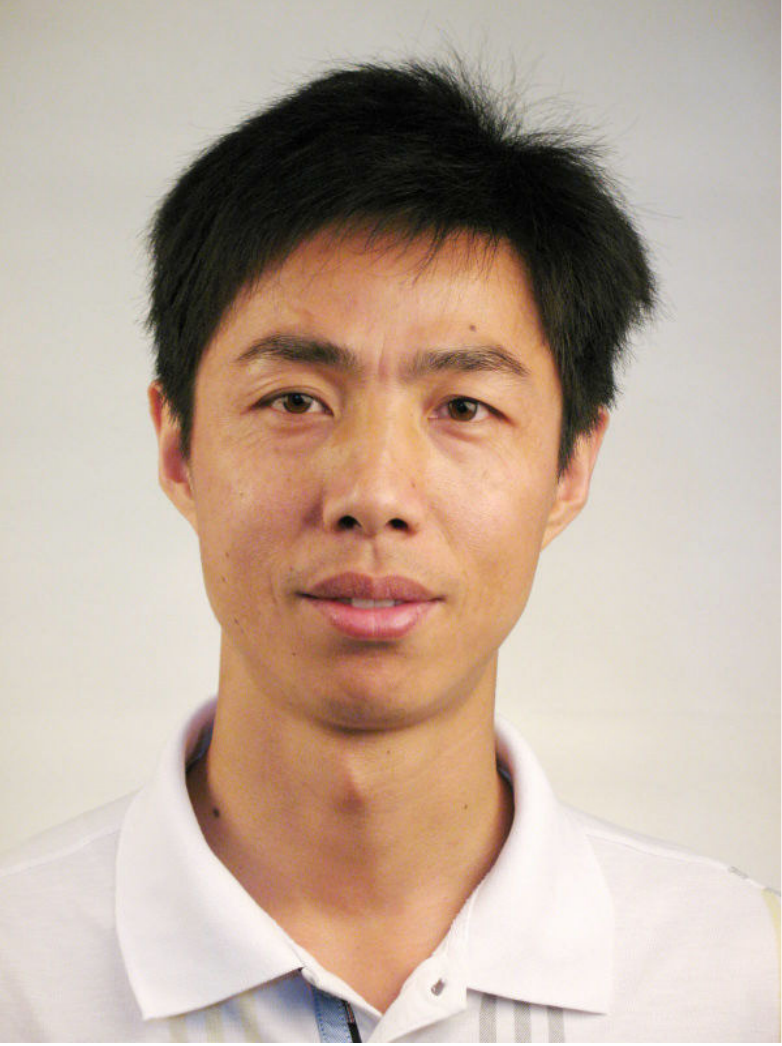}}]
{Bowen An}
received the M.S. degree in communication and information engineering from Wuhan University, Wuhan, China, in April 2004,
the Ph.D. degree in circuits and systems with Shanghai Institute of Technical Physics, the institute of Chinese Academy of Sciences (CAS),
Shanghai, China, in July 2006.

He is currently a full professor with Shanghai Maritime University, China.
His research interests include photoelectric signal acquisition and remote sensing image processing.
\end{IEEEbiography}

\begin{IEEEbiography}[{\includegraphics[width=1in,height=1.25in,clip,keepaspectratio]{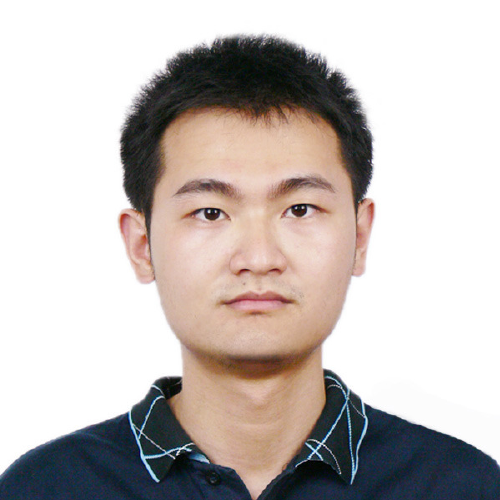}}]
{Yongpeng Wu} (S'08--M'13)
received the B.S. degree in telecommunication engineering from Wuhan University, Wuhan, China, in July 2007,
the Ph.D. degree in communication and signal processing
with the National Mobile Communications Research Laboratory,
Southeast University, Nanjing, China, in November 2013.

He is currently a Humboldt post-doc research fellow with Institute for Digital Communications, Universit\"{a}t Erlangen-N\"{u}rnberg,
Germany.  During his doctoral studies, he has conducted cooperative
research at the Department of Electrical Engineering, Missouri University of Science and Technology, USA. His
research interests include MIMO systems, signal processing for wireless communications, and
multivariate statistical theory.
\end{IEEEbiography}

\begin{IEEEbiography}[{\includegraphics[width=1in,height=1.25in,clip,keepaspectratio]{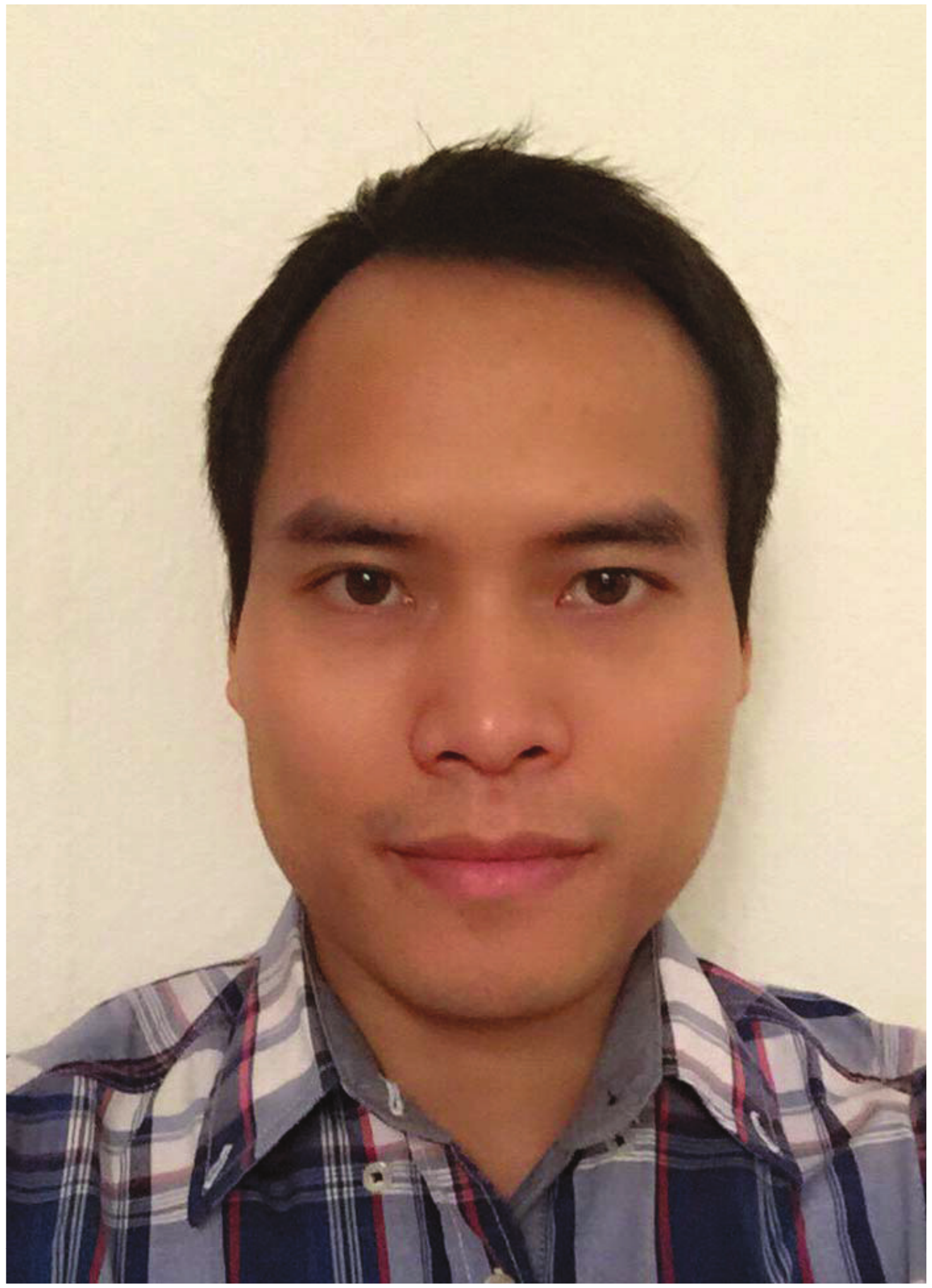}}]
{Huynh Van Luong}
received the M.Sc. degree incomputer engineering from the University of Ulsan,Ulsan, Korea, in 2009 and the Ph.D. degree from the Technical University of Denmark, KongensLyngby, Denmark, in 2013. He is currently a Humboldt-research fellow from the Alexander von Humboldt foundation with the Multimedia Communications and Signal Processing, University of Erlangen-Nuremberg, Erlangen, Germany.

His research interests include signal recovery, compressed sensing, image and video processing, distributed source coding, and visual communications.
\end{IEEEbiography}

\begin{IEEEbiography}[{\includegraphics[width=1in,height=1.25in,clip,keepaspectratio]{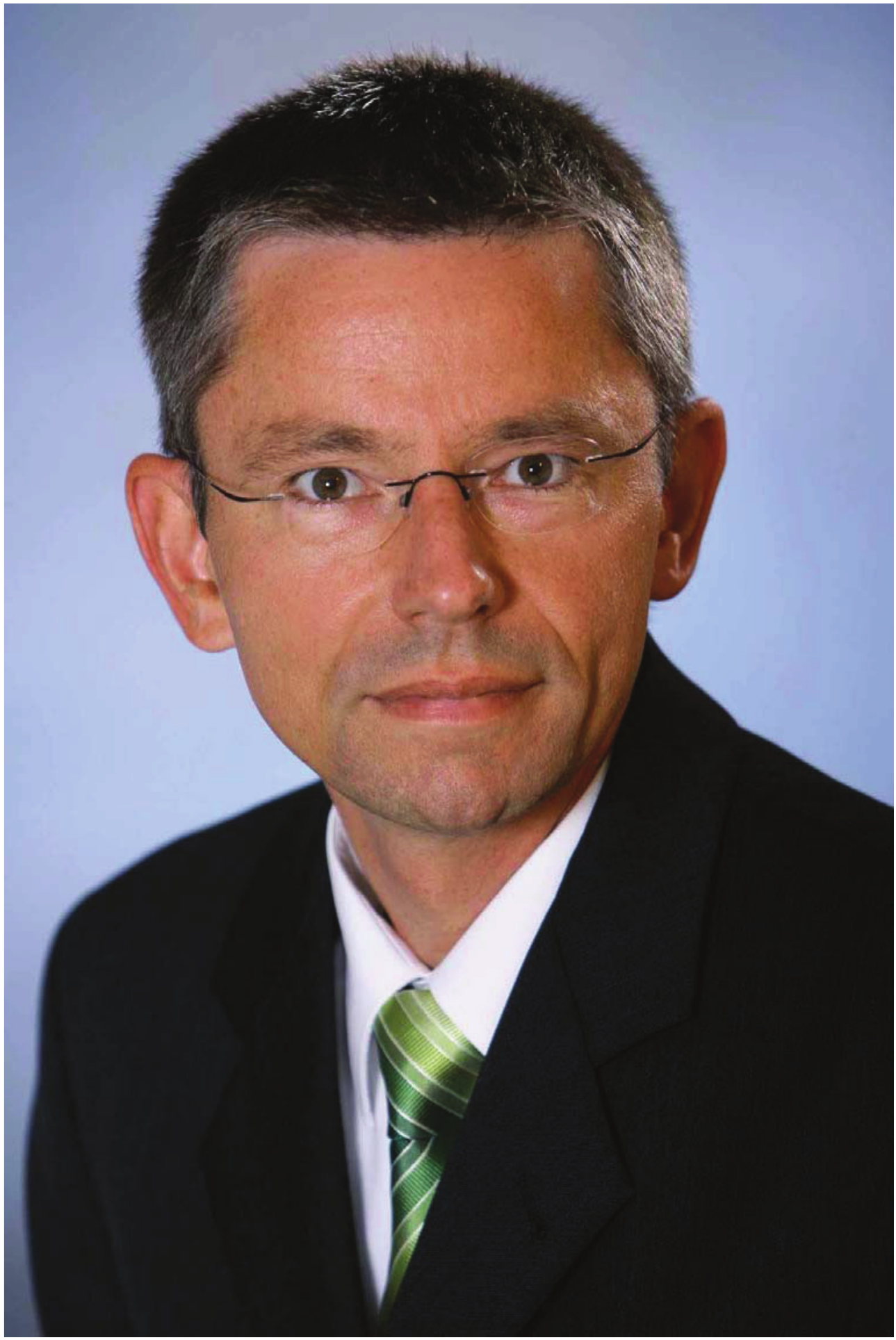}}]
{Andr\'{e} Kaup} (M'96--SM'99--F'13)
received the Dipl.-Ing. and Dr.-Ing. degrees in electrical engineering from Rheinisch-Westf\"{a}lische Technische Hochschule (RWTH) Aachen University, Aachen, Germany, in 1989 and 1995, respectively.

He was with the Institute for Communication Engineering, RWTH Aachen University, from 1989 to 1995. He joined the Networks and Multimedia Communications Department, Siemens Corporate Technology, Munich, Germany, in 1995 and became Head of the Mobile Applications and Services Group in 1999. Since 2001 he has been a Full Professor and the Head of the Chair of Multimedia Communications and Signal Processing, University of Erlangen-Nuremberg, Erlangen, Germany. From 1997 to 2001 he was the Head of the German MPEG delegation. From 2005 to 2007 he was a Vice Speaker of the DFG Collaborative Research Center 603. Since 2015 he serves as Head of the Department of Electrical Engineering and Vice Dean of the Faculty of Engineering. He has authored more than 300 journal and conference papers and has over 50 patents granted and patent applications published. His research interests include image and video signal processing and coding, and multimedia communication.

Dr. Kaup is a member of the IEEE Multimedia Signal Processing Technical Committee, a member of the scientific advisory board of the German VDE/ITG, and a Fellow of the IEEE. He serves as an Associate Editor for IEEE Transaction on Circuits and Systems for Video Technology and was a Guest Editor for IEEE Journal of Selected Topics in Signal Processing. From 1998 to 2001 he served as an Adjunct Professor with the Technical University of Munich, Munich. He was a Siemens Inventor of the Year 1998 and received the 1999 ITG Award. He has received several best paper awards, including the Paul Dan Cristea Special Award from the International Conference on Systems, Signals, and Image Processing in 2013. His group won the Grand Video Compression Challenge at the Picture Coding Symposium 2013 and he received the teaching award of the Faculty of Engineering in 2015.

\end{IEEEbiography}

\newpage

\end{document}